\RequirePackage{lineno}
\documentclass[onecolunm,superscriptaddress,preprintnumbers,reprint,amsmath,amssymb,prc]{revtex4}  
\usepackage{graphicx}
\usepackage{dcolumn}
\usepackage{bm}
\usepackage{float}
\usepackage{xcolor}
\usepackage{CJK}
\usepackage[colorlinks,linkcolor=blue,urlcolor=blue,citecolor=blue]{hyperref}
\usepackage{isotope}
\usepackage[normalem]{ulem}
\renewcommand{\sout}{\bgroup \color{red} \ULdepth=-0.5ex \ULset}

\usepackage{amsmath}
\usepackage{cases}
\usepackage{amssymb}
\usepackage{eqnalign}

\begin{document}

\title{Effect of transverse momentum conservation and flow on symmetric cumulants $sc_{2,3} \left \{ 4 \right \}$ and $sc_{2,3,4} \left \{ 6 \right \}$ }

\author{Jia-Lin Pei}
\affiliation{Key Laboratory of Nuclear Physics and Ion-beam Application~(MOE), Institute of Modern Physics, Fudan University, Shanghai $200433$, China}
\affiliation{Shanghai Research Center for Theoretical Nuclear Physics, NSFC and Fudan University, Shanghai $200438$, China}

\author{Guo-Liang Ma}
\email{glma@fudan.edu.cn}
\affiliation{Key Laboratory of Nuclear Physics and Ion-beam Application~(MOE), Institute of Modern Physics, Fudan University, Shanghai $200433$, China}
\affiliation{Shanghai Research Center for Theoretical Nuclear Physics, NSFC and Fudan University, Shanghai $200438$, China}

\author{Adam Bzdak}
\email{bzdak@fis.agh.edu.pl}
\affiliation{AGH University of Science and Technology,\\
Faculty of Physics and Applied Computer Science,
30-059 Krak\'ow, Poland}

\begin{abstract}
Symmetric cumulants can improve our understanding of the joint probability distribution function $ P\left ( v_{m},v_{n},v_{k}, \dots,\Psi _{m},\Psi _{n},\Psi _{k},\dots \right )$, potentially offering new insights into the nature of the fluctuations of the quark-gluon plasma produced in relativistic heavy-ion collisions. In this work, the four-particle symmetric cumulants $sc_{2,3} \left \{ 4 \right \}$, six-particle symmetric cumulants $sc_{2,3,4} \left \{ 6 \right \}$, and the normalized cumulants $nsc_{2,3} \left \{ 4 \right \}$ and $nsc_{2,3,4} \left \{ 6 \right \}$ originating from transverse momentum conservation, collective flow, and the interplay between the two effects are calculated. Our results on $sc_{2,3} \left \{ 4 \right \}$ are consistent with the ATLAS data using the subevent cumulant method, facilitating a more profound understanding of the origins of the symmetric cumulant in small systems. Our results on $sc_{2,3,4} \left \{ 6 \right \}$ serve as theoretical predictions for future experimental measurements in small systems.
\end{abstract}
\maketitle

\section{Introduction}
The new state of matter, quark-gluon plasma (QGP), can be produced by relativistic heavy ion collisions at Relativistic Heavy-Ion Collider and the Large Hadron Collider (LHC) due to the asymptotical freedom nature of strong interactions \cite{LHC1,LHC2,LHC3,LHC4,LHC5}. The experimental discovery of collective flow (i.e., the collective correlation of azimuthal angles in the transverse plane between outgoing particles) suggests that the QGP matter is a nearly perfect fluid, resulting from the expansion of the geometrical asymmetric shape driven by the pressure gradient \cite{flow1,flow2,flow3,flow4}. In this way, the study of collective flow not only contributes to the exploration of the nature of the QGP matter \cite{property1,property2}, but also helps to investigate the properties of the nuclei, especially the deformation and skin of the nuclei \cite{skin1,skin2}.  
For the initial state,  the transverse positions of participant nucleons in the overlap region can fluctuate since the nuclear density of the nuclei is described by the wave function with the Woods-Saxon distribution \cite{wood1}. Besides, the CGC effective field theory also suggests that the fluctuating substructure of the proton is necessary to understand the collectivity of small systems \cite{sub}.
These fluctuations result in collective flow up to at least sixth order in the final state, which can be quantified by the Fourier expansion of the azimuthal distribution of final particles, $\frac{\mathrm{d} N}{\mathrm{d} \phi } 
\propto 
1+ {\textstyle \sum_{n}} 
v_{n}\cos[n(\phi -\Psi _{n} )] $, where the collective flow coefficients $v_{2}$, $v_{3}$, and $v_{4}$ are elliptic, triangular, and quadrangular flows with respect to their corresponding $\Psi _{n}$, respectively \cite{6th1,6th2,6th3}.

Because the number of participating nucleons is finite, the initial energy density event-by-event (EbyE) fluctuates, resulting in EbyE fluctuations of the geometry eccentricities $\vec{\varepsilon } _{n}$ in high energy nucleus-nucleus (\textit{A}+\textit{A}) collisions \cite{ebye1,ebye2,ebye3,ebye4}. According to hydrodynamics, flow harmonics $\vec{v}_{n}$ exhibit a relationship with the eccentricities $\vec{\epsilon }_{n}$. Specifically, for lower orders of flow harmonics $\vec{v}_{n}$ ($n\le 3$), there is a linear response between the eccentricities $\vec{\epsilon }_{n}$ and $\vec{v}_{n}$, i.e., $\vec{v}_{n} \propto \vec{\epsilon }_{n}$. However, for the higher orders of flow harmonics $\vec{v}_{n}$ (for $n > 3$), there is not only the linear response to the eccentricity of the corresponding order but also a nonlinear response to lower order $\vec{\epsilon }_{2}$ or/and $\vec{\epsilon }_{3}$ \cite{response1,response2,response3}. Consequently, the EbyE fluctuations of $\vec{\epsilon }_{n}$ lead to EbyE fluctuations and correlations of $\vec{v }_{n}$ \cite{response4,response5,response6}. 
For this reason, the joint distribution function $ P\left ( v_{m},v_{n},v_{k}, \dots,\Psi _{m},\Psi _{n},\Psi _{k},\dots \right )$ is an ideal experimental observable because it carries not only the information about the initial energy density fluctuations but also the mechanisms of final state interactions \cite{joint1,joint2,joint3,joint4}. For example, the measurements of the integrated joint distribution function, like the probability distribution of a single flow harmonic $P(v_2)$ based on multiparticle cumulants, indicate that the ratio of the shear viscosity to the entropy density $\eta /s$ of hot and dense QGP matter is close to the minimum value of $1/4\pi$ \cite{fluid1,fluid2,fluid3}.  
Furthermore, the experimental observables based on the joint distribution function are more advanced for more precise measurements. For instance, previous studies have found that symmetric cumulants $sc_{m,n}\left \{ 4 \right \} $ representing joint distribution function between different order flow $P(v_m,v_n)$ can be used as a sensitive probe to study both the initial condition of the QGP and the temperature dependence of   $\eta /s$ \cite{scprobe1,scprobe2,scprobe3}.
The first experimental measurements of the $sc_{m,n}\left \{ 4 \right \} $ can be found in Ref. \cite{first1}.  
For example, experimental researches on $sc_{2,3}\left \{ 4 \right \} $ have demonstrated that the sign and magnitude of $sc_{2,3}\left \{ 4 \right \} $ align with those of the kurtosis in triangular flow and negative quantities suggest that the energy distribution around a given point is positively skewed \cite{response1,first1,sc231}. Furthermore, higher orders of symmetric cumulants such as $sc_{2,3,4} \left \{ 6 \right \}$ can offer additional and independent constraints on the initial conditions and characteristics of QGP in high-energy nuclear collisions \cite{threevn1}. 
This is because since the elliptic shape only can provide the correlation between the even orders of flow harmonics, $v_2$ and $v_4$, any nonzero value of $sc_{2,3,4} \left \{ 6 \right \}$ indicates the presence of additional sources of fluctuations in the system that couple all the three flow harmonics $v_2$, $v_3$, and $v_4$ \cite{threevn2}. In this sense, further exploration of $sc_{2,3}\left \{ 4 \right \} $ and $sc_{2,3,4} \left \{ 6 \right \}$ can enhance our comprehension of the joint probability distribution function and thus could provide fresh perspectives on the nature of the fluctuations of the QGP generated by heavy-ion collisions.

However, understanding of the long-range (in rapidity) azimuthal correlations of particles in small systems (\textit{p}+\textit{p} or \textit{p}+\textit{A}) still awaits further deepening \cite{small1,small2}. At present, theoretical investigations concerning the origin of long-range correlations can be primarily divided into two categories. The first category is thought to result from initial effects. Based on the color glass condensation (CGC) framework, it is found that two-particle long-range correlations and Fourier $v_n$ coefficients can be generated by the emission mechanism of two gluons in the glasma Feynman diagrams \cite{cgc1,cgc2,cgc3}. The second category is considered to be the result of final-state interactions. For example, hydrodynamic models convert an initial geometric asymmetry into a final-state momentum anisotropic flow through a pressure gradient that satisfies the energy-momentum conservation equation of the fluid created by small collisions \cite{final1,final2,final3}. Nevertheless, the applicability of hydrodynamics in small collision systems is presently contentious, given their notably smaller size and shorter lifetimes relative to those in large collision systems. Another example is a multiphase transport model (AMPT) based on the Boltzmann transport equation, which simulates the final-state interactions by parton scattering and describes well the experimental results in small and large collision systems \cite{ampt1,ampt2}.  Moreover, physical conservation laws are an obvious source for generating azimuthal correlations between particles such as transverse momentum conservation (TMC) \cite{law1,law2,law3,law4,law5}, since it provides a natural constraint to the momentum space for the produced particles. Our previous studies have calculated multiparticle cumulants from TMC indicating that multiparticle cumulants are inversely proportional to the number of particles $N$, reflecting the property of TMC \cite{pretmc1,pretmc2,pretmc3}. In our recent study, the four-particle symmetric cumulant $sc_{2,4} \left \{ 4 \right \}$ and three-particle asymmetric cumulant $ac_{2} \left \{ 3 \right \}$ from TMC and flow have been calculated, and by comparing them with the experimental data, it has been shown that when the number of particles is small, the correlation comes from the TMC, and when the number of particles is large, the collective flow is dominant, which is in good agreement with the ATLAS measurement in \textit{p}+\textit{p} collisions at the LHC \cite{pretmc4}. Similarly, there are also experimental measurements regarding $sc_{2,3} \left \{ 4 \right \}$ as well as the model prediction concerning $sc_{2,3,4} \left \{ 6 \right \}$. Consequently, one can also utilize $sc_{2,3} \left \{ 4 \right \}$ and $sc_{2,3,4} \left \{ 6 \right \}$ as more probes to further enhance our understanding of the origins of the particle correlations in small systems.

In this paper, we calculate the four-particle symmetric cumulants $sc_{2,3} \left \{ 4 \right \}$, six-particle symmetric cumulants $sc_{2,3,4} \left \{ 6 \right \}$, and the normalized cumulants $nsc_{2,3} \left \{ 4 \right \}$ and $nsc_{2,3,4} \left \{ 6 \right \}$ based on transverse momentum conservation and collective flow. Compared to the recent ATLAS experimental measurements with the subevent method suppressing nonflow
contribution from jets \cite{data1}, we conclude that the symmetric cumulant $sc_{2,3} \left \{ 4 \right \}$ in small systems originate from the combined effects of TMC and collective flow. Given the absence of direct experimental measurements for $sc_{2,3,4} \left \{ 6 \right \}$ in small systems, our computational results are presented as theoretical predictions.

\section{$sc_{2,3} \left \{ 4 \right \}$ and $sc_{2,3,4} \left \{ 6 \right \}$ from transverse momentum conservation}

In this section, we denote by $\vec{p }_{1},...,\vec{p }_{N}$ the transverse momenta of $N$ particles produced in the collision. For simplicity, we only consider transverse momentum conservation, i.e., $\vec{p }_{1}+\dots +\vec{p }_{N}=0$, as the only source of azimuthal correlation between the final particles.  
Then we determine how the TMC affects the joint probability distributions of the final particles. The $k$-particle ($k< N$)  probability distribution under the TMC constraint is defined as \cite{law3,tmcdef1,law4,tmcdef2,tmcdef3}:
\begin{equation}
f_{k}(\vec{p}_{1},\dots,\vec{p}_{k})\equiv 
\frac{ \left ( \displaystyle \prod_{i=1}^{k}f\left ( \vec{p}_{i} \right ) \right )    
\displaystyle\int  \delta ^{2}\left ( \vec{p }_{1}+\dots +\vec{p }_{N} \right )
\displaystyle \prod_{i=k+1}^{N}\left (f\left ( \vec{p}_{i} \right )d^{2}\vec{p}_{i}  \right ) }
{\displaystyle\int  \delta ^{2}\left ( \vec{p }_{1}+\dots +\vec{p }_{N} \right )
\displaystyle \prod_{i=1}^{N}\left (f\left ( \vec{p}_{i} \right )d^{2}\vec{p}_{i}  \right )} , 
\label{eq1}
\end{equation}
where $f\left ( \vec{p}_{i} \right )$ is the single-particle transverse momentum distribution. Obviously, the Dirac $\delta$ in the equation represents the restriction of TMC and without it, the joint probability distribution would simply factorize. Furthermore, the average value of transverse momentum $\vec{p}$ over the full phase space $F$ is clearly zero:
\begin{equation}
\left \langle \vec{p}\right \rangle_{F} \equiv 
\int \vec{p}f(\vec{p})d^{2}\vec{p}= 0.
\end{equation}

To obtain a concrete form of the joint distribution function $f(\vec{p}_{1},\dots,\vec{p}_{k})$ defined in Eq.~(\ref{eq1}), we define a new variable $\vec{P}\equiv \sum_{i=1}^{M}\vec{p}_{i}$, the sum of the momenta of the $M$ particles in the system, which satisfies the Gaussian distribution according to the central limit theorem, i.e.:
\begin{align}
F_{M}(\vec{P})&\equiv \displaystyle\int  \delta ^{2}\left ( -\vec{P}+\sum_{i=1}^{M}\vec{p}_{i} \right )
\displaystyle \prod_{i=1}^{M}\left (f\left ( \vec{p}_{i} \right )d^{2}\vec{p}_{i}  \right ) \notag\\
&= \frac{1}{\pi\sigma ^{2} }\text{exp}\bigg( -\frac{\vec{P}^{2} }{\sigma ^{2}} \bigg) ,
\label{eq3}
\end{align}
where the mean and variance are
\begin{align}
\left \langle \vec{P} \right \rangle_{F} &=\left \langle \sum_{i=1}^{M}\vec{p}_{i} \right \rangle_{F}
= \sum_{i=1}^{M}\left \langle \vec{p}_{i} \right \rangle_{F} =0 ,\notag\\
\sigma ^{2}&=M(\left \langle \left | \vec{p} \right |^{2}   \right \rangle_{F}-\left \langle \vec{p} \right \rangle^{2}_{F})
= M\left \langle \left | \vec{p} \right |^{2}   \right \rangle_{F},
\label{eq4}
\end{align}
and
\begin{equation}
   \left \langle \left | \vec{p} \right | ^{2}\right \rangle_{F} =
\frac{\int \left | \vec{p} \right |^{2}f(\vec{p})d^{2}\vec{p}}{\int f(\vec{p})d^{2}\vec{p}}. 
\end{equation}

Based on Eqs.~\eqref{eq3} and~\eqref{eq4}, Eq.~\eqref{eq1} can be rewritten as follows:
\begin{align}
f_{k}(\vec{p}_{1},\dots,\vec{p}_{k})&= \left ( \displaystyle \prod_{i=1}^{k}f\left ( \vec{p}_{i} \right ) \right )
\frac{F_{N-k}\displaystyle\left (- \sum_{i=1}^{k}\vec{p}_{i} \right )   }
{F_{N}\displaystyle\left ( \sum_{i=1}^{N}\vec{p}_{i} \right )} \notag\\
&=\left ( \displaystyle \prod_{i=1}^{k}f\left ( \vec{p}_{i} \right ) \right )
\frac{N}{N-k} \text{exp}\left ( -\frac{ \left ( \sum_{i=1}^{k}\vec{p}_{i}  \right )^{2} }
{(N-k)\left \langle \left | \vec{p} \right |^{2}  \right \rangle_{F} }  \right ).
\label{eq6}
\end{align}

Our aim is to calculate the four-particle symmetric cumulant $sc_{2,3}\left \{ 4 \right \} $ and the six-particle symmetric cumulant $sc_{2,3,4}\left \{ 6 \right \} $. The four-particle symmetric cumulant and six-particle symmetric cumulant are defined as follows: 
\begin{align}
 sc_{2,3}\left \{ 4 \right \}&= \left \langle \left \langle e^{i2\left ( \phi _{1}-\phi _{2}  \right )+i3\left ( \phi _{3}-\phi _{4}  \right ) }  \right \rangle \right \rangle - 
\left \langle \left \langle e^{i2\left ( \phi _{1}-\phi _{2}  \right ) }  \right \rangle \right \rangle \left \langle \left \langle e^{i3\left ( \phi _{3}-\phi _{4}  \right ) }  \right \rangle \right \rangle ,
\label{eq7}
\end{align}
\begin{align}
\hspace{1.5cm}sc_{2,3,4}\left \{ 6 \right \}&=\left \langle \left \langle e^{i(2 \phi _{1}+3\phi _{2}  +4\phi _{3})
-i(2 \phi _{4}+3\phi _{5}  +4\phi _{6}) }  \right \rangle \right \rangle \notag \\
&-\left \langle \left \langle e^{i(2 \phi _{1}+3\phi _{2} )
-i(2 \phi _{3}+3\phi _{4}  ) }  \right \rangle \right \rangle
\left \langle \left \langle e^{i(4 \phi _{5}-4\phi _{6} )
 }  \right \rangle \right \rangle \notag \\
&-\left \langle \left \langle e^{i(2 \phi _{1}+4\phi _{2} )
-i(2 \phi _{5}+4\phi _{6}  ) }  \right \rangle \right \rangle
\left \langle \left \langle e^{i(3 \phi _{3}-3\phi _{4} )
 }  \right \rangle \right \rangle \notag \\
&-\left \langle \left \langle e^{i(3 \phi _{3}+4\phi _{4} )
-i(3 \phi _{5}+4\phi _{6}  ) }  \right \rangle \right \rangle
\left \langle \left \langle e^{i(2 \phi _{1}-2\phi _{2} )
 }  \right \rangle \right \rangle \notag \\
 &+2\left \langle \left \langle e^{i2\left ( \phi _{1}-\phi _{2}  \right ) }  \right \rangle \right \rangle
 \left \langle \left \langle e^{i3\left ( \phi _{3}-\phi _{4}  \right ) }  \right \rangle \right \rangle
\left \langle \left \langle e^{i4\left ( \phi _{5}-\phi _{6}  \right ) }  \right \rangle \right \rangle.
\label{eq8}
\end{align}

\subsection{$sc_{2,3} \left \{ 4 \right \}$}
For four particles, using Eq.~(\ref{eq6}) we have
\begin{equation}
f_{4}(\vec{p}_{1},\dots,\vec{p}_{4})=f(\vec{p}_{1})\cdots f(\vec{p}_{4})
\frac{N}{N-4}
\mathrm{exp}\bigg(-\frac{p_{1}^{2}+p_{2}^{2}+p_{3}^{2}+p_{4}^{2} }{(N-4)\left \langle p^{2} \right \rangle_{F}} \bigg)\mathrm{exp}(-\Phi ),
\end{equation}
where
\begin{equation}
\Phi =\frac{2}{(N-4)\left \langle p^{2}  \right \rangle_{F} }
\sum_{i,j=1;i< j}^{4}
p_{i}p_{j}\cos (\phi_{i}-\phi_{j} ),
\end{equation}
and $p_{i}=|\vec{p}_{i}|$.

Thus we can use the following integral to calculate  $\left \langle e^{i2\left ( \phi _{1}-\phi _{2}  \right )+i3\left ( \phi _{3}-\phi _{4}  \right ) }  \right \rangle$ at a given transverse momenta $p_{1},p_{2},p_{3},$ and $p_{4}$:
\begin{equation}
\left \langle e^{i2\left ( \phi _{1}-\phi _{2}  \right )+i3\left ( \phi _{3}-\phi _{4}  \right ) }  \right \rangle
|p_{1},p_{2},p_{3},p_{4} = 
\frac{\int_{0}^{2\pi}e^{i2\left ( \phi _{1}-\phi _{2}  \right )+i3\left ( \phi _{3}-\phi _{4}  \right ) } 
\mathrm{exp}(-\Phi )d\phi _{1}\cdots d\phi _{4} }
{\int_{0}^{2\pi}
\mathrm{exp}(-\Phi )d\phi _{1}\cdots d\phi _{4}}.
\label{eq11}
\end{equation}
To calculate this integral, we carry out a Taylor expansion of $\mathrm{exp}(-\Phi )$. For the numerator, we expand it to the seventh order, and all higher-order terms are omitted since they are diminished by the higher powers of $N$.
For the denominator, we simply take the first term, i.e., $\mathrm{exp}(-\Phi)\approx 1$, in which case the integral value is $\left ( 2\pi  \right )^{4}$. For simplicity, we assume that all transverse momenta $p_{i}$ are equal. 
Accordingly, we obtain
\begin{equation}
\left \langle e^{i2\left ( \phi _{1}-\phi _{2}  \right )+i3\left ( \phi _{3}-\phi _{4}  \right ) }  \right \rangle
|p \approx -\frac{5p^{10}}{6(N-4)^{5}\left \langle p^{2} \right \rangle_{F}^{5}}+\frac{5p^{12}}{2(N-4)^{6}\left \langle p^{2} \right \rangle_{F}^{6}}-\frac{827p^{14}}{144(N-4)^{7}\left \langle p^{2} \right \rangle_{F}^{7}}.
\label{eq12}
\end{equation}
Carrying out calculations by analogy, we obtain
\begin{equation}
\left \langle e^{i2\left ( \phi _{1}-\phi _{2}  \right ) }  \right \rangle
|p \approx
\frac{p^{4}}{2(N-2)^{2}\left \langle p^{2} \right \rangle_{F}^{2}},
\label{eq13}
\end{equation}
and
\begin{equation}
\left \langle e^{i3\left ( \phi _{3}-\phi _{4}  \right ) }  \right \rangle
|p \approx
-\frac{p^{6}}{6(N-2)^{3}\left \langle p^{2} \right \rangle_{F}^{3}}.
\label{eq14}
\end{equation}
Using Eqs.~(\ref{eq7}), ~(\ref{eq12}), ~(\ref{eq13}), and~(\ref{eq14}) we find
\begin{equation}
sc_{2,3} \left \{ 4 \right \}\approx 
-\frac{5p^{10}}{6(N-4)^{5}\left \langle p^{2} \right \rangle_{F}^{5}}+\frac{5p^{12}}{2(N-4)^{6}\left \langle p^{2} \right \rangle_{F}^{6}}-\frac{827p^{14}}{144(N-4)^{7}\left \langle p^{2} \right \rangle_{F}^{7}}+\frac{p^{10}}{12(N-2)^{5}\left \langle p^{2} \right \rangle_{F}^{5}}.
\label{eq15}
\end{equation}

\subsection{$sc_{2,3,4} \left \{ 6 \right \}$}
For six particles, using Eq.~(\ref{eq6}) we have
\begin{equation}
f_{6}(\vec{p}_{1},\dots,\vec{p}_{6})=f(\vec{p}_{1})\cdots f(\vec{p}_{6})
\frac{N}{N-6}
 \mathrm{exp}\bigg(-\frac{p_{1}^{2}+p_{2}^{2}+p_{3}^{2}+p_{4}^{2}+p_{5}^{2}+p_{6}^{2} }{(N-6)\left \langle p^{2} \right \rangle_{F}} \bigg)\mathrm{exp}(-\Phi ),
\end{equation}
where
\begin{equation}
\Phi =\frac{2}{(N-6)\left \langle p^{2}  \right \rangle_{F}  }
\sum_{i,j=1;i< j}^{6}
p_{i}p_{j}\cos (\phi_{i}-\phi_{j} ).
\end{equation}
In the same way, we receive
\begin{align}
  \left \langle e^{i(2 \phi _{1}+3\phi _{2}  +4\phi _{3})
-i(2 \phi _{4}+3\phi _{5}  +4\phi _{6}) }  \right \rangle |p & \approx -\frac{35p^{18}}{8(N-6)^{9}\left \langle p^{2} \right \rangle_{F}^{9}} + \frac{1015p^{20}}{24(N-6)^{10}\left \langle p^{2} \right \rangle_{F}^{10}}-\frac{470037645p^{22}}{1995844(N-6)^{11}\left \langle p^{2} \right \rangle_{F}^{11}} \notag \\
&+\frac{92573p^{24}}{96(N-6)^{12}\left \langle p^{2} \right \rangle_{F}^{12}}-\frac{4601851p^{26}}{1440(N-6)^{13}\left \langle p^{2} \right \rangle_{F}^{13}}+\frac{155770819p^{28}}{17280(N-6)^{14}\left \langle p^{2} \right \rangle_{F}^{14}} \notag \\
&-\frac{514265807p^{30}}{23040(N-6)^{15}\left \langle p^{2} \right \rangle_{F}^{15}},
\label{eq18}
\end{align}
\begin{equation}
\left \langle e^{i2\left ( \phi _{1}-\phi _{2}  \right )+i4\left ( \phi _{3}-\phi _{4}  \right ) }  \right \rangle
|p \approx \frac{5p^{12}}{16(N-4)^{6}\left \langle p^{2} \right \rangle_{F}^{6}} -\frac{13p^{14}}{12(N-4)^{7}\left \langle p^{2} \right \rangle_{F}^{7}}+\frac{907p^{16}}{360(N-4)^{8}\left \langle p^{2} \right \rangle_{F}^{8}},
\label{eq19}
\end{equation}
\begin{equation}
   \left \langle e^{i3\left ( \phi _{1}-\phi _{2}  \right )+i4\left ( \phi _{3}-\phi _{4}  \right ) }  \right \rangle
|p \approx -\frac{35p^{14}}{144(N-4)^{7}\left \langle p^{2} \right \rangle_{F}^{7}}+\frac{7p^{16}}{9(N-4)^{8}\left \langle p^{2} \right \rangle_{F}^{8}}-\frac{2443p^{18}}{1440(N-4)^{9}\left \langle p^{2} \right \rangle_{F}^{9}}, 
\label{eq20}
\end{equation}
\begin{equation}
  \left \langle e^{i4\left ( \phi _{1}-\phi _{2}  \right ) }  \right \rangle
|p \approx
\frac{p^{8}}{24(N-2)^{4}\left \langle p^{2} \right \rangle_{F}^{4}}.  
\label{eq21}
\end{equation}
Using Eqs.~(\ref{eq8}), ~(\ref{eq12}), ~(\ref{eq13}), ~(\ref{eq14}), ~(\ref{eq18}), ~(\ref{eq19}), ~(\ref{eq20}), and~(\ref{eq21}) we obtain
\begin{align}
sc_{2,3,4} \left \{ 6 \right \}&\approx 
-\frac{35p^{18}}{8(N-6)^{9}\left \langle p^{2} \right \rangle_{F}^{9}} + \frac{1015p^{20}}{24(N-6)^{10}\left \langle p^{2} \right \rangle_{F}^{10}}-\frac{470037645p^{22}}{1995844(N-6)^{11}\left \langle p^{2} \right \rangle_{F}^{11}} +\frac{92573p^{24}}{96(N-6)^{12}\left \langle p^{2} \right \rangle_{F}^{12}}\notag \\ &-\frac{4601851p^{26}}{1440(N-6)^{13}\left \langle p^{2} \right \rangle_{F}^{13}}+\frac{155770819p^{28}}{17280(N-6)^{14}\left \langle p^{2} \right \rangle_{F}^{14}} -\frac{514265807p^{30}}{23040(N-6)^{15}\left \langle p^{2} \right \rangle_{F}^{15}}-\frac{p^{18}}{144(N-4)^{9}\left \langle p^{2} \right \rangle_{F}^{9}}\notag \\
&-\frac{p^{8}}{24(N-2)^{4}\left \langle p^{2} \right \rangle_{F}^{4}}\left ( -\frac{5p^{10}}{6(N-4)^{5}\left \langle p^{2} \right \rangle_{F}^{5}}+\frac{5p^{12}}{2(N-4)^{6}\left \langle p^{2} \right \rangle_{F}^{6}}-\frac{827p^{14}}{144(N-4)^{7}\left \langle p^{2} \right \rangle_{F}^{7}} \right )  \notag \\
& +\frac{p^{6}}{6(N-2)^{3}\left \langle p^{2} \right \rangle_{F}^{3}}\left ( \frac{5p^{12}}{16(N-4)^{6}\left \langle p^{2} \right \rangle_{F}^{6}} -\frac{13p^{14}}{12(N-4)^{7}\left \langle p^{2} \right \rangle_{F}^{7}}+\frac{907p^{16}}{360(N-4)^{8}\left \langle p^{2} \right \rangle_{F}^{8}} \right )  \notag \\
&-\frac{p^{4}}{2(N-2)^{2}\left \langle p^{2} \right \rangle_{F}^{2}}\left ( -\frac{35p^{14}}{144(N-4)^{7}\left \langle p^{2} \right \rangle_{F}^{7}}+\frac{7p^{16}}{9(N-4)^{8}\left \langle p^{2} \right \rangle_{F}^{8}}-\frac{2443p^{18}}{1440(N-4)^{9}\left \langle p^{2} \right \rangle_{F}^{9}} \right ).
\label{eq22}
\end{align}

\section{$sc_{2,4} \left \{ 4 \right \}$ and $sc_{2,3,4} \left \{ 6 \right \}$ from transverse momentum conservation and flow}

Next, we take collective flow into account and subsequently determine the effect of TMC on the four-particle symmetric cumulant $sc_{2,4} \left \{ 4 \right \}$ and the six-particle symmetric cumulant $sc_{2,3,4} \left \{ 6 \right \}$. Under this condition, the probability distribution of a single particle in the azimuth angle $\phi$ is described by the Fourier expansion in the following way:
 \begin{align}
 f\left ( p ,\phi \right ) &=\frac{g\left ( p \right ) }{2\pi}\left ( 1+\sum_{n}2v_{n} \left (  p\right ) \cos \left [ n\left ( \phi -\Psi _{n}  \right )  \right ]   \right ) \notag \\
 &\approx \frac{g\left ( p \right ) }{2\pi}\left ( 1+2v_{2}
 \left (  p\right ) \cos \left [ 2\left ( \phi -\Psi _{2}  \right )  \right ]
+2v_{3}
 \left (  p\right ) \cos \left [ 3\left ( \phi -\Psi _{3}  \right )  \right ]
+2v_{4}
 \left (  p\right ) \cos \left [ 4\left ( \phi -\Psi _{4}  \right )  \right ]   \right ),
 \label{eq23}
 \end{align}
where $v_{n}$ and $\Psi _{n}$ denote the $n^{th}$-order flow coefficient and the reaction plane angle respectively, and we consider up to the fourth order, i.e., $v_{2}$, $v_{3}$, and $v_{4}$.

\subsection{$sc_{2,3} \left \{ 4 \right \}$}
According to Eqs.~(\ref{eq6}) and~(\ref{eq23}), the four-particle probability distribution with TMC and flow can be expressed as
\begin{equation}
f_{4} \left ( p_{1},\phi _{1}  ,\dots,p_{4},\phi _{4}  \right )  =f\left ( p_{1},\phi _{1}   \right )\cdots f\left ( p_{4},\phi _{4}   \right )\frac{N}{N- 4}
\mathrm{exp}\left ( - \frac{\left ( p_{1,x}+  \dots +p_{4,x}    \right )^{2}  }{2\left (  N-4\right )\left \langle p_{x}^{2}     \right   \rangle _{F}  }  - \frac{\left ( p_{1,y}+  \dots +p_{4,y}    \right )^{2}  }{2\left (  N-4\right )\left \langle p_{y}^{2}     \right   \rangle _{F}  }\right ),
\end{equation}
where
\begin{equation}
p_{x} = p\cos \left (  \phi\right ),
\hspace{0.5cm}p_{y} = p\sin \left (  \phi\right ),
\end{equation}
\begin{equation}
\left \langle p_{x}^{2}   \right \rangle_{F}  = \frac{1}{2} \left \langle p^{2}  \right \rangle _{F}\left ( 1+  v_{2F}   \right ),
\hspace{0.5cm}\left \langle p_{y}^{2}   \right \rangle_{F}  = \frac{1}{2} \left \langle p^{2}  \right \rangle _{F}\left ( 1-  v_{2F}   \right ),
\end{equation}
\begin{equation}
v_{2F}=   \frac{\int_{F} v_{2}\left ( p \right )g(p)p^{2}d^{2}p    }{\int_{F} g(p)p^{2}d^{2}p}.
\end{equation}

By applying
\begin{equation}
\left \langle e^{i2\left ( \phi _{1}-\phi _{2}  \right )+i3\left ( \phi _{3}-\phi _{4}  \right ) }  \right \rangle\mid p_{1},p_{2},p_{3},p_{4} =\frac{\int_{0}^{2\pi } e^{i2\left ( \phi _{1}-\phi _{2}  \right )+i4\left ( \phi _{3}-\phi _{4}  \right ) }f_{4} \left ( p_{1},\phi _{1}  ,\dots,p_{4},\phi _{4}  \right )d\phi _{1} \dots d\phi _{4} }{\int_{0}^{2\pi } f_{4} \left ( p_{1},\phi _{1}  ,\dots,p_{4},\phi _{4}  \right )d\phi _{1} \dots d\phi _{4}},
\label{eq28}
\end{equation}
\begin{equation}
p_{1} =p_{2} =p_{3} =p_{4} =p,
\end{equation}
and extending the exponential term in Eq.~(\ref{eq28}) to the fifth order, i.e., until the first TMC term appears,
we obtain
\begin{equation}
\left \langle e^{i2\left ( \phi _{1}-\phi _{2}  \right )+i3\left ( \phi _{3}-\phi _{4}  \right ) }  \right \rangle\mid p\approx A_{0}+A_{1}Y_{A}+\frac{1}{2} A_{2}Y_{A}^{2}+\frac{1}{6} A_{3}Y_{A}^{3}+\frac{1}{24} A_{4}Y_{A}^{4}+\frac{1}{120} A_{5}Y_{A}^{5},
\label{eq30}
\end{equation}
where
\begin{equation}
Y _{A}=-\frac{p^{2} }{(N-4) \left \langle p^{2}  \right \rangle _{F}(1-v_{2F}^{2} )},
\end{equation}
and
\begin{align}
A_{0}&=v_{2}^{2} v_{3}^{2}, \notag \\
A_{1}&=v_2^4 + 6 v_2^2 v_3^2 + v_3^4 + v_2^2 v_4^2 - v_2 v_{2F} v_3^2 \cos(2 \Psi_2) + 2 v_2 v_3^2 v_4 \cos(2 \Psi_2 - 6 \Psi_3 + 4 \Psi_4)
,\notag \\
A_{2} &=12 v_2^4 + v_3^2 + 56 v_2^2 v_3^2 + \frac{v_{2F}^2 v_3^2}{2} + 12 v_3^4 + 16 v_2^2 v_4^2 + 9 v_3^2 v_4^2 
- 8 v_2^3 v_{2F} \cos(2 \Psi_2) - 28 v_2 v_{2F} v_3^2 \cos(2 \Psi_2) 
\notag\\ &- 6 v_2 v_{2F} v_4^2 \cos(2 \Psi_2) + 4 v_2^2 v_4 \cos(4 \Psi_2 - 4 \Psi_4) 
- 6 v_{2F} v_3^2 v_4 \cos(6 \Psi_3 - 4 \Psi_4) 
+ 28 v_2 v_3^2 v_4 \cos(2 \Psi_2 - 6 \Psi_3 + 4 \Psi_4)
,\notag\\
A_{3} &=16 v_2^2 + 141 v_2^4 + 24 v_2^2 v_{2F}^2 + 30 v_3^2 + 610 v_2^2 v_3^2 
+ 45 v_{2F}^2 v_3^2 + 141 v_3^4
+ 9 v_4^2 + 234 v_2^2 v_4^2 
+ \frac{27}{2} v_{2F}^2 v_4^2 + 180 v_3^2 v_4^2 \notag \\ &+ 9 v_4^4 
- 234 v_2^3 v_{2F} \cos(2 \Psi_2) 
- 606 v_2 v_{2F} v_3^2 \cos(2 \Psi_2)- 216 v_2 v_{2F} v_4^2 \cos(2 \Psi_2) 
+ 15 v_2^2 v_{2F}^2 \cos(4 \Psi_2)\notag \\ &- 48 v_2 v_{2F} v_4 \cos(2 \Psi_2 - 4 \Psi_4) 
+ 90 v_2^2 v_4 \cos(4 \Psi_2 - 4 \Psi_4) 
- 180 v_{2F} v_3^2 v_4 \cos(6 \Psi_3 - 4 \Psi_4) \notag\\
&+ 366 v_2 v_3^2 v_4 \cos(2 \Psi_2 - 6 \Psi_3 + 4 \Psi_4)
,\notag\\
A_{4} &=480 v_2^2 + 1720 v_2^4 + 1440 v_2^2 v_{2F}^2 + 652 v_3^2 
+ 7213 v_2^2 v_3^2 + 1956 v_{2F}^2 v_3^2 + 1720 v_3^4 + 288 v_4^2 
+ 3328 v_2^2 v_4^2+ 864 v_{2F}^2 v_4^2\notag\\ &
+ 2860 v_3^2 v_4^2 + 240 v_4^4 
- 160 v_2 v_{2F} \cos(2 \Psi_2) 
- 5068 v_2^3 v_{2F} \cos(2 \Psi_2) 
- 120 v_2 v_{2F}^3 \cos(2 \Psi_2)- 11716 v_2 v_{2F} v_3^2 \cos(2 \Psi_2)\notag\\ &
- 5300 v_2 v_{2F} v_4^2 \cos(2 \Psi_2) 
+ 750 v_2^2 v_{2F}^2 \cos(4 \Psi_2) 
- 1860 v_2 v_{2F} v_4 \cos(2 \Psi_2 - 4 \Psi_4) 
+ 1600 v_2^2 v_4 \cos(4 \Psi_2 - 4 \Psi_4)\notag\\ &
- 3932 v_{2F} v_3^2 v_4 \cos(6 \Psi_3 - 4 \Psi_4) 
+ 180 v_{2F}^2 v_4 \cos(4 \Psi_4) 
+ 4808 v_2 v_3^2 v_4 \cos(2 \Psi_2 - 6 \Psi_3 + 4 \Psi_4)
,\notag\\
A_{5}& =100 + 10225 v_2^2 + 21796 v_2^4 + 500 v_{2F}^2 + 51125 v_2^2 v_{2F}^2 
+ \frac{375}{2} v_{2F}^4 + 12200 v_3^2 + 89906 v_2^2 v_3^2 + 61000 v_{2F}^2 v_3^2 + 21761 v_3^4 \notag\\ &
+ 6400 v_4^2 + 47025 v_2^2 v_4^2 
+ 32000 v_{2F}^2 v_4^2 + 42560 v_3^2 v_4^2 + 4600 v_4^4 - 7300 v_2 v_{2F} \cos(2 \Psi_2) 
- 98380 v_2^3 v_{2F} \cos(2 \Psi_2)\notag\\ &- 10950 v_2 v_{2F}^3 \cos(2 \Psi_2) 
- 213785 v_2 v_{2F} v_3^2 \cos(2 \Psi_2) 
- 111320 v_2 v_{2F} v_4^2 \cos(2 \Psi_2) 
+ 24535 v_2^2 v_{2F}^2 \cos(4 \Psi_2)\notag\\&- 48580 v_2 v_{2F} v_4 \cos(2 \Psi_2 - 4 \Psi_4) 
+ 26490 v_2^2 v_4 \cos(4 \Psi_2 - 4 \Psi_4) 
- 76170 v_{2F} v_3^2 v_4 \cos(6 \Psi_3 - 4 \Psi_4) \notag\\&+ 9975 v_{2F}^2 v_4 \cos(4 \Psi_4) 
+ 64170 v_2 v_3^2 v_4 \cos(2 \Psi_2 - 6 \Psi_3 + 4 \Psi_4).
\label{eq32}
\end{align}
We show the terms only up to $(v_{2,3,4})^4$. The full results of Eq.~(\ref{eq32}) are shown in Eq.~(\ref{eqA1}) in the Appendix (the ``full results" here refer to all results obtained by expanding to the corresponding order, specifically the order at which the first nonzero TMC term appears).
Similarly, we have
\begin{align}
\left \langle e^{i2\left ( \phi _{1}-\phi _{2}  \right ) }  \right \rangle
|p \approx  B_{0}+B_{1}Y_{B}+\frac{1}{2} B_{2}Y_{B}^{2},
\label{eq33}
\end{align}
where
\begin{equation}
Y_{B}=-\frac{p^{2} }{(N-2) \left \langle p^{2}  \right \rangle _{F}(1-v_{2F}^{2} )},
\end{equation}
and
\begin{align}
B_{0} &=v_{2}^{2},\notag\\
B_{1} &=2 v_{2}^{2} + v_{3}^{2} - v_{2} v_{2F} \cos(2 \Psi_{2}) - v_{2} v_{2F} v_{4} \cos(2 \Psi_{2} - 4 \Psi_{4}),\notag\\
B_{2} &=1 + 6 v_{2}^{2} + \frac{{v_{2F}^{2}}}{2} + 3 v_{2}^{2} v_{2F}^{2} + 4 v_{3}^{2} + 2 v_{2F}^{2} v_{3}^{2} + v_{4}^{2} + \frac{{v_{2F}^{2} v_{4}^{2}}}{2} - 8 v_{2} v_{2F} \cos(2 \Psi_{2}) + \frac{{1}}{{2}} v_{2}^{2} v_{2F}^{2} \cos(4 \Psi_{2}) \notag\\ &- 8 v_{2} v_{2F} v_{4} \cos(2 \Psi_{2} - 4 \Psi_{4}) + 3 v_{2F}^{2} v_{4} \cos(4 \Psi_{4})
.\notag\\
\label{eq35}
\end{align}
Furthermore
\begin{equation}
\left \langle e^{i3\left ( \phi _{3}-\phi _{4}  \right ) }  \right \rangle
|p \approx 
C_{0}+C_{1}Y_{C}+\frac{1}{2} C_{2}Y_{C}^{2}+\frac{1}{6} C_{3}Y_{C}^{3},
\label{eq36}
\end{equation}
where
\begin{equation}
Y_{C}=-\frac{p^{2} }{(N-2) \left \langle p^{2}  \right \rangle _{F}(1-v_{2F}^{2} )},
\end{equation}
and
\begin{align}
C_{0} &=v_{3}^{2},\notag\\
C_{1} &=v_2^2 + 2 v_3^2 + v_4^2 - 2 v_2 v_{2F} v_4 \cos(2 \Psi_2 - 4 \Psi_4)
,\notag\\
C_{2} &=4 v_2^2 + 2 v_2^2 v_{2F}^2 + 6 v_3^2 + 3 v_{2F}^2 v_3^2 + 4 v_4^2 + 2 v_{2F}^2 v_4^2 
- 2 v_2 v_{2F} \cos(2 \Psi_2) 
- 12 v_2 v_{2F} v_4 \cos(2 \Psi_2 - 4 \Psi_4) 
+ 2 v_{2F}^2 v_4 \cos(4 \Psi_4)
,\notag\\
C_{3}& =1 + 15 v_2^2 + \frac{3 v_{2F}^2}{2} + \frac{45 v_2^2 v_{2F}^2}{2}
 + 20 v_3^2 + 30 v_{2F}^2 v_3^2 + 15 v_4^2 + \frac{45 v_{2F}^2 v_4^2}{2}- 18 v_2 v_{2F} \cos(2 \Psi_2) 
- \frac{9}{2} v_2 v_{2F}^3 \cos(2 \Psi_2) 
\notag\\ &+ \frac{3}{2} v_2^2 v_{2F}^2 \cos(4 \Psi_2) - \frac{1}{4} v_{2F}^3 v_3^2 \cos(6 \Psi_3) 
- 60 v_2 v_{2F} v_4 \cos(2 \Psi_2 - 4 \Psi_4) 
- 15 v_2 v_{2F}^3 v_4 \cos(2 \Psi_2 - 4 \Psi_4) \notag\\ &+ \frac{45}{2} v_{2F}^2 v_4 \cos(4 \Psi_4) 
- \frac{3}{2} v_2 v_{2F}^3 v_4 \cos(2 \Psi_2 + 4 \Psi_4).
\label{eq38}
\end{align}

Although the symmetric cumulant $sc_{2,3} \left \{ 4 \right \}$ reflects the nature of the correlation between $v_2$ and $v_3$, their magnitudes also depend on the square of single flow harmonics $v_2$ and $v_3$. The dependence on the single-flow harmonics can be scaled out via the normalized cumulant, i.e.,
\begin{align}
nsc_{2,3}\left \{ 4 \right \}&=\frac{sc_{2,3}\left \{ 4 \right \}}{v_{2}\left \{ 2 \right \}^{2} v_{3}\left \{ 2 \right \}^{2}  } ,
\label{eq39}
\end{align}
where $v_{2} \left \{ 2 \right \}$ and $v_{3} \left \{ 2 \right \}$ are given by Eqs.~(\ref{eq33}) and~(\ref{eq36}), respectively.

\subsection{$sc_{2,3,4} \left \{ 6 \right \}$}
According to Eqs.~(\ref{eq6}) and~(\ref{eq23}), the six-particle probability distribution with TMC and flow can be expressed as
\begin{equation}
f_{6} \left ( p_{1},\phi _{1}  ,\dots,p_{6},\phi _{6}  \right )  =f\left ( p_{1},\phi _{1}   \right )\cdots f\left ( p_{6},\phi _{6}   \right )\frac{N}{N-6}
\mathrm{exp}\left ( - \frac{\left ( p_{1,x}+  \dots +p_{6,x}    \right )^{2}  }{2\left (  N-6\right )\left \langle p_{x}^{2}     \right   \rangle _{F}  }  - \frac{\left ( p_{1,y}+  \dots +p_{6,y}    \right )^{2}  }{2\left (  N-6\right )\left \langle p_{y}^{2}     \right   \rangle _{F}  }\right ).
\end{equation}
So, we have
\begin{align}
\left \langle e^{i(2 \phi _{1}+3\phi _{2}  +4\phi _{3})
-i(2 \phi _{4}+3\phi _{5}  +4\phi _{6}) }  \right \rangle |p &=\frac{\int_{0}^{2\pi } e^{i(2 \phi _{1}+3\phi _{2}  +4\phi _{3})
-i(2 \phi _{4}+3\phi _{5}  +4\phi _{6}) }f_{6} \left ( p_{1},\phi _{1}  ,\dots,p_{6},\phi _{6}  \right )d\phi _{1} \dots d\phi _{6} }{\int_{0}^{2\pi } f_{6} \left ( p_{1},\phi _{1}  ,\dots,p_{6},\phi _{6}  \right )d\phi _{1} \dots d\phi _{6}}\notag\\
&\approx 
\sum_{n=0}^{9} \frac{Y_{D}^{n}  }{n!}D_{n},
\label{eq41}
\end{align}
where
\begin{equation}
Y_{D}=-\frac{p^{2} }{(N-6) \left \langle p^{2}  \right \rangle _{F}(1-v_{2F}^{2} )},
\end{equation}
and the main results for each order $D$ are given in Eq.~(\ref{eqA2}) in the Appendix. The full results are not shown in the Appendix because the total number of items is very large (about 3600), so only the main results are shown (about 560 terms), where the minimum percentage of the main results to the full results is $94\%$. However, what is presented in Fig.~\ref{sc234} is still the result of all terms.

In addition, for two-particle and four-particle correlations of different orders, we obtain the following results:
\begin{equation}
\left \langle e^{i4\left ( \phi _{1}-\phi _{2}  \right ) }  \right \rangle
|p \approx 
E_{0}+E_{1}Y_{E}+\frac{1}{2} E_{2}Y_{E}^{2}+\frac{1}{6} E_{3}Y_{E}^{3}+\frac{1}{24} E_{4}Y_{E}^{4},
\label{eq43}
\end{equation}
\begin{equation}
  \left \langle e^{i2\left ( \phi _{1}-\phi _{2}  \right )+i4\left ( \phi _{3}-\phi _{4}  \right ) }  \right \rangle\mid p\approx F_{0}+F_{1}Y_{F}+\frac{1}{2} F_{2}Y_{F}^{2}+\frac{1}{6} F_{3}Y_{F}^{3}+\frac{1}{24} F_{4}Y_{A}^{4}+\frac{1}{120} F_{5}Y_{F}^{5}+\frac{1}{720} F_{6}Y_{F}^{6}, 
  \label{eq44}
\end{equation}
\begin{equation}
  \left \langle e^{i3\left ( \phi _{1}-\phi _{2}  \right )+i4\left ( \phi _{3}-\phi _{4}  \right ) }  \right \rangle\mid p\approx G_{0}+G_{1}Y_{G}+\frac{1}{2} G_{2}Y_{G}^{2}+\frac{1}{6} G_{3}Y_{G}^{3}+\frac{1}{24} G_{4}Y_{G}^{4}+\frac{1}{120} G_{5}Y_{G}^{5}+\frac{1}{720} G_{6}Y_{G}^{6}+\frac{1}{5040} G_{7}Y_{G}^{7}, 
\label{eq45}
\end{equation}
where
\begin{equation}
Y_{E}=-\frac{p^{2} }{(N-2) \left \langle p^{2}  \right \rangle _{F}(1-v_{2F}^{2} )},
\end{equation}
\begin{equation}
Y_{F}=Y_{G}=-\frac{p^{2} }{(N-4) \left \langle p^{2}  \right \rangle _{F}(1-v_{2F}^{2} )},
\end{equation}
\begin{align}
E_{0} &=v_{4}^{2},\notag\\
E_{1} &=v_{3}^{2}+2v_{4}^{2}-v_{2} v_{2F} v_{4} \cos(2 \Psi_{2} - 4 \Psi_{4}),\notag\\
E_{2} &=v_{2}^{2} + \frac{v_{2}^{2} v_{2F}^{2}}{2} + 4 v_{3}^{2} + 2 v_{2F}^{2} v_{3}^{2} + 6 v_{4}^{2} + 3 v_{2F}^{2} v_{4}^{2} - 8 v_{2} v_{2F} v_{4} \cos(2 \Psi_{2} - 4 \Psi_{4}) + \frac{1}{2} v_{2F}^{2} v_{4} \cos(4 \Psi_{4}),\notag\\
E_{3} &=6 v_{2}^{2} + 9 v_{2}^{2} v_{2F}^{2} + 15 v_{3}^{2} + \frac{45 v_{2F}^{2} v_{3}^{2}}{2} + 20 v_{4}^{2} + 30 v_{2F}^{2} v_{4}^{2} - 3 v_{2} v_{2F} \cos(2 \Psi_{2}) - \frac{3}{4} v_{2} v_{2F}^{3} \cos(2 \Psi_{2}) \notag\\ &- 45 v_{2} v_{2F} v_{4} \cos(2 \Psi_{2} - 4 \Psi_{4}) - \frac{45}{4} v_{2} v_{2F}^{3} v_{4} \cos(2 \Psi_{2} - 4 \Psi_{4}) + 9 v_{2F}^{2} v_{4} \cos(4 \Psi_{4}) - \frac{1}{4} v_{2} v_{2F}^{3} v_{4} \cos(2 \Psi_{2} + 4 \Psi_{4}),\notag\\
E_{4} &= 1 + 28 v_{2}^{2} + 3 v_{2F}^{2} + 84 v_{2}^{2} v_{2F}^{2} + \frac{3 v_{2F}^{4}}{8} + \frac{21 v_{2}^{2} v_{2F}^{4}}{2} + 56 v_{3}^{2} + 168 v_{2F}^{2} v_{3}^{2} + 21 v_{2F}^{4} v_{3}^{2} + 70 v_{4}^{2} + 210 v_{2F}^{2} v_{4}^{2} \notag\\ &+ \frac{105 v_{2F}^{4} v_{4}^{2}}{4} 
 - 32 v_{2} v_{2F} \cos(2 \Psi_{2}) - 24 v_{2} v_{2F}^{3} \cos(2 \Psi_{2}) + 3 v_{2}^{2} v_{2F}^{2} \cos(4 \Psi_{2}) + \frac{1}{2} v_{2}^{2} v_{2F}^{4} \cos(4 \Psi_{2}) - v_{2F}^{3} v_{3}^{2} \cos(6 \Psi_{3}) 
 \notag\\ &- 224 v_{2} v_{2F} v_{4} \cos(2 \Psi_{2} - 4 \Psi_{4}) - 168 v_{2} v_{2F}^{3} v_{4} \cos(2 \Psi_{2} - 4 \Psi_{4}) + 84 v_{2F}^{2} v_{4} \cos(4 \Psi_{4}) + 14 v_{2F}^{4} v_{4} \cos(4 \Psi_{4}) 
 \notag\\ &+ \frac{1}{8} v_{2F}^{4} v_{4}^{2} \cos(8 \Psi_{4}) - 8 v_{2} v_{2F}^{3} v_{4} \cos(2 \Psi_{2} + 4 \Psi_{4}),
\end{align}
and the full results for each order of $F$ in Eq.~(\ref{eq44}) were presented in Ref.\cite{pretmc4} as well as the full results for each order of $G$ in Eq.~(\ref{eq45}) are presented in Eq.~(\ref{eqA3}) in the Appendix.

Similarly, the normalized higher-order symmetric cumulant $nsc_{2,3,4} \left \{ 6 \right \}$ is defined as follows,
\begin{align}
nsc_{2,3,4}\left \{ 6 \right \}=
\frac{sc_{2,3,4}\left \{ 6 \right \}}
{v_{2}\left \{ 2 \right \}^{2} v_{3}\left \{ 2 \right \}^{2}v_{4}\left \{ 2 \right \}^{2}  }, 
\label{eq49}
\end{align}
where $v_{2} \left \{ 2 \right \}$, $v_{3} \left \{ 2 \right \}$, and $v_{4} \left \{ 2 \right \}$ are from Eqs.~(\ref{eq33}), Eqs.~(\ref{eq36}), and~(\ref{eq43}), respectively.

\section{Results}

\begin{figure}[H]
\centering
\includegraphics[scale=0.4]{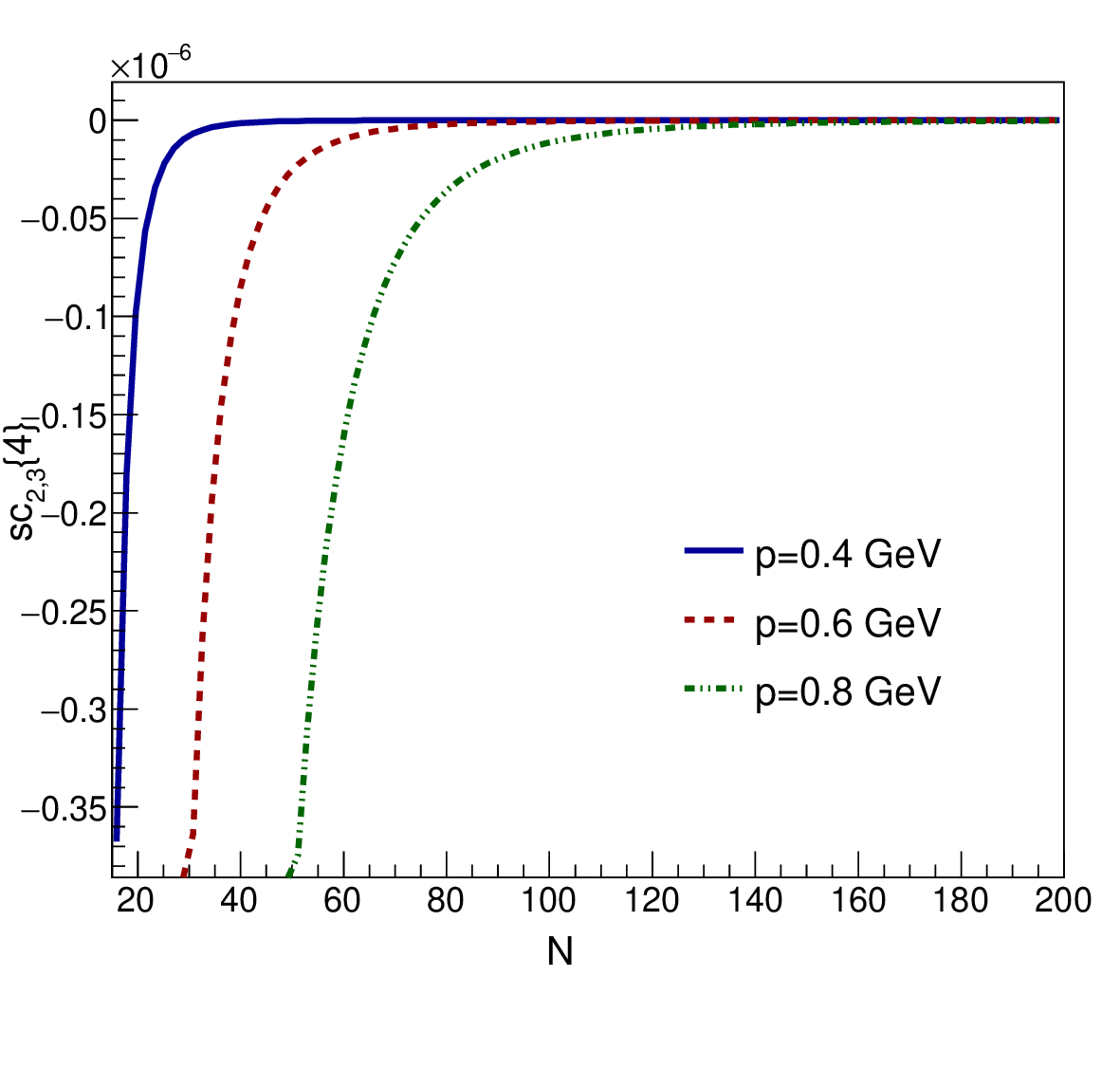}
\includegraphics[scale=0.4]{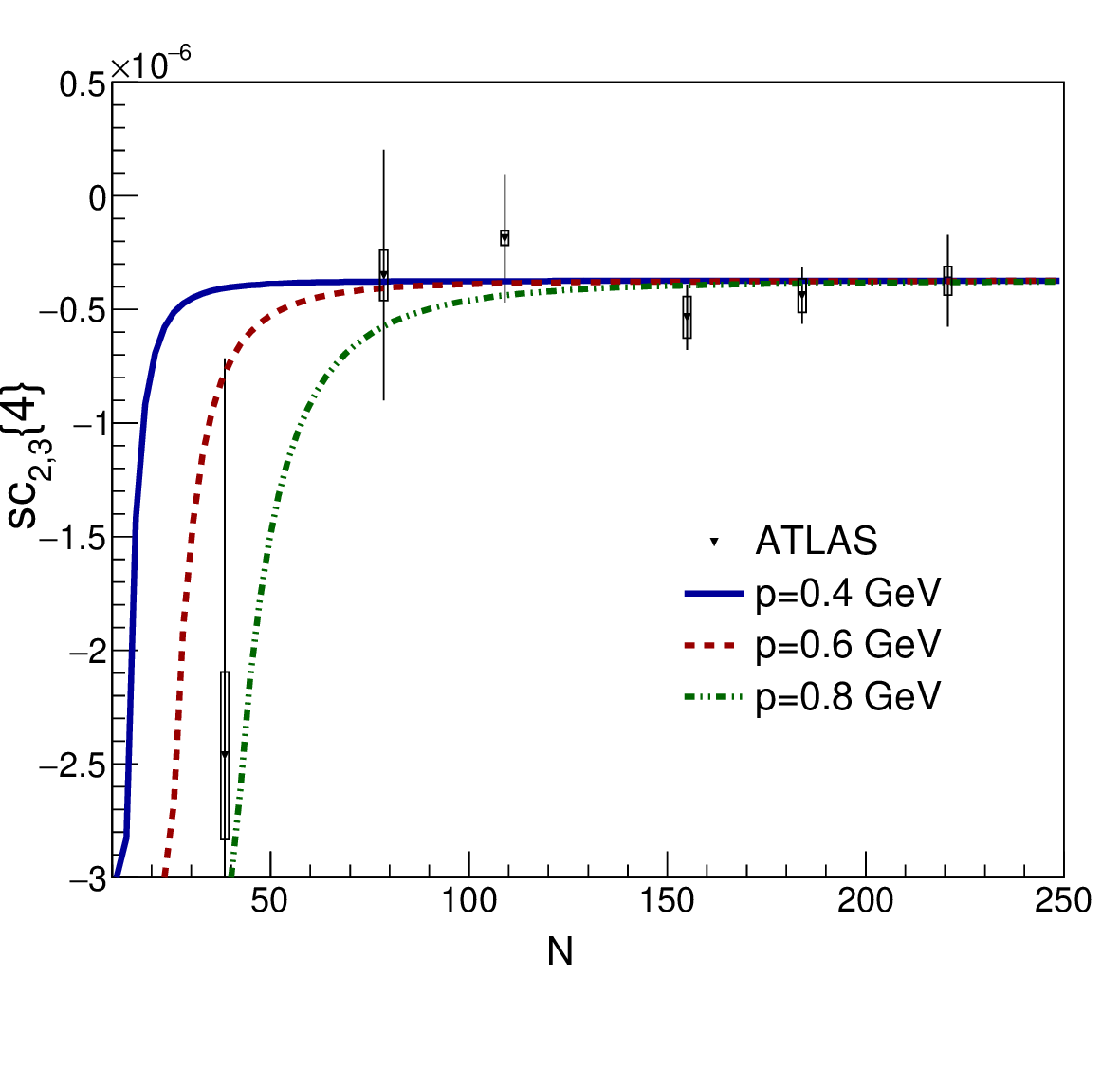}
\includegraphics[scale=0.4]{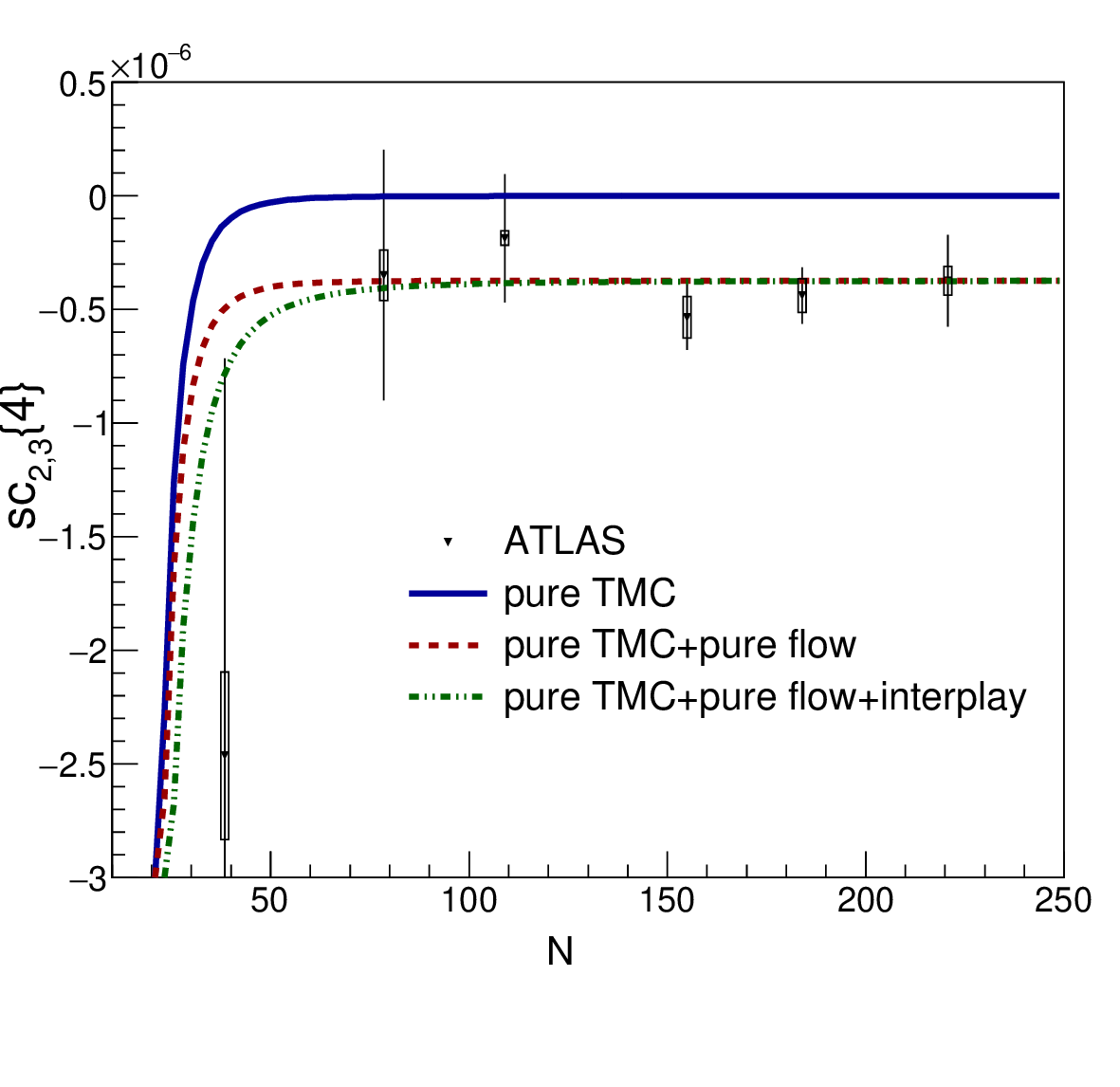}
\includegraphics[scale=0.4]{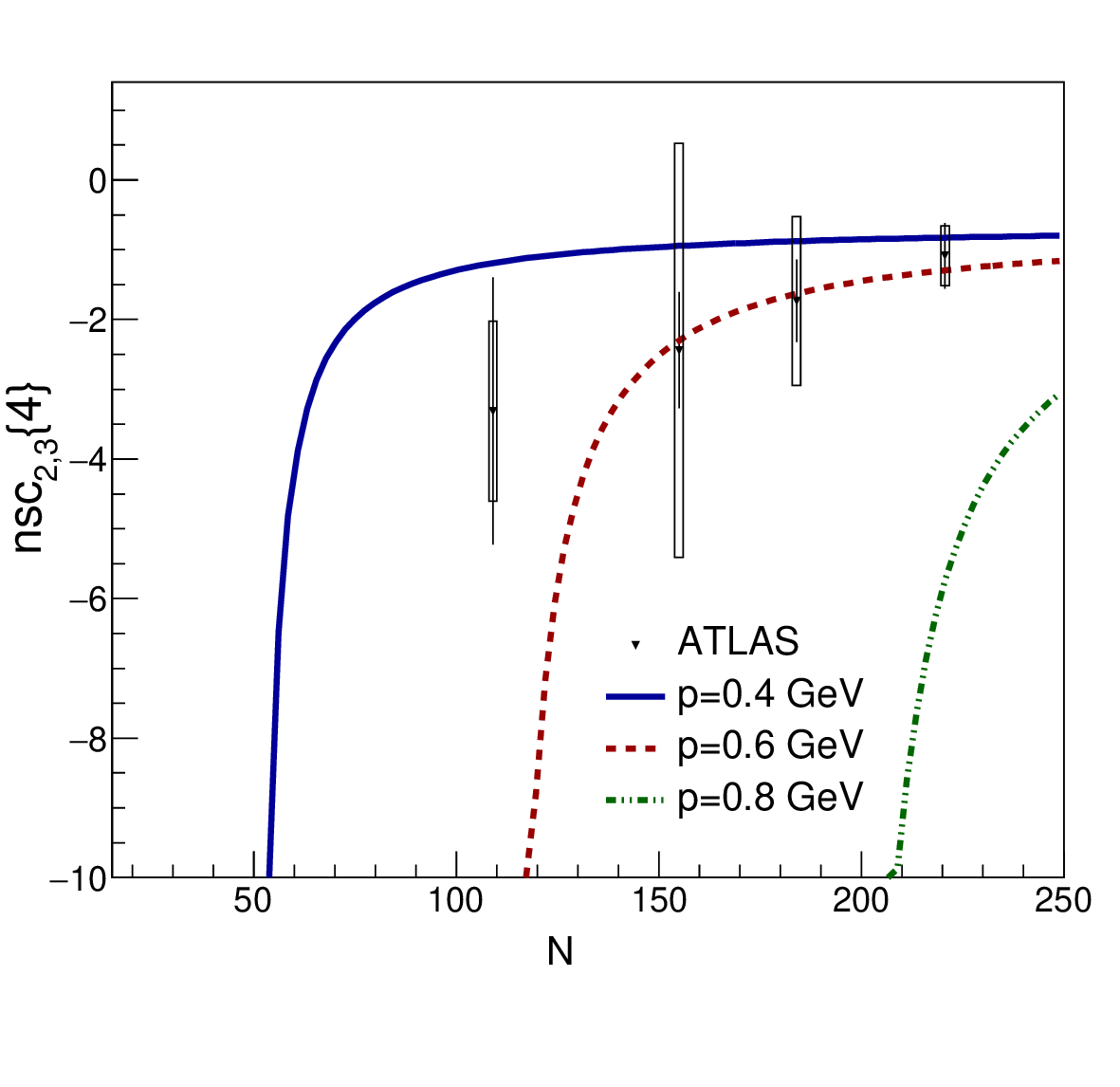}
\caption{The four-particle symmetric cumulant $sc_{2,3} \left \{ 4 \right \}$ from ``pure TMC'' (top-left), from ``pure TMC+pure flow+interplay'' (top-right), and the normalized cumulants  $nsc_{2,3} \left \{ 4 \right \}$ from ``pure TMC+pure flow+interplay'' (bottom-right) as a function of the number of particles $N$ for various values of transverse momenta $p$; and $sc_{2,3} \left \{ 4 \right \}$ from ``pure TMC,'' ``pure TMC+pure collective flow,'' and ``pure TMC+pure flow+interplay'' as a function of the number of particles $N$ for momentum $p=0.6$ GeV (bottom-left). The ATLAS data for $0.3< p_{T} < 3$ GeV in \textit{p}+\textit{p} collisions at 13 TeV using the four-subevent cumulant method are shown for comparisons, where the error bars and boxes represent the statistical and systematic uncertainties, respectively \cite{data1}.}
\label{sc23}
\end{figure}
{In the top-left panel of Fig.~\ref{sc23}, based on Eq.~(\ref{eq15}), we present the four-particle symmetric cumulant $sc_{2,3} \left \{ 4 \right \}$ from transverse momentum conservation only as a function of the number of particles $N$ for various values of transverse momenta $p=0.4, 0.6, 0.8$ GeV, where $\left \langle p^{2}  \right \rangle _{F}$= 0.24 $\rm GeV^{2}$. We see that there is a negative correlation between $v_2$ and $v_3$ that tends towards zero as $N$ increases. Furthermore, we observe an enhancement as the transverse momenta $p$ increases for smaller $N$, reflecting the property of TMC. In the top-right panel, based on Eqs.~(\ref{eq30}),~(\ref{eq33}), and~(\ref{eq36}), which are obtained from the
full results given in the Appendix, we show $sc_{2,3} \left \{ 4 \right \}$ from transverse momentum conservation and flow as a function of the number of particles $N$ for various values of transverse momenta $p=0.4, 0.6, 0.8$ GeV. In our calculation, we set $v_{2}= 0.05$, $v_{3}= 0.006$, $v_{4}= 0.0075$, $\left \langle p^{2}  \right \rangle _{F}= 0.24 $ GeV$^{2}$, $v_{2F}= 0.025$, $\Psi _{2} = 0$, $\cos(4[\Psi _{4}-\Psi _{2} ])=0.8$, $\cos(6[\Psi _{2}-\Psi _{3} ])=0.02$, which are taken from the experimental results \cite{plane,vnATLAS}, and other event plane correlations that are not directly measured experimentally are obtained indirectly by solving $\cos(4[\Psi _{4}-\Psi _{2} ])=0.8$ and $\cos(6[\Psi _{2}-\Psi _{3} ])=0.02$. After taking the collective flow into account, our theoretical results show that a negative correlation between $v_{2}$ and $v_{3}$ still exists and we can essentially describe the experimental data. Note that since the multiplicity $N$ refers to the number of particles under the influence of the TMC rather than the number of experimentally detected charged particles, we multiply the experimental number of charged particles by 1.5 to obtain the total number of particles $N$ for the experimental data in all figures. In addition, we present the results of $sc_{2,3} \left \{ 4 \right \}$ at $p=0.6$ GeV under different $v_n$, as shown in Fig.~\ref{changevn} of the Appendix. It can be seen that the variation of $v_n$ does not affect our conclusions. The bottom-left panel presents the respective contributions from the TMC only (denoted as ``pure TMC''), the pure TMC and pure collective flow (denoted as ``pure TMC+pure flow''), and plus interplay (denoted as ``pure TMC+pure flow+interplay'') for $p=0.6$ GeV. Here ``pure TMC'' refers to the terms that depend only on $N$ and $p$, ``pure flow" refers to the terms that depend only on $v_{n}$ and $\Psi _{n}$, and ``interplay" refers to terms that depend on both $N$, $p$, $v_{n}$, and $\Psi _{n}$ in Eqs.~(\ref{eq32}),~(\ref{eq35}), and~(\ref{eq38}). We can conclude that after taking collective flow into account, the curve based on TMC is lower, i.e., the negative correlation is enhanced. When $N$ is small, the correlation comes from the TMC, but when $N$ is large, the contribution from collective flow becomes significant, which is consistent with findings in Ref. \cite{pretmc4}. In the bottom-right panel, based on Eq.~(\ref{eq39}), our results on normalized cumulants  $nsc_{2,3} \left \{ 4 \right \}$ from ``pure TMC+pure flow+interplay'' can basically describe the experimental data. The correlation between $v_{2}$ and $v_{3}$ will become weaker as $N$ increases and will eventually tend to -1. In addition, it also suggests that our results on the two-particle $v_{2}\left \{ 2 \right \}$ and $v_{3}\left \{ 2 \right \}$ from $c_{2}\left \{ 2 \right \}$ and $c_{3}\left \{ 2 \right \}$ should be in agreement with the experimental data.

\begin{figure}[H]
\centering
\includegraphics[scale=0.4]{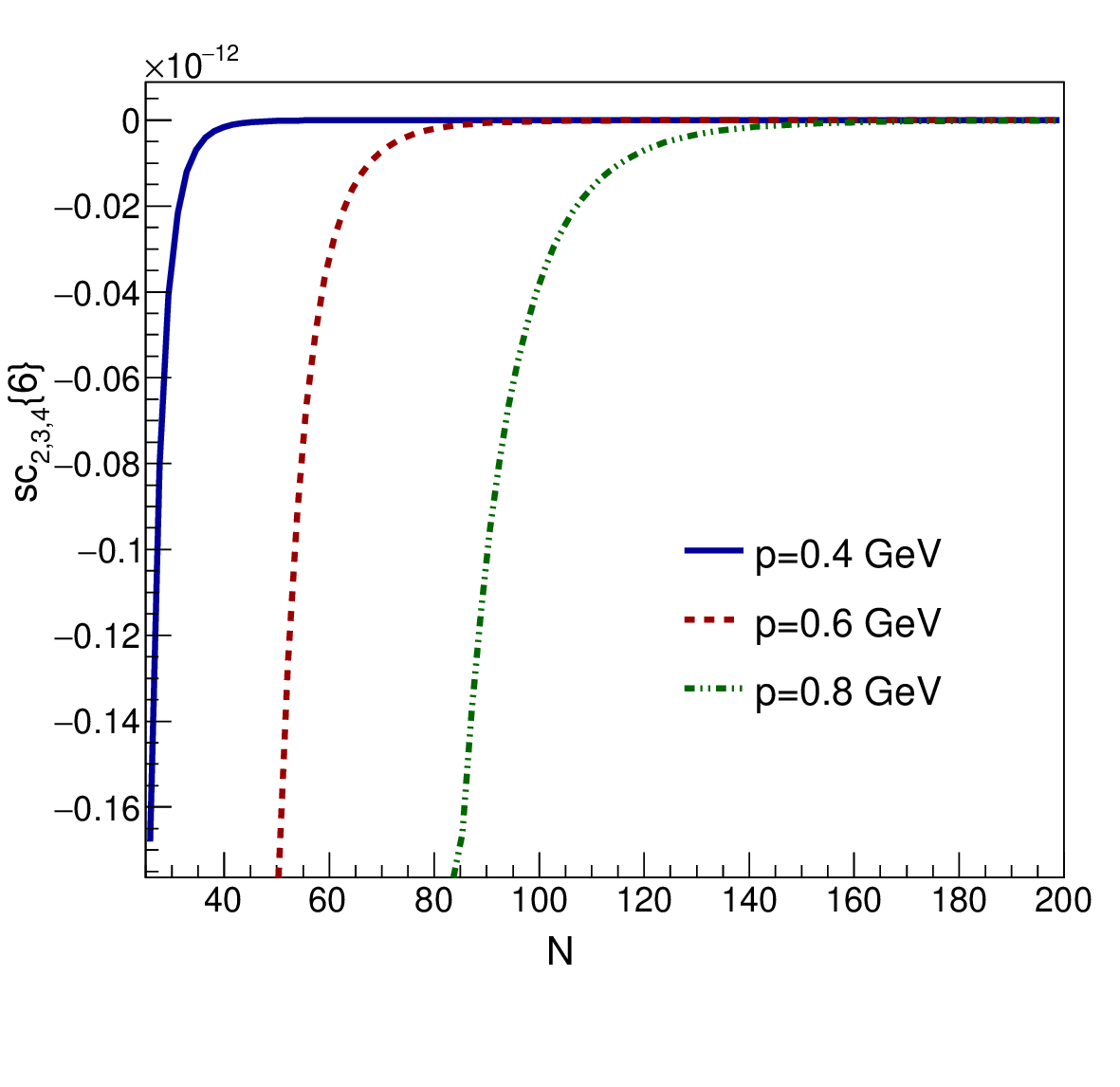}
\includegraphics[scale=0.4]{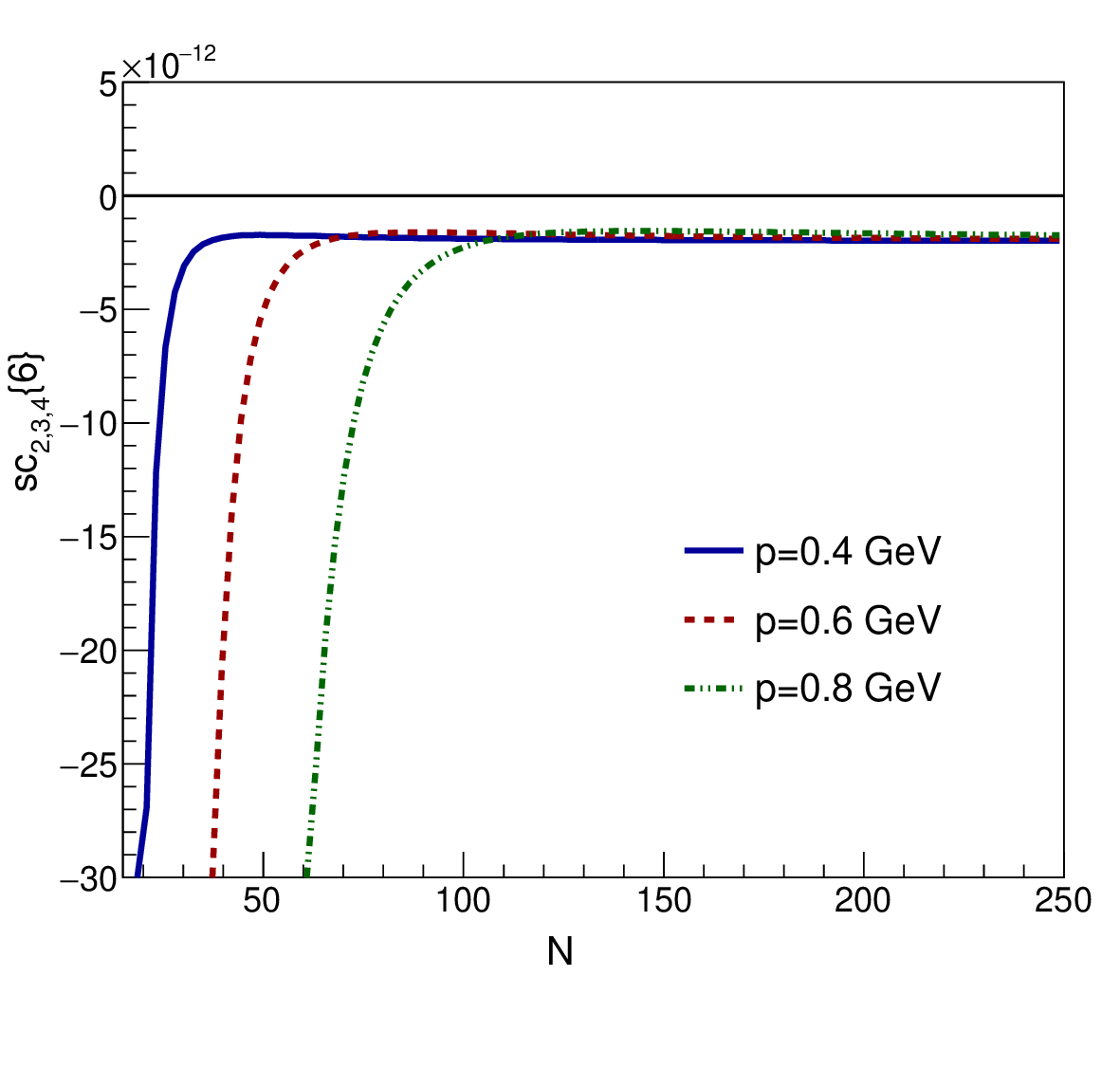}
\includegraphics[scale=0.4]{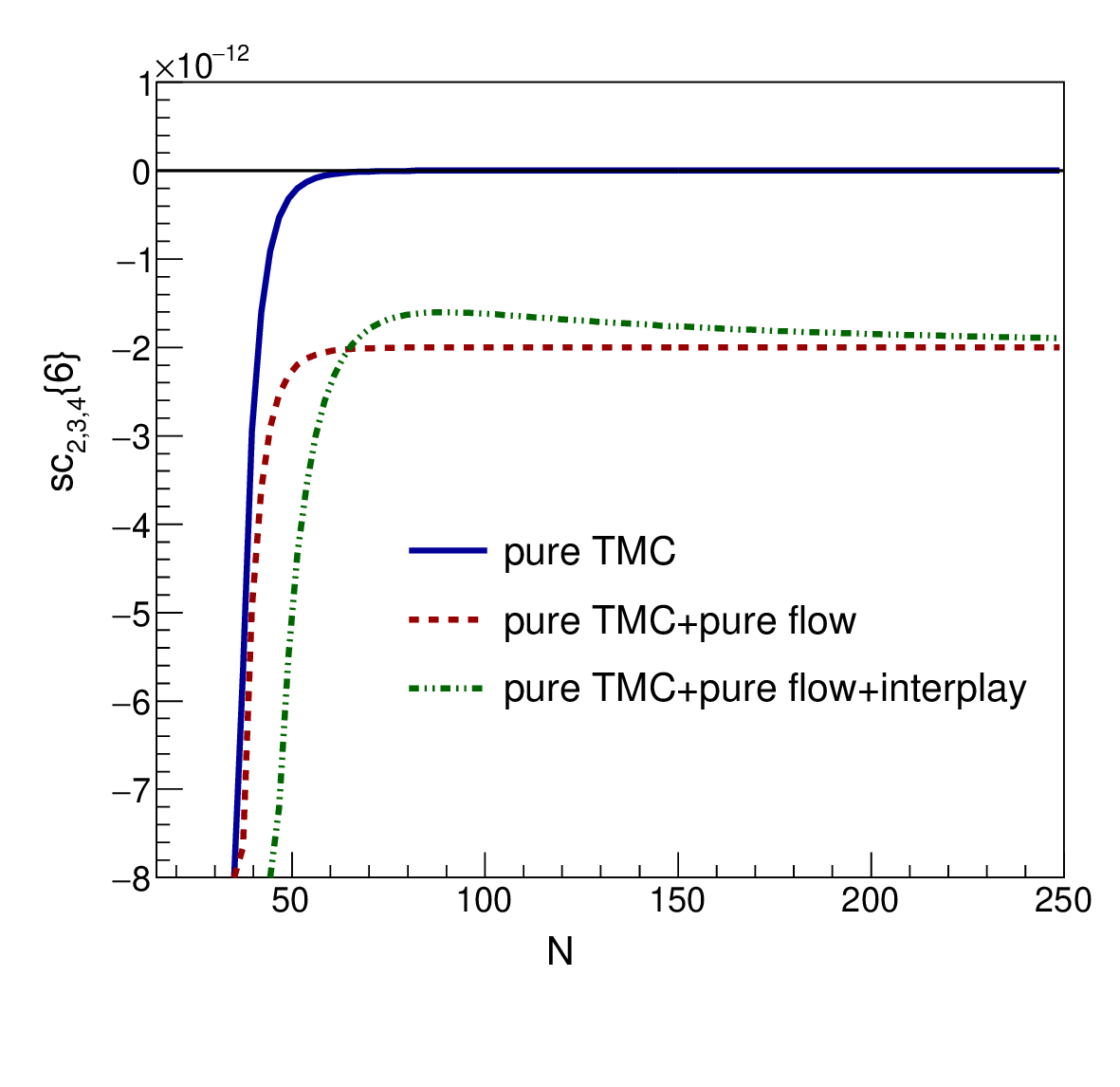}
\includegraphics[scale=0.4]{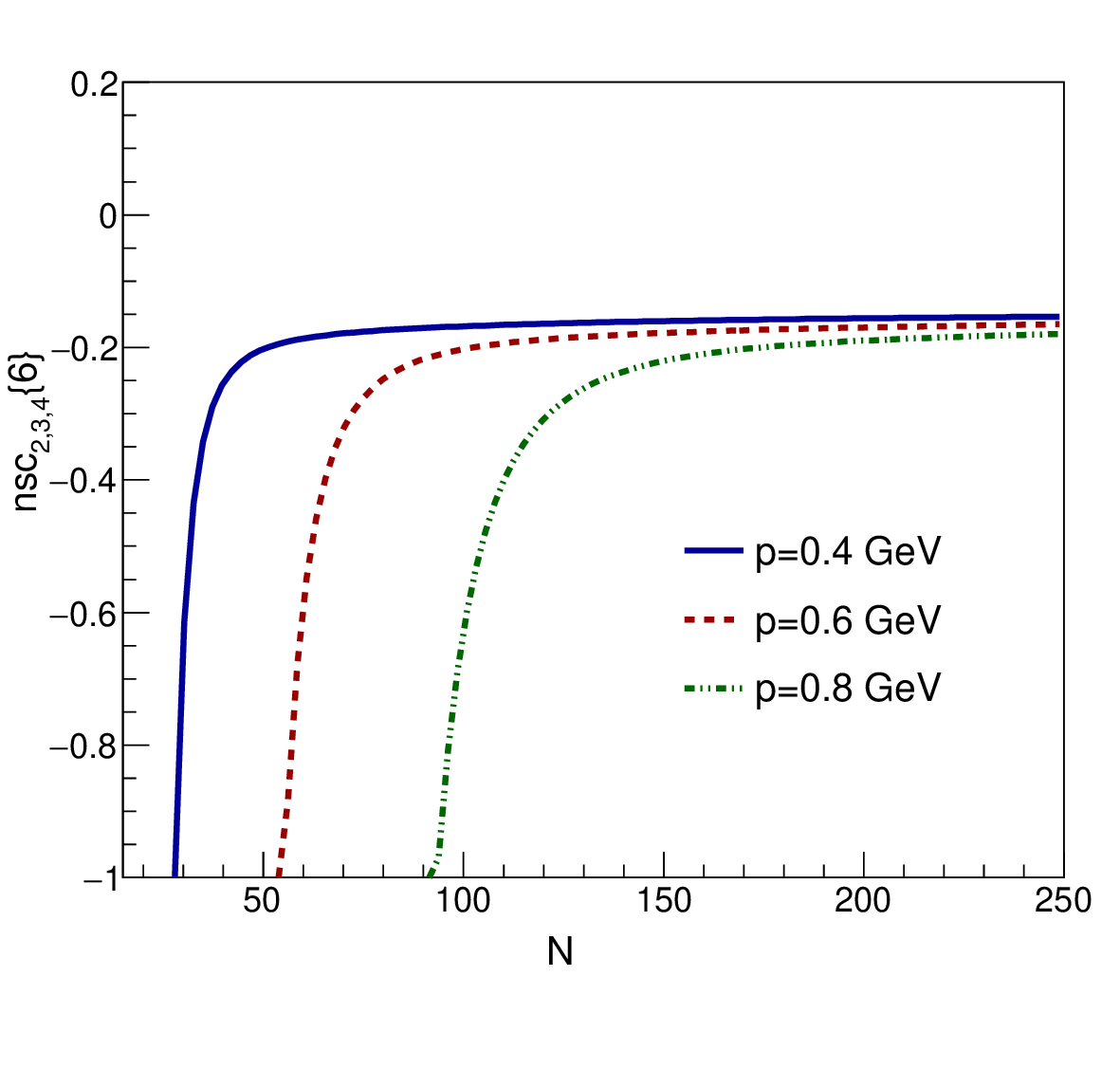}
\caption{The six-particle symmetric cumulant $sc_{2,3,4} \left \{ 6 \right \}$ from ``pure TMC'' (top-left), from ``pure TMC+pure flow+interplay'' (top-right), normalized cumulants  $nsc_{2,3,4} \left \{ 6 \right \}$ from ``pure TMC+pure flow+interplay'' as a function of the number of particles $N$ for various values of transverse momenta $p$ (bottom-right), and $sc_{2,3,4} \left \{ 6 \right \}$ from ``pure TMC,'' ``pure TMC and pure flow,'' and `pure TMC+pure flow+interplay'' as a function of the number of particles $N$ for momentum $p=0.6$ GeV (bottom-left).}
\label{sc234}
\end{figure}
In Fig.~\ref{sc234}, all the parameters are set in the same way as in Fig.~\ref{sc23}. In the top-left panel, as expected, based on Eq.~(\ref{eq22}), the influence of TMC on the correlation between $v_{2}$, $v_{3}$, and $v_{4}$ is more significant for particles with higher momenta $p$ and a smaller number of particles $N$. The negative correlation between these $v_{n}$ gradually approaches zero as $N$ increases. In top-right panel, based on Eqs.~(\ref{eq30}),~(\ref{eq33}),~(\ref{eq36}),~(\ref{eq41}),~(\ref{eq43}),~(\ref{eq44}), and~(\ref{eq45}), which are obtained from the full results as given in the Appendix, the addition of flow enhances the negative correlation and it does not eventually approach zero. In addition, we present the results of $sc_{2,3,4} \left \{ 6 \right \}$ at $p=0.6$ GeV under different $v_n$, as shown in Fig.~\ref{changevn} of the Appendix. In the bottom-left panel, it can be clearly seen that ``pure flow'' enhances the negative correlation, while ``interplay'' weakens the negative correlation. In the bottom-right panel, based on Eq.~(\ref{eq49}), our results on the normalized cumulants $nsc_{2,3,4} \left \{ 6 \right \}$ from ``pure TMC+pure flow+interplay,'' which are independent of the value of $v_{n}$, show the same trend as $sc_{2,3,4} \left \{ 6 \right \}$. Given that we have excluded the lower-order fewer-particle correlations from the calculation, the nonzero outcome of $nsc_{2,3,4} \left \{ 6 \right \}$ indicates the presence of true correlations among the three flow coefficients, providing additional insights for determining $ P\left ( v_{m},v_{n},v_{k}, \dots,\Psi _{m},\Psi _{n},\Psi _{k},\dots \right )$, which cannot be derived from measurements of correlations between just two flow coefficients.

\section{Conclusions}
In this paper, we calculated the four-particle symmetric cumulant $sc_{2,3} \left \{ 4 \right \}$, six-particle symmetric cumulant $sc_{2,3,4} \left \{ 6 \right \}$, and the normalized cumulants $nsc_{2,3} \left \{ 4 \right \}$ and $nsc_{2,3,4} \left \{ 6 \right \}$, originating from the transverse momentum conservation, collective flow, and the interplay between the two effects. We observe that when the number of particles $N$ is small, these correlations originate from the TMC, while with increasing $N$, the collective flow becomes more dominant. Compared to the basis of the TMC, the additional effect of flow generation in small systems is that it leads to the enhancement of the correlations $sc_{2,3} \left \{ 4 \right \}$ and $sc_{2,3,4} \left \{ 6 \right \}$. Our results on $sc_{2,3} \left \{ 4 \right \}$ are consistent with the ATLAS data using the subevent cumulant method, facilitating a more profound understanding of the origins of the symmetric cumulant in small systems. Our results on $sc_{2,3,4} \left \{ 6 \right \}$ can serve as theoretical predictions for future measurements of the correlation of the flow coefficients $v_2$, $ v_3$, and $v_4$ in small systems.

\section*{ACKNOWLEDGMENTS}
This work is partially supported by the National Natural Science Foundation of China under Grants No.12147101 and No. 12325507, the National Key Research and Development Program of China under Grant No. 2022YFA1604900, and the Guangdong Major Project of Basic and Applied Basic Research under Grant No. 2020B0301030008 (J.P. and G.M.), the Ministry of Science and Higher Education (PL), and the National Science Centre (PL), Grant No. 2023/51/B/ST2/01625 (A. B.).

\newpage
\setcounter{section}{0} 
\section*{APPENDIX:FULL EQUATIONS AND PARAMETER SENSITIVITY ANALYSIS}
\renewcommand{\thesection}{A} 
\setcounter{equation}{0} 
\renewcommand{\theequation}{\thesection.\arabic{equation}} 
The full results of Eq.~(\ref{eq32}) are as follows:
\begin{align*}
A_{0}&=v_2^2 v_3^2,\notag \\
A_{1}&=v_2^4 + 6 v_2^2 v_3^2 + v_3^4 + v_2^2 v_4^2 - v_2 v_{2F} v_3^2 \cos(2 \Psi_2) 
- 2 v_2^2 v_{2F} v_3^2 \cos(4 \Psi_2 - 6 \Psi_3)- 2 v_2^3 v_{2F} v_4 \cos(2 \Psi_2 - 4 \Psi_4) 
\notag \\ &- 3 v_2 v_{2F} v_3^2 v_4 \cos(2 \Psi_2 - 4 \Psi_4)+ 2 v_2 v_3^2 v_4 \cos(2 \Psi_2 - 6 \Psi_3 + 4 \Psi_4),\notag \\
A_{2}&=12 v_2^4 + 6 v_2^4 v_{2F}^2 + v_3^2 + 56 v_2^2 v_3^2 + \frac{v_{2F}^2 v_3^2}{2} 
+ 12 v_3^4 + 6 v_{2F}^2 v_3^4 + 16 v_2^2 v_4^2 + 8 v_2^2 v_{2F}^2 v_4^2 + 9 v_3^2 v_4^2 + \frac{9}{2} v_{2F}^2 v_3^2 v_4^2\notag \\ &- 8 v_2^3 v_{2F} \cos(2 \Psi_2) - 28 v_2 v_{2F} v_3^2 \cos(2 \Psi_2) 
- 6 v_2 v_{2F} v_4^2 \cos(2 \Psi_2) + \frac{1}{2} v_2^2 v_{2F}^2 v_3^2 \cos(4 \Psi_2)+ 8 v_2 v_{2F}^2 v_3^2 \cos(2 \Psi_2 - 6 \Psi_3)\notag \\ &- 36 v_2^2 v_{2F} v_3^2 \cos(4 \Psi_2 - 6 \Psi_3) 
+ 6 v_2^2 v_{2F}^2 v_4^2 \cos(4 \Psi_2 - 8 \Psi_4) 
- 6 v_{2F} v_3^2 v_4^2 \cos(6 \Psi_3 - 8 \Psi_4) - 42 v_2^3 v_{2F} v_4 \cos(2 \Psi_2 - 4 \Psi_4)\notag \\ &- 74 v_2 v_{2F} v_3^2 v_4 \cos(2 \Psi_2 - 4 \Psi_4) 
- 6 v_2 v_{2F} v_4^3 \cos(2 \Psi_2 - 4 \Psi_4) 
+ 4 v_2^2 v_4 \cos(4 \Psi_2 - 4 \Psi_4) 
+ 2 v_2^2 v_{2F}^2 v_4 \cos(4 \Psi_2 - 4 \Psi_4)\notag \\ &- 6 v_{2F} v_3^2 v_4 \cos(6 \Psi_3 - 4 \Psi_4) 
+ 8 v_2^2 v_{2F}^2 v_4 \cos(4 \Psi_4) 
+ 9 v_{2F}^2 v_3^2 v_4 \cos(4 \Psi_4)+ 28 v_2 v_3^2 v_4 \cos(2 \Psi_2 - 6 \Psi_3 + 4 \Psi_4)\notag \\ &+ 14 v_2 v_{2F}^2 v_3^2 v_4 \cos(2 \Psi_2 - 6 \Psi_3 + 4 \Psi_4),\notag \\ 
A_{3}&= 16 v_2^2 + 141 v_2^4 + 24 v_2^2 v_{2F}^2 + \frac{423}{2} v_2^4 v_{2F}^2 + 30 v_3^2 + 610 v_2^2 v_3^2 + 45 v_{2F}^2 v_3^2 + 915 v_2^2 v_{2F}^2 v_3^2  + 141 v_3^4 + \frac{423}{2} v_{2F}^2 v_3^4 \\&+ 9 v_4^2 + 234 v_2^2 v_4^2 + \frac{27}{2} v_{2F}^2 v_4^2 + 351 v_2^2 v_{2F}^2 v_4^2  + 180 v_3^2 v_4^2 + 270 v_{2F}^2 v_3^2 v_4^2 + 9 v_4^4 + \frac{27}{2} v_{2F}^2 v_4^4  - 234 v_2^3 v_{2F} \cos(2 \Psi_2) \\&- \frac{117}{2} v_2^3 v_{2F}^3 \cos(2 \Psi_2) - 606 v_2 v_{2F} v_3^2 \cos(2 \Psi_2)  - \frac{303}{2} v_2 v_{2F}^3 v_3^2 \cos(2 \Psi_2) - 216 v_2 v_{2F} v_4^2 \cos(2 \Psi_2) - 54 v_2 v_{2F}^3 v_4^2 \cos(2 \Psi_2) \\
& + 15 v_2^2 v_{2F}^2 \cos(4 \Psi_2) + 9 v_2^4 v_{2F}^2 \cos(4 \Psi_2) + \frac{69}{2} v_2^2 v_{2F}^2 v_3^2 \cos(4 \Psi_2) + \frac{15}{2} v_2^2 v_{2F}^2 v_4^2 \cos(4 \Psi_2)  + \frac{585}{2} v_2 v_{2F}^2 v_3^2 \cos(2 \Psi_2 - 6 \Psi_3) \\&- 564 v_2^2 v_{2F} v_3^2 \cos(4 \Psi_2 - 6 \Psi_3)  - 141 v_2^2 v_{2F}^3 v_3^2 \cos(4 \Psi_2 - 6 \Psi_3) - 15 v_{2F}^3 v_3^2 \cos(6 \Psi_3) - \frac{43}{4} v_2^2 v_{2F}^3 v_3^2 \cos(6 \Psi_3) \\
& - \frac{3}{2} v_{2F}^3 v_3^4 \cos(6 \Psi_3) - 60 v_2 v_{2F}^3 v_4^2 \cos(2 \Psi_2 - 8 \Psi_4)  + 210 v_2^2 v_{2F}^2 v_4^2 \cos(4 \Psi_2 - 8 \Psi_4) - 168 v_{2F} v_3^2 v_4^2 \cos(6 \Psi_3 - 8 \Psi_4) \\
& - 42 v_{2F}^3 v_3^2 v_4^2 \cos(6 \Psi_3 - 8 \Psi_4) - 48 v_2 v_{2F} v_4 \cos(2 \Psi_2 - 4 \Psi_4)  - 732 v_2^3 v_{2F} v_4 \cos(2 \Psi_2 - 4 \Psi_4) - 12 v_2 v_{2F}^3 v_4 \cos(2 \Psi_2 - 4 \Psi_4) \\
& - 183 v_2^3 v_{2F}^3 v_4 \cos(2 \Psi_2 - 4 \Psi_4) - 1365 v_2 v_{2F} v_3^2 v_4 \cos(2 \Psi_2 - 4 \Psi_4)  - \frac{1365}{4} v_2 v_{2F}^3 v_3^2 v_4 \cos(2 \Psi_2 - 4 \Psi_4) \\&- 204 v_2 v_{2F} v_4^3 \cos(2 \Psi_2 - 4 \Psi_4)  - 51 v_2 v_{2F}^3 v_4^3 \cos(2 \Psi_2 - 4 \Psi_4) + 90 v_2^2 v_4 \cos(4 \Psi_2 - 4 \Psi_4)  + 135 v_2^2 v_{2F}^2 v_4 \cos(4 \Psi_2 - 4 \Psi_4) \\&- 6 v_2^3 v_{2F} v_4 \cos(6 \Psi_2 - 4 \Psi_4)  - \frac{3}{2} v_2^3 v_{2F}^3 v_4 \cos(6 \Psi_2 - 4 \Psi_4) - 180 v_{2F} v_3^2 v_4 \cos(6 \Psi_3 - 4 \Psi_4) - 45 v_{2F}^3 v_3^2 v_4 \cos(6 \Psi_3 - 4 \Psi_4) \\&+ 9 v_2 v_{2F}^2 v_3^2 v_4 \cos(2 \Psi_2 + 6 \Psi_3 - 4 \Psi_4)  + \frac{705}{2} v_2^2 v_{2F}^2 v_4 \cos(4 \Psi_4) + 375 v_{2F}^2 v_3^2 v_4 \cos(4 \Psi_4) + 45 v_{2F}^2 v_4^3 \cos(4 \Psi_4) \\
& - 15 v_2 v_{2F}^3 v_4 \cos(2 \Psi_2 + 4 \Psi_4) - 9 v_2^3 v_{2F}^3 v_4 \cos(2 \Psi_2 + 4 \Psi_4)  - 13 v_2 v_{2F}^3 v_3^2 v_4 \cos(2 \Psi_2 + 4 \Psi_4) \\&+ 366 v_2 v_3^2 v_4 \cos(2 \Psi_2 - 6 \Psi_3 + 4 \Psi_4)  + 549 v_2 v_{2F}^2 v_3^2 v_4 \cos(2 \Psi_2 - 6 \Psi_3 + 4 \Psi_4),\\
A_{4}&=
480 v_2^2 + 1720 v_2^4 + 1440 v_2^2 v_{2F}^2 + 5160 v_2^4 v_{2F}^2 + 180 v_2^2 v_{2F}^4 + 645 v_2^4 v_{2F}^4 + 652 v_3^2 + 7213 v_2^2 v_3^2 + 1956 v_{2F}^2 v_3^2 \\&+ 21639 v_2^2 v_{2F}^2 v_3^2 + \frac{489}{2} v_{2F}^4 v_3^2 + \frac{21639}{8} v_2^2 v_{2F}^4 v_3^2 + 1720 v_3^4 + 5160 v_{2F}^2 v_3^4 + 645 v_{2F}^4 v_3^4 + 288 v_4^2 + 3328 v_2^2 v_4^2 + 864 v_{2F}^2 v_4^2 \\&+ 9984 v_2^2 v_{2F}^2 v_4^2 + 108 v_{2F}^4 v_4^2 + 1248 v_2^2 v_{2F}^4 v_4^2 + 2860 v_3^2 v_4^2 + 8580 v_{2F}^2 v_3^2 v_4^2 + \frac{2145}{2} v_{2F}^4 v_3^2 v_4^2 + 240 v_4^4 + 720 v_{2F}^2 v_4^4 \\&+ 90 v_{2F}^4 v_4^4 - 160 v_2 v_{2F} \cos(2 \Psi_2) - 5068 v_2^3 v_{2F} \cos(2 \Psi_2) - 120 v_2 v_{2F}^3 \cos(2 \Psi_2) - 3801 v_2^3 v_{2F}^3 \cos(2 \Psi_2) \\&- 11716 v_2 v_{2F} v_3^2 \cos(2 \Psi_2) - 8787 v_2 v_{2F}^3 v_3^2 \cos(2 \Psi_2) - 5300 v_2 v_{2F} v_4^2 \cos(2 \Psi_2) - 3975 v_2 v_{2F}^3 v_4^2 \cos(2 \Psi_2) \\&+ 750 v_2^2 v_{2F}^2 \cos(4 \Psi_2) + 492 v_2^4 v_{2F}^2 \cos(4 \Psi_2) + 125 v_2^2 v_{2F}^4 \cos(4 \Psi_2) + 82 v_2^4 v_{2F}^4 \cos(4 \Psi_2) \\
&+ 1416 v_2^2 v_{2F}^2 v_3^2 \cos(4 \Psi_2) + 236 v_2^2 v_{2F}^4 v_3^2 \cos(4 \Psi_2) + 510 v_2^2 v_{2F}^2 v_4^2 \cos(4 \Psi_2) + 85 v_2^2 v_{2F}^4 v_4^2 \cos(4 \Psi_2) - 56 v_2^3 v_{2F}^3 \cos(6 \Psi_2) \\
&+ 7479 v_2 v_{2F}^2 v_3^2 \cos(2 \Psi_2 - 6 \Psi_3) + \frac{2493}{2} v_2 v_{2F}^4 v_3^2 \cos(2 \Psi_2 - 6 \Psi_3) - 8600 v_2^2 v_{2F} v_3^2 \cos(4 \Psi_2 - 6 \Psi_3) \\
&- 6450 v_2^2 v_{2F}^3 v_3^2 \cos(4 \Psi_2 - 6 \Psi_3) - 875 v_{2F}^3 v_3^2 \cos(6 \Psi_3) - 757 v_2^2 v_{2F}^3 v_3^2 \cos(6 \Psi_3) - 119 v_{2F}^3 v_3^4 \cos(6 \Psi_3) \\
&- 70 v_{2F}^3 v_3^2 v_4^2 \cos(6 \Psi_3) + \frac{133}{2} v_2 v_{2F}^4 v_3^2 \cos(2 \Psi_2 + 6 \Psi_3) - 2975 v_2 v_{2F}^3 v_4^2 \cos(2 \Psi_2 - 8 \Psi_4)  
\end{align*}

\newpage
\begin{align*}
\\
&+ 5130 v_2^2 v_{2F}^2 v_4^2 \cos(4 \Psi_2 - 8 \Psi_4) + 855 v_2^2 v_{2F}^4 v_4^2 \cos(4 \Psi_2 - 8 \Psi_4) - 3420 v_{2F} v_3^2 v_4^2 \cos(6 \Psi_3 - 8 \Psi_4) \\
&- 2565 v_{2F}^3 v_3^2 v_4^2 \cos(6 \Psi_3 - 8 \Psi_4) - 1860 v_2 v_{2F} v_4 \cos(2 \Psi_2 - 4 \Psi_4) - 12020 v_2^3 v_{2F} v_4 \cos(2 \Psi_2 - 4 \Psi_4) \\
&- 1395 v_2 v_{2F}^3 v_4 \cos(2 \Psi_2 - 4 \Psi_4) - 9015 v_2^3 v_{2F}^3 v_4 \cos(2 \Psi_2 - 4 \Psi_4) - 23004 v_2 v_{2F} v_3^2 v_4 \cos(2 \Psi_2 - 4 \Psi_4) \\
&- 17253 v_2 v_{2F}^3 v_3^2 v_4 \cos(2 \Psi_2 - 4 \Psi_4) - 4620 v_2 v_{2F} v_4^3 \cos(2 \Psi_2 - 4 \Psi_4) - 3465 v_2 v_{2F}^3 v_4^3 \cos(2 \Psi_2 - 4 \Psi_4) \\
&+ 1600 v_2^2 v_4 \cos(4 \Psi_2 - 4 \Psi_4) + 4800 v_2^2 v_{2F}^2 v_4 \cos(4 \Psi_2 - 4 \Psi_4) + 600 v_2^2 v_{2F}^4 v_4 \cos(4 \Psi_2 - 4 \Psi_4)\\
    &- 276 v_2^3 v_{2F} v_4 \cos(6 \Psi_2 - 4 \Psi_4) - 207 v_2^3 v_{2F}^3 v_4 \cos(6 \Psi_2 - 4 \Psi_4) + \frac{163}{2} v_2 v_{2F}^4 v_3^2 v_4 \cos(2 \Psi_2 - 6 \Psi_3 - 4 \Psi_4) \\
&- 3932 v_{2F} v_3^2 v_4 \cos(6 \Psi_3 - 4 \Psi_4) - 2949 v_{2F}^3 v_3^2 v_4 \cos(6 \Psi_3 - 4 \Psi_4) + 549 v_2 v_{2F}^2 v_3^2 v_4 \cos(2 \Psi_2 + 6 \Psi_3 - 4 \Psi_4) \\
&+ \frac{183}{2} v_2 v_{2F}^4 v_3^2 v_4 \cos(2 \Psi_2 + 6 \Psi_3 - 4 \Psi_4) + 180 v_{2F}^2 v_4 \cos(4 \Psi_4) + 10116 v_2^2 v_{2F}^2 v_4 \cos(4 \Psi_4) + 30 v_{2F}^4 v_4 \cos(4 \Psi_4) \\
&+ 1686 v_2^2 v_{2F}^4 v_4 \cos(4 \Psi_4) + 10431 v_{2F}^2 v_3^2 v_4 \cos(4 \Psi_4) + \frac{3477}{2} v_{2F}^4 v_3^2 v_4 \cos(4 \Psi_4) + 1980 v_{2F}^2 v_4^3 \cos(4 \Psi_4) \\
&+ 330 v_{2F}^4 v_4^3 \cos(4 \Psi_4) + 210 v_{2F}^4 v_4^2 \cos(8 \Psi_4) + 134 v_2^2 v_{2F}^4 v_4^2 \cos(8 \Psi_4) + \frac{63}{2} v_{2F}^4 v_3^2 v_4^2 \cos(8 \Psi_4) \\
&- 1085 v_2 v_{2F}^3 v_4 \cos(2 \Psi_2 + 4 \Psi_4) - 708 v_2^3 v_{2F}^3 v_4 \cos(2 \Psi_2 + 4 \Psi_4) - 961 v_2 v_{2F}^3 v_3^2 v_4 \cos(2 \Psi_2 + 4 \Psi_4) \\
&- 112 v_2 v_{2F}^3 v_4^3 \cos(2 \Psi_2 + 4 \Psi_4) + 56 v_2^2 v_{2F}^4 v_4 \cos(4 \Psi_2 + 4 \Psi_4) + 4808 v_2 v_3^2 v_4 \cos(2 \Psi_2 - 6 \Psi_3 + 4 \Psi_4) \\
&+ 14424 v_2 v_{2F}^2 v_3^2 v_4 \cos(2 \Psi_2 - 6 \Psi_3 + 4 \Psi_4) + 1803 v_2 v_{2F}^4 v_3^2 v_4 \cos(2 \Psi_2 - 6 \Psi_3 + 4 \Psi_4),\\
A_{5}&= 100 + 10225 v_2^2 + 21796 v_2^4 + 500 v_{2F}^2 + 51125 v_2^2 v_{2F}^2 + 108980 v_2^4 v_{2F}^2  + \frac{375}{2} v_{2F}^4 + \frac{153375}{8} v_2^2 v_{2F}^4 + \frac{81735}{2} v_2^4 v_{2F}^4 + 12200 v_3^2 \\
& + 89906 v_2^2 v_3^2 + 61000 v_{2F}^2 v_32 + 449530 v_2^2 v_{2F}^2 v_3^2 + 22875 v_{2F}^4 v_3^2  + \frac{674295}{4} v_2^2 v_{2F}^4 v_3^2 + 21761 v_3^4 + 108805 v_{2F}^2 v_3^4 + \frac{326415}{8} v_{2F}^4 v_{3}^4 \\
& + 6400 v_4^2 + 47025 v_2^2 v_4^2 + 32000 v_{2F}^2 v_4^2 + 235125 v_2^2 v_{2F}^2 v_4^2 + 12000 v_{2F}^4 v_4^2  + \frac{705375}{8} v_2^2 v_{2F}^4 v_4^2 + 42560 v_3^2 v_4^2 + 212800 v_{2F}^2 v_3^2 v_4^2\\
&+79800 v_{2F}^4 v_3^2 v_4^2  + 4600 v_4^4 + 23000 v_{2F}^2 v_4^4 + 8625 v_{2F}^4 v_4^4  - 7300 v_2 v_{2F} \cos(2\Psi_2) - 98380 v_2^3 v_{2F} \cos(2\Psi_2) \\
&- 10950 v_2 v_{2F}^3 \cos(2\Psi_2)  - 147570 v_2^3 v_{2F}^3 \cos(2\Psi_2) - \frac{1825}{2} v_2 v_{2F}^5 \cos(2\Psi_2) - \frac{24595}{2} v_2^3 v_{2F}^5 \cos(2\Psi_2) \\
& - 213785 v_{2} v_{2F} v_3^2 \cos(2\Psi_2) - \frac{641355}{2} v_2 v_{2F}^3 v_3^2 \cos(2\Psi_2)  - \frac{213785}{8} v_2 v_{2F}^5 v_3^2 \cos(2\Psi_2) - 111320 v_2 v_{2F} v_4^2 \cos(2\Psi_2) \\
& - 166980 v_2 v_{2F}^3 v_4^2 \cos(2\Psi_2) - 13915 v_2 v_{2F}^5 v_4^2 \cos(2\Psi_2)  + 24535 v_2^2 v_{2F}^2 \cos(4\Psi_2) + 16920 v_2^4 v_{2F}^2 \cos(4\Psi_2) \\&+ \frac{24535}{2} v_2^2 v_{2F}^4 \cos(4\Psi_2)  + 8460 v_2^4 v_{2F}^4 \cos(4\Psi_2) + 44585 v_2^2 v_{2F}^2 v_3^2 \cos(4\Psi_2) + \frac{44585}{2} v_2^2 v_{2F}^4 v_3^2 \cos(4\Psi_2)\\
& + 19995 v_2^2 v_{2F}^2 v_4^2 \cos(4\Psi_2) + \frac{19995}{2} v_2^2 v_{2F}^4 v_4^2 \cos(4\Psi_2)  - 4410 v_2^3 v_{2F}^3 \cos(6\Psi_2) - \frac{2205}{4} v_2^3 v_{2F}^5 \cos(6\Psi_2) \\
& + \frac{315}{4} v_2^4 v_{2F}^4 \cos(8\Psi_2) + 165285 v_2 v_{2F}^2 v_3^2 \cos(2\Psi_2 - 6\Psi_3)  + \frac{165285}{2} v_2 v_{2F}^4 v_3^2 \cos(2\Psi_2 - 6\Psi_3) \\&- 130595 v_2^2 v_{2F} v_3^2 \cos(4\Psi_2 - 6\Psi_3)  - \frac{391785}{2} v_2^2 v_{2F}^3 v_3^2 \cos(4\Psi_2 - 6\Psi_3) - \frac{130595}{8} v_2^2 v_{2F}^5 v_3^2 \cos(4\Psi_2 - 6\Psi_3) \\
& - 31430 v_{2F}^3 v_3^2 \cos(6\Psi_3) - \frac{64385}{2} v_2^2 v_{2F}^3 v_3^2 \cos(6\Psi_3)  - \frac{15715}{4} v_{2F}^5 v_3^2 \cos(6\Psi_3) - \frac{64385}{16} v_2^2 v_{2F}^5 v_3^2 \cos(6\Psi_3) \\
& - 5660 v_{2F}^3 v_3^4 \cos(6\Psi_3) - \frac{1415}{2} v_{2F}^5 v_3^4 \cos(6\Psi_3)  - 5310 v_{2F}^3 v_3^2 v_4^2 \cos(6\Psi_3) - \frac{2655}{4} v_{2F}^5 v_3^2 v_4^2 \cos(6\Psi_3) \\
& + 6585 v_2 v_{2F}^4 v_3^2 \cos(2\Psi_2 + 6\Psi_3) - \frac{861}{8} v_2^2 v_{2F}^5 v_3^2 \cos(4\Psi_2 + 6\Psi_3)  - 180 v_2 v_{2F}^5 v_4^3 \cos(2\Psi_2 - 12\Psi_4) \\&- 95780 v_2 v_{2F}^3 v_4^2 \cos(2\Psi_2 - 8\Psi_4)  - \frac{23945}{2} v_2 v_{2F}^5 v_4^2 \cos(2\Psi_2 - 8\Psi_4) + 108360 v_2^2 v_{2F}^2 v_4^2 \cos(4\Psi_2 - 8\Psi_4) \\
& + 54180 v_2^2 v_{2F}^4 v_4^2 \cos(4\Psi_2 - 8\Psi_4) - 61920 v_{2F} v_3^2 v_4^2 \cos(6\Psi_3 - 8\Psi_4)  - 92880 v_{2F}^3 v_3^2 v_4^2 \cos(6\Psi_3 - 8\Psi_4) \\&- 7740 v_{2F}^5 v_3^2 v_4^2 \cos(6\Psi_3 - 8\Psi_4)  - 48580 v_2 v_{2F} v_4 \cos(2\Psi_2 - 4\Psi_4) - 192660 v_2^3 v_{2F} v_4 \cos(2\Psi_2 - 4\Psi_4)\\&
 - 72870 v_2 v_{2F}^3 v_4 \cos(2\Psi_2 - 4\Psi_4) - 288990 v_2^3 v_{2F}^3 v_4 \cos(2\Psi_2 - 4\Psi_4)  - \frac{12145}{2} v_2 v_{2F}^5 v_4 \cos(2\Psi_2 - 4\Psi_4) 
\end{align*}

\newpage
\begin{align}
&- \frac{48165}{2} v_2^3 v_{2F}^5 v_4 \cos(2\Psi_2 - 4\Psi_4)  - 373555 v_2 v_{2F} v_3^2 v_4 \cos(2\Psi_2 - 4\Psi_4) - \frac{1120665}{2} v_2 v_{2F}^3 v_3^2 v_4 \cos(2\Psi_2 - 4\Psi_4)\notag
\\& - \frac{373555}{8} v_2 v_{2F}^5 v_3^2 v_4 \cos(2\Psi_2 - 4\Psi_4) - 89520 v_2 v_{2F} v_4^3 \cos(2\Psi_2 - 4\Psi_4)  - 134280 v_2 v_{2F}^3 v_4^3 \cos(2\Psi_2 - 4\Psi_4) \notag\\&- 11190 v_2 v_{2F}^5 v_4^3 \cos(2\Psi_2 - 4\Psi_4)+ 26490 v_2^2 v_4 \cos(4\Psi_2 - 4\Psi_4)  + 132450 v_2^2 v_{2F}^2 v_4 \cos(4\Psi_2 - 4\Psi_4)\notag\\&
 + \frac{198675}{4} v_2^2 v_{2F}^4 v_4 \cos(4\Psi_2 - 4\Psi_4)  - 8220 v_2^3 v_{2F} v_4 \cos(6\Psi_2 - 4\Psi_4)  - 12330 v_2^3 v_{2F}^3 v_4 \cos(6\Psi_2 - 4\Psi_4) \notag\\
& - \frac{2055}{2} v_2^3 v_{2F}^5 v_4 \cos(6\Psi_2 - 4\Psi_4)  + \frac{55185}{8} v_2 v_{2F}^4 v_3^2 v_4 \cos(2\Psi_2 - 6\Psi_3 - 4\Psi_4) - 76170 v_{2F} v_3^2 v_4 \cos(6\Psi_3 - 4\Psi_4)\notag \\
& - 114255 v_{2F}^3 v_3^2 v_4 \cos(6\Psi_3 - 4\Psi_4)  - \frac{38085}{4} v_{2F}^5 v_3^2 v_4 \cos(6\Psi_3 - 4\Psi_4)  + 20645 v_2 v_{2F}^2 v_3^2 v_4 \cos(2\Psi_2 + 6\Psi_3 - 4\Psi_4)
\notag \\& + \frac{20645}{2} v_2 v_{2F}^4 v_3^2 v_4 \cos(2\Psi_2 + 6\Psi_3 - 4\Psi_4)  + 9975 v_{2F}^2 v_4 \cos(4\Psi_4) + 242115 v_2^2 v_{2F}^2 v_4 \cos(4\Psi_4) \notag\\
& + \frac{9975}{2} v_{2F}^4 v_4 \cos(4\Psi_4) + \frac{242115}{2} v_2^2 v_{2F}^4 v_4 \cos(4\Psi_4)  + 243595 v_{2F}^2 v_3^2 v_4 \cos(4\Psi_4) + \frac{243595}{2} v_{2F}^4 v_3^2 v_4 \cos(4\Psi_4) \notag\\
& + 57330 v_{2F}^2 v_4^3 \cos(4\Psi_4) + 28665 v_{2F}^4 v_4^3 \cos(4\Psi_4)  + \frac{26775}{2} v_{2F}^4 v_4^2 \cos(8\Psi_4) + \frac{80865}{8} v_2^2 v_{2F}^4 v_4^2 \cos(8\Psi_4) \notag\\
& + 3180 v_{2F}^4 v_3^2 v_4^2 \cos(8\Psi_4) + \frac{315}{2} v_{2F}^4 v_4^4 \cos(8\Psi_4)  - 44240 v_2 v_{2F}^3 v_4 \cos(2\Psi_2 + 4\Psi_4) - 31385 v_2^3 v_{2F}^3 v_4 \cos(2\Psi_2 + 4\Psi_4)\notag \\
& - 5530 v_2 v_{2F}^5 v_4 \cos(2\Psi_2 + 4\Psi_4) - \frac{31385}{8} v_2^3 v_{2F}^5 v_4 \cos(2\Psi_2 + 4\Psi_4)  - \frac{85365}{2} v_2 v_{2F}^3 v_3^2 v_4 \cos(2\Psi_2 + 4\Psi_4)\notag \\&- \frac{85365}{16} v_2 v_{2F}^5 v_3^2 v_4 \cos(2\Psi_2 + 4\Psi_4)  - 7620 v_2 v_{2F}^3 v_4^3 \cos(2\Psi_2 + 4\Psi_4) - \frac{1905}{2} v_2 v_{2F}^5 v_4^3 \cos(2\Psi_2 + 4\Psi_4)\notag \\
& + \frac{25185}{4} v_2^2 v_{2F}^4 v_4 \cos(4\Psi_2 + 4\Psi_4) - \frac{315}{4} v_2^3 v_{2F}^5 v_4 \cos(6\Psi_2 + 4\Psi_4)  + 64170 v_2 v_3^2 v_4 \cos(2\Psi_2 - 6\Psi_3 + 4\Psi_4)\notag \\&+ 320850 v_2 v_{2F}^2 v_3^2 v_4 \cos(2\Psi_2 - 6\Psi_3 + 4\Psi_4)  + \frac{481275}{4} v_2 v_{2F}^4 v_3^2 v_4 \cos(2\Psi_2 - 6\Psi_3 + 4\Psi_4)\notag \\&- \frac{3135}{4} v_{2F}^5 v_3^2 v_4 \cos(6\Psi_3 + 4\Psi_4)  - 1335 v_2 v_{2F}^5 v_4^2 \cos(2\Psi_2 + 8\Psi_4).
\label{eqA1}
\end{align}

\newpage
The main results of Eq.~(\ref{eq41}) are as follows:
\begin{align*}
D_{0} &= v_{2}^{2}v_{3}^{2}v_{4}^{2},\notag\\
D_{1}& =v_2^2 v_3^4 + v_2^4 v_4^2 + 10 v_2^2 v_3^2 v_4^2 + v_3^4 v_4^2 + v_2^2 v_4^4 
- v_2 v_{2F} v_3^2 v_4^2 \cos(2 \Psi_2)+ 2 v_2^3 v_3^2 v_4 \cos(2 \Psi_2 - 6 \Psi_3 + 4 \Psi_4) 
\notag\\ &+ 2 v_2 v_3^4 v_4 \cos(2 \Psi_2 - 6 \Psi_3 + 4 \Psi_4)+ 2 v_2 v_3^2 v_4^3 \cos(2 \Psi_2 - 6 \Psi_3 + 4 \Psi_4)  ,\notag\\
D_{2}& =9 v_2^4 v_3^2 + 32 v_2^2 v_3^4 + 4 v_3^6 + 24 v_2^4 v_4^2 + v_3^2 v_4^2 + 172 v_2^2 v_3^2 v_4^2+ \frac{1}{2} v_{2F}^2 v_3^2 v_4^2 + 32 v_3^4 v_4^2+ 24 v_2^2 v_4^4 + 9 v_3^2 v_4^4 
- 6 v_2 v_{2F} v_3^4 \cos(2 \Psi_2)\notag\\ &
- 8 v_2^3 v_{2F} v_4^2 \cos(2 \Psi_2) 
- 52 v_2 v_{2F} v_3^2 v_4^2 \cos(2 \Psi_2) - 6 v_2 v_{2F} v_4^4 \cos(2 \Psi_2)+ 4 v_2 v_3^2 v_4^2 \cos(2 \Psi_2 + 6 \Psi_3 - 8 \Psi_4) 
\notag\\ &+ 6 v_2^2 v_3^2 v_4 \cos(4 \Psi_2 - 4 \Psi_4)+ 4 v_2^2 v_4^3 \cos(4 \Psi_2 - 4 \Psi_4) 
- 8 v_2^2 v_{2F} v_3^2 v_4 \cos(6 \Psi_3 - 4 \Psi_4) 
- 6 v_{2F} v_3^4 v_4 \cos(6 \Psi_3 - 4 \Psi_4)\notag\\ &- 6 v_{2F} v_3^2 v_4^3 \cos(6 \Psi_3 - 4 \Psi_4) 
+ 56 v_2^3 v_3^2 v_4 \cos(2 \Psi_2 - 6 \Psi_3 + 4 \Psi_4) 
+ 56 v_2 v_3^4 v_4 \cos(2 \Psi_2 - 6 \Psi_3 + 4 \Psi_4)\notag\\ &+ 56 v_2 v_3^2 v_4^3 \cos(2 \Psi_2 - 6 \Psi_3 + 4 \Psi_4) 
- 6 v_2^2 v_{2F} v_3^2 v_4 \cos(4 \Psi_2 - 6 \Psi_3 + 4 \Psi_4)   ,\notag\\
D_{3}& =9 v_2^6 + 396 v_2^4 v_3^2 + 9 v_3^4 + 936 v_2^2 v_3^4 + \frac{27}{2} v_{2F}^2 v_3^4 + 144 v_3^6 + 16 v_2^2 v_4^2 + 594 v_2^4 v_4^2 + 24 v_2^2 v_{2F}^2 v_4^2 + 66 v_3^2 v_4^2+ 3792 v_2^2 v_3^2 v_4^2 \notag\\ &+ 99 v_{2F}^2 v_3^2 v_4^2+ 936 v_3^4 v_4^2 
+ 9 v_4^4 + 594 v_2^2 v_4^4 + \frac{27}{2} v_{2F}^2 v_4^4 + 396 v_3^2 v_4^4 + 9 v_4^6- 207 v_2^3 v_{2F} v_3^2 \cos(2 \Psi_2) - 450 v_2 v_{2F} v_3^4 \cos(2 \Psi_2)\notag\\ &- 516 v_2^3 v_{2F} v_4^2 \cos(2 \Psi_2) - 2289 v_2 v_{2F} v_3^2 v_4^2 \cos(2 \Psi_2) 
- 360 v_2 v_{2F} v_4^4 \cos(2 \Psi_2) + 15 v_2^2 v_{2F}^2 v_4^2 \cos(4 \Psi_2)\notag\\ &+ 18 v_2^3 v_3^2 \cos(6 \Psi_2 - 6 \Psi_3) 
- 36 v_2^3 v_{2F} v_4^2 \cos(6 \Psi_2 - 8 \Psi_4) 
- 36 v_{2F} v_3^2 v_4^2 \cos(6 \Psi_3 - 8 \Psi_4) 
+ 216 v_2 v_3^2 v_4^2 \cos(2 \Psi_2 + 6 \Psi_3 - 8 \Psi_4)\notag\\ &- 69 v_2 v_{2F} v_3^2 v_4 \cos(2 \Psi_2 - 4 \Psi_4) 
- 48 v_2 v_{2F} v_4^3 \cos(2 \Psi_2 - 4 \Psi_4) 
+ 24 v_2^4 v_4 \cos(4 \Psi_2 - 4 \Psi_4) 
+ 330 v_2^2 v_3^2 v_4 \cos(4 \Psi_2 - 4 \Psi_4) \notag\\ &+ 180 v_2^2 v_4^3 \cos(4 \Psi_2 - 4 \Psi_4) 
- 612 v_2^2 v_{2F} v_3^2 v_4 \cos(6 \Psi_3 - 4 \Psi_4) 
- 408 v_{2F} v_3^4 v_4 \cos(6 \Psi_3 - 4 \Psi_4) 
- 432 v_{2F} v_3^2 v_4^3 \cos(6 \Psi_3 - 4 \Psi_4)\notag\\ &+ 15 v_2 v_{2F}^2 v_3^2 v_4 \cos(2 \Psi_2 + 6 \Psi_3 - 4 \Psi_4) 
+ 24 v_2 v_3^2 v_4 \cos(2 \Psi_2 - 6 \Psi_3 + 4 \Psi_4) 
+ 1494 v_2^3 v_3^2 v_4 \cos(2 \Psi_2 - 6 \Psi_3 + 4 \Psi_4)\notag\\ &+ 36 v_2 v_{2F}^2 v_3^2 v_4 \cos(2 \Psi_2 - 6 \Psi_3 + 4 \Psi_4) 
+ 1488 v_2 v_3^4 v_4 \cos(2 \Psi_2 - 6 \Psi_3 + 4 \Psi_4) 
+ 1494 v_2 v_3^2 v_4^3 \cos(2 \Psi_2 - 6 \Psi_3 + 4 \Psi_4) 
\notag\\ &- 378 v_2^2 v_{2F} v_3^2 v_4 \cos(4 \Psi_2 - 6 \Psi_3 + 4 \Psi_4)   ,\notag\\
D_{4}& = 528 v_2^6 + 529 v_2^2 v_3^2 + 13684 v_2^4 v_3^2 + 1587 v_2^2 v_{2F}^2 v_3^2 + 
648 v_3^4 + 26688 v_2^2 v_3^4 + 1944 v_{2F}^2 v_3^4 + 4224 v_3^6 + 
1216 v_2^2 v_4^2 \notag\\ &+ 15792 v_2^4 v_4^2 + 3648 v_2^2 v_{2F}^2 v_4^2+ 
3208 v_3^2 v_4^2 + 95151 v_2^2 v_3^2 v_4^2 + 9624 v_{2F}^2 v_3^2 v_4^2+ 
26688 v_3^4 v_4^2 + 576 v_4^4 + 15792 v_2^2 v_4^4 \notag\\ &+ 1728 v_{2F}^2 v_4^4+ 
13684 v_3^2 v_4^4 + 528 v_4^6 - 540 v_2^5 v_{2F} \cos(2 \Psi_2) - 
16996 v_2^3 v_{2F} v_3^2 \cos(2 \Psi_2) - 
23004 v_2 v_{2F} v_3^4 \cos(2 \Psi_2) \notag\\ &-160 v_2 v_{2F} v_4^2 \cos(2 \Psi_2) - 
25152 v_2^3 v_{2F} v_4^2 \cos(2 \Psi_2) - 
120 v_2 v_{2F}^3 v_4^2 \cos(2 \Psi_2) - 
94672 v_2 v_{2F} v_3^2 v_4^2 \cos(2 \Psi_2)\notag\\&-16520 v_2 v_{2F} v_4^4 \cos(2 \Psi_2) + 
675 v_2^2 v_{2F}^2 v_3^2 \cos(4 \Psi_2) + 
1740 v_2^2 v_{2F}^2 v_4^2 \cos(4 \Psi_2) - 
540 v_2^2 v_{2F} v_3^2 \cos(4 \Psi_2 - 6 \Psi_3)\notag\\&+ 
1176 v_2^3 v_3^2 \cos(6 \Psi_2 - 6 \Psi_3) + 
810 v_2^2 v_{2F}^2 v_4^2 \cos(4 \Psi_2 - 8 \Psi_4) - 
2760 v_2^3 v_{2F} v_4^2 \cos(6 \Psi_2 - 8 \Psi_4) + 
24 v_2^4 v_4^2 \cos(8 \Psi_2 - 8 \Psi_4)\notag\\&- 
3060 v_{2F} v_3^2 v_4^2 \cos(6 \Psi_3 - 8 \Psi_4) + 
8480 v_2 v_3^2 v_4^2 \cos(2 \Psi_2 + 6 \Psi_3 - 8 \Psi_4) - 
720 v_2^3 v_{2F} v_4 \cos(2 \Psi_2 - 4 \Psi_4) \notag\\&- 
7352 v_2 v_{2F} v_3^2 v_4 \cos(2 \Psi_2 - 4 \Psi_4)- 
4200 v_2 v_{2F} v_4^3 \cos(2 \Psi_2 - 4 \Psi_4) + 
1616 v_2^4 v_4 \cos(4 \Psi_2 - 4 \Psi_4) \notag\\&+ 
13586 v_2^2 v_3^2 v_4 \cos(4 \Psi_2 - 4 \Psi_4)+ 
6616 v_2^2 v_4^3 \cos(4 \Psi_2 - 4 \Psi_4) - 
120 v_2^3 v_{2F} v_4 \cos(6 \Psi_2 - 4 \Psi_4)- 
120 v_{2F} v_3^2 v_4 \cos(6 \Psi_3 - 4 \Psi_4)\notag\\& - 
32088 v_2^2 v_{2F} v_3^2 v_4 \cos(6 \Psi_3 - 4 \Psi_4) - 
90 v_{2F}^3 v_3^2 v_4 \cos(6 \Psi_3 - 4 \Psi_4)- 
19332 v_{2F} v_3^4 v_4 \cos(6 \Psi_3 - 4 \Psi_4) \notag\\&- 
21152 v_{2F} v_3^2 v_4^3 \cos(6 \Psi_3 - 4 \Psi_4) + 
2100 v_2 v_{2F}^2 v_3^2 v_4 \cos(2 \Psi_2 + 6 \Psi_3 - 4 \Psi_4) + 
225 v_{2F}^2 v_3^2 v_4 \cos(4 \Psi_4) + 
180 v_{2F}^2 v_4^3 \cos(4 \Psi_4) \notag\\&+ 
1776 v_2 v_3^2 v_4 \cos(2 \Psi_2 - 6 \Psi_3 + 4 \Psi_4) + 
41040 v_2^3 v_3^2 v_4 \cos(2 \Psi_2 - 6 \Psi_3 + 4 \Psi_4) + 
5328 v_2 v_{2F}^2 v_3^2 v_4 \cos(2 \Psi_2 - 6 \Psi_3 + 4 \Psi_4) \notag\\&+ 
40080 v_2 v_3^4 v_4 \cos(2 \Psi_2 - 6 \Psi_3 + 4 \Psi_4)+ 
41040 v_2 v_3^2 v_4^3 \cos(2 \Psi_2 - 6 \Psi_3 + 4 \Psi_4) - 
17984 v_2^2 v_{2F} v_3^2 v_4 \cos(4 \Psi_2 - 6 \Psi_3 + 4 \Psi_4),\notag\\
D_{5}& = 2025 v_2^4 + 22000 v_2^6 + 10125 v_2^4 v_{2F}^2 
+ 43930 v_2^2 v_3^2 + 438960 v_2^4 v_3^2 + 219650 v_2^2 v_{2F}^2 v_3^2+ 32145 v_3^4 + 760175 v_2^2 v_3^4 \notag\\&+ 160725 v_{2F}^2 v_3^4 + 118800 v_3^6 + 100 v_4^2 + 64060 v_2^2 v_4^2 + 443286 v_2^4 v_4^2 + 500 v_{2F}^2 v_4^2 
+ 320300 v_2^2 v_{2F}^2 v_4^2+ \frac{375}{2} v_{2F}^4 v_4^2 \notag\\&+ 135980 v_3^2 v_4^2
 + 2563698 v_2^2 v_3^2 v_4^2 + 679900 v_{2F}^2 v_3^2 v_4^2 + 760151 v_3^4 v_4^2+ 26500 v_4^4 + 443275 v_2^2 v_4^4 + 132500 v_{2F}^2 v_4^4 + 22000 v_4^6 \notag\\&+ 438960 v_3^2 v_4^4 - 51060 v_2^5 v_{2F} \cos(2 \Psi_2) - 8625 v_2 v_{2F} v_3^2 \cos(2 \Psi_2) 
- 937315 v_2^3 v_{2F} v_3^2 \cos(2 \Psi_2) - \frac{25875}{2} v_2 v_{2F}^3 v_3^2 \cos(2 \Psi_2)\notag\\&- 1004520 v_2 v_{2F} v_3^4 \cos(2 \Psi_2) 
- 20200 v_2 v_{2F} v_4^2 \cos(2 \Psi_2) - 1105640 v_2^3 v_{2F} v_4^2 \cos(2 \Psi_2) 
- 30300 v_2 v_{2F}^3 v_4^2 \cos(2 \Psi_2)\notag\\&- 3757720 v_2 v_{2F} v_3^2 v_4^2 \cos(2 \Psi_2) 
- 689180 v_2 v_{2F} v_4^4 \cos(2 \Psi_2) 
+ 5250 v_2^4 v_{2F}^2 \cos(4 \Psi_2) + 87045 v_2^2 v_{2F}^2 v_3^2 \cos(4 \Psi_2)\notag\\&+ 133840 v_2^2 v_{2F}^2 v_4^2 \cos(4 \Psi_2) 
+ 5250 v_2 v_{2F}^2 v_3^2 \cos(2 \Psi_2 - 6 \Psi_3) 
- 57810 v_2^2 v_{2F} v_3^2 \cos(4 \Psi_2 - 6 \Psi_3) 
+ 53970 v_2^3 v_3^2 \cos(6 \Psi_2 - 6 \Psi_3) \notag\\&- 7000 v_2 v_{2F}^3 v_4^2 \cos(2 \Psi_2 - 8 \Psi_4) 
+ 94080 v_2^2 v_{2F}^2 v_4^2 \cos(4 \Psi_2 - 8 \Psi_4) 
- 146550 v_2^3 v_{2F} v_4^2 \cos(6 \Psi_2 - 8 \Psi_4) 
\end{align*}

\newpage
\begin{align*}
&+ 2180 v_2^4 v_4^2 \cos(8 \Psi_2 - 8 \Psi_4) - 175260 v_{2F} v_3^2 v_4^2 \cos(6 \Psi_3 - 8 \Psi_4)
 + 300480 v_2 v_3^2 v_4^2 \cos(2 \Psi_2 + 6 \Psi_3 - 8 \Psi_4)
\notag \\&- 79780 v_2^3 v_{2F} v_4 \cos(2 \Psi_2 - 4 \Psi_4) 
- 482835 v_2 v_{2F} v_3^2 v_4 \cos(2 \Psi_2 - 4 \Psi_4) 
- 248140 v_2 v_{2F} v_4^3 \cos(2 \Psi_2 - 4 \Psi_4)\notag\\&+ 900 v_2^2 v_4 \cos(4 \Psi_2 - 4 \Psi_4) + 76420 v_2^4 v_4 \cos(4 \Psi_2 - 4 \Psi_4) 
+ 4500 v_2^2 v_{2F}^2 v_4 \cos(4 \Psi_2 - 4 \Psi_4)+ 504010 v_2^2 v_3^2 v_4 \cos(4 \Psi_2 - 4 \Psi_4)\notag\\
&+ 230760 v_2^2 v_4^3 \cos(4 \Psi_2 - 4 \Psi_4) 
- 13200 v_2^3 v_{2F} v_4 \cos(6 \Psi_2 - 4 \Psi_4) 
- 14700 v_{2F} v_3^2 v_4 \cos(6 \Psi_3 - 4 \Psi_4) \notag\\&- 1439945 v_2^2 v_{2F} v_3^2 v_4 \cos(6 \Psi_3 - 4 \Psi_4) 
- 22050 v_{2F}^3 v_3^2 v_4 \cos(6 \Psi_3 - 4 \Psi_4)
 - 797970 v_{2F} v_3^4 v_4 \cos(6 \Psi_3 - 4 \Psi_4)\notag\\&- 895470 v_{2F} v_3^2 v_4^3 \cos(6 \Psi_3 - 4 \Psi_4) 
+ 172410 v_2 v_{2F}^2 v_3^2 v_4 \cos(2 \Psi_2 + 6 \Psi_3 - 4 \Psi_4) 
+ 7000 v_2^2 v_{2F}^2 v_4 \cos(4 \Psi_4) \notag\\&+ 38640 v_{2F}^2 v_3^2 v_4 \cos(4 \Psi_4) + 24150 v_{2F}^2 v_4^3 \cos(4 \Psi_4) 
+ 90840 v_2 v_3^2 v_4 \cos(2 \Psi_2 - 6 \Psi_3 + 4 \Psi_4) 
\notag\\&+ 1160560 v_2^3 v_3^2 v_4 \cos(2 \Psi_2 - 6 \Psi_3 + 4 \Psi_4)+ 454200 v_2 v_{2F}^2 v_3^2 v_4 \cos(2 \Psi_2 - 6 \Psi_3 + 4 \Psi_4) 
\notag\\&+ 1103030 v_2 v_3^4 v_4 \cos(2 \Psi_2 - 6 \Psi_3 + 4 \Psi_4)+ 1160510 v_2 v_3^2 v_4^3 \cos(2 \Psi_2 - 6 \Psi_3 + 4 \Psi_4)\notag\\&- 771020 v_2^2 v_{2F} v_3^2 v_4 \cos(4 \Psi_2 - 6 \Psi_3 + 4 \Psi_4),\\
D_{6}&= 192240 v_2^4 + 805320 v_2^6 + 1441800 v_2^4 v_{2F}^2 + 5625 v_3^2 + 2413386 v_2^2 v_3^2 + 13704259 v_2^4 v_3^2 + \frac{84375}{2} v_{2F}^2 v_3^2 + 18100395 v_2^2 v_{2F}^2 v_3^2\\& + \frac{253125}{8} v_{2F}^4 v_3^2 + 1364940 v_3^4  + 21845160 v_2^2 v_3^4 + 10237050 v_{2F}^2 v_3^4 + 3336156 v_3^6 + 13200 v_4^2 + 2893520 v_2^2 v_4^2 \\
& + 12901176 v_2^4 v_4^2 + 99000 v_{2F}^2 v_4^2 + 21701400 v_2^2 v_{2F}^2 v_4^2 + 74250 v_{2F}^4 v_4^2  + 5345910 v_3^2 v_4^2 + 72096207 v_2^2 v_3^2 v_4^2 + 40094325 v_{2F}^2 v_3^2 v_4^2 \\&+ 21841308 v_3^4 v_4^2  + 1076880 v_4^4 + 12898728 v_2^2 v_4^4 + 8076600 v_{2F}^2 v_4^4 + 13701604 v_3^2 v_4^4 + 805320 v_4^6  - 94500 v_2^3 v_{2F} \cos(2\Psi_2)\\&
 - 3129000 v_2^5 v_{2F} \cos(2\Psi_2) - 236250 v_2^3 v_{2F}^3 \cos(2\Psi_2)  - 1050840 v_2 v_{2F} v_3^2 \cos(2\Psi_2) - 43761864 v_2^3 v_{2F} v_3^2 \cos(2\Psi_2) \\
& - 2627100 v_2 v_{2F}^3 v_3^2 \cos(2\Psi_2) - 40502862 v_2 v_{2F} v_3^4 \cos(2\Psi_2)  - 1578360 v_2 v_{2F} v_4^2 \cos(2\Psi_2) - 45981984 v_2^3 v_{2F} v_4^2 \cos(2\Psi_2) \\
& - 3945900 v_2 v_{2F}^3 v_4^2 \cos(2\Psi_2) - 144827712 v_2 v_{2F} v_3^2 v_4^2 \cos(2\Psi_2)  - 27408450 v_2 v_{2F} v_4^4 \cos(2\Psi_2) + 715260 v_2^4 v_{2F}^2 \cos(4\Psi_2) \\
& + 6884325 v_2^2 v_{2F}^2 v_3^2 \cos(4\Psi_2) + 8492820 v_2^2 v_{2F}^2 v_4^2 \cos(4\Psi_2)  - 18900 v_2^3 v_{2F}^3 \cos(6\Psi_2) + 820260 v_2 v_{2F}^2 v_3^2 \cos(2\Psi_2 - 6\Psi_3)\\
& - 3857280 v_2^2 v_{2F} v_3^2 \cos(4\Psi_2 - 6\Psi_3) + 2161740 v_2^3 v_3^2 \cos(6\Psi_2 - 6\Psi_3)  - 18900 v_{2F}^3 v_3^2 \cos(6\Psi_3) \\&- 1135890 v_2 v_{2F}^3 v_4^2 \cos(2\Psi_2 - 8\Psi_4)  + 6922020 v_2^2 v_{2F}^2 v_4^2 \cos(4\Psi_2 - 8\Psi_4) - 6717900 v_2^3 v_{2F} v_4^2 \cos(6\Psi_2 - 8\Psi_4) \\
& + 129420 v_2^4 v_4^2 \cos(8\Psi_2 - 8\Psi_4) - 8489040 v_{2F} v_3^2 v_4^2 \cos(6\Psi_3 - 8\Psi_4)  + 10231002 v_2 v_3^2 v_4^2 \cos(2\Psi_2 + 6\Psi_3 - 8\Psi_4) \\&- 21000 v_2 v_{2F} v_4 \cos(2\Psi_2 - 4\Psi_4)  - 5518380 v_2^3 v_{2F} v_4 \cos(2\Psi_2 - 4\Psi_4) - 52500 v_2 v_{2F}^3 v_4 \cos(2\Psi_2 - 4\Psi_4) \\
& - 25612140 v_2 v_{2F} v_3^2 v_4 \cos(2\Psi_2 - 4\Psi_4) - 12396360 v_2 v_{2F} v_4^3 \cos(2\Psi_2 - 4\Psi_4)  + 102120 v_2^2 v_4 \cos(4\Psi_2 - 4\Psi_4)\\&
 + 3149832 v_2^4 v_4 \cos(4\Psi_2 - 4\Psi_4) + 765900 v_2^2 v_{2F}^2 v_4 \cos(4\Psi_2 - 4\Psi_4)  + 17841576 v_2^2 v_3^2 v_4 \cos(4\Psi_2 - 4\Psi_4) \\&+ 7908780 v_2^2 v_4^3 \cos(4\Psi_2 - 4\Psi_4)  - 925260 v_2^3 v_{2F} v_4 \cos(6\Psi_2 - 4\Psi_4) - 1110060 v_{2F} v_3^2 v_4 \cos(6\Psi_3 - 4\Psi_4) \\
& - 59619936 v_2^2 v_{2F} v_3^2 v_4 \cos(6\Psi_3 -4\Psi_4) - 2775150 v_{2F}^3 v_3^2 v_4 \cos6\Psi_3 - 4\Psi_4)  - 30828702 v_{2F} v_3^4 v_4 \cos(6\Psi_3 - 4\Psi_4) \\&- 35350170 v_{2F} v_3^2 v_4^3 \cos(6\Psi_3 - 4\Psi_4)  + 11038230 v_2 v_{2F}^2 v_3^2 v_4 \cos(2\Psi_2 + 6\Psi_3 - 4\Psi_4) + 1135680 v_2^2 v_{2F}^2 v_4 \cos(4\Psi_4) \\
& + 3639510 v_{2F}^2 v_3^2 v_4 \cos(4\Psi_4)+ 2005080 v_{2F}^2 v_4^3 \cos(4\Psi_4) + 23625 v_{2F}^4 v_4^2 \cos(8\Psi_4)- 25200 v_2 v_{2F}^3 v_4 \cos(2\Psi_2 + 4\Psi_4)\\&
 + 3979890 v_2 v_3^2 v_4 \cos(2\Psi_2 - 6\Psi_3 + 4\Psi_4) + 33576176 v_2^3 v_3^2 v_4 \cos(2\Psi_2 - 6\Psi_3 + 4\Psi_4) \\
& + 29849175 v_2 v_{2F}^2 v_3^2 v_4 \cos(2\Psi_2 - 6\Psi_3 + 4\Psi_4) + 31011132 v_2 v_3^4 v_4 \cos(2\Psi_2 - 6\Psi_3 + 4\Psi_4) \\
& + 33568400 v_2 v_3^2 v_4^3 \cos(2\Psi_2 - 6\Psi_3 + 4\Psi_4) - 31272960 v_2^2 v_{2F} v_3^2 v_4 \cos(4\Psi_2 - 6\Psi_3 + 4\Psi_4),\\
D_{7}&= 122500 v_2^2 + 11687774 v_2^4 + 27751444 v_2^6 + 1286250 v_2^2 v_{2F}^2 + 122721627 v_2^4 v_{2F}^2  + \frac{3215625}{2} v_2^2 v_{2F}^4 + 683550 v_3^2 \\&+ 111134380 v_2^2 v_3^2 + 424167758 v_2^4 v_3^2  + 7177275 v_{2F}^2 v_3^2 + 1166910990 v_2^2 v_{2F}^2 v_3^2 + \frac{35886375}{4} v_{2F}^4 v_3^2  + 53454618 v_3^4 \\&+ 635623787 v_2^2 v_3^4 + 561273489 v_{2F}^2 v_3^4 + 94714536 v_3^6  + 1045660 v_4^2 + 120215256 v_2^2 v_4^2 + 384736184 v_2^4 v_4^2 + 10979430 v_{2F}^2 v_4^2 \\
& + 1262260188 v_2^2 v_{2F}^2 v_4^2 + \frac{27448575}{2} v_{2F}^4 v_4^2 + 200853016 v_3^2 v_4^2  + 2086298754 v_2^2 v_3^2 v_4^2 + 2108956668 v_{2F}^2 v_3^2 v_4^2 + 635260787 v_3^4 v_4^2 \\
& + 41165915 v_4^4 + 384465110 v_2^2 v_4^4 + \frac{864484215}{2} v_{2F}^2 v_4^4 + 423775408 v_3^2 v_4^4  + 27733804 v_4^6-12296060 v_2^3 v_{2F} \cos(2\Psi_2)\\&
 -158684008 v_2^5 v_{2F} \cos(2\Psi_2) - 46110225 v_2^3 v_{2F}^3 \cos(2\Psi_2) - 78825516 v_2 v_{2F} v_3^2 \cos(2\Psi_2)  - 1871926644 v_2^3 v_{2F} v_3^2 \cos(2\Psi_2) \\&- 295595685 v_2 v_{2F}^3 v_3^2 \cos(2\Psi_2) - 1562069460 v_2 v_{2F} v_3^4 \cos(2\Psi_2)  - 97854960 v_2 v_{2F} v_4^2 \cos(2\Psi_2) 
\end{align*}

\newpage
\begin{align*}
&- 1845374104 v_2^3 v_{2F} v_4^2 \cos(2\Psi_2) - 366956100 v_2 v_{2F}^3 v_4^2 \cos(2\Psi_2)  - 5466399603 v_2 v_{2F} v_3^2 v_4^2 \cos(2\Psi_2) \\&- 1058771000 v_2 v_{2F} v_4^4 \cos(2\Psi_2) + 595350 v_2^2 v_{2F}^2 \cos(4\Psi_2)  + 59746680 v_2^4 v_{2F}^2 \cos(4\Psi_2) + 992250 v_2^2 v_{2F}^4 \cos(4\Psi_2) \\&+ 434680407 v_2^2 v_{2F}^2 v_3^2 \cos(4\Psi_2)  + 478840824 v_2^2 v_{2F}^2 v_4^2 \cos(4\Psi_2) - 3461850 v_2^3 v_{2F}^3 \cos(6\Psi_2)\\
& +73856475 v_2 v_{2F}^2 v_3^2 \cos(2\Psi_2 - 6\Psi_3) - 208028632 v_2^2 v_{2F} v_3^2 \cos(4\Psi_2 - 6\Psi_3)  + 81108244 v_2^3 v_3^2 \cos(6\Psi_2 - 6\Psi_3) 
\\&- 4013100 v_{2F}^3 v_3^2 \cos(6\Psi_3) - 109757550 v_2 v_{2F}^3 v_4^2 \cos(2\Psi_2 - 8\Psi_4)  + 412711110 v_2^2 v_{2F}^2 v_4^2 \cos(4\Psi_2 - 8\Psi_4) \\&- 285501048 v_2^3 v_{2F} v_4^2 \cos(6\Psi_2 - 8\Psi_4)  + 6363840 v_2^4 v_4^2 \cos(8\Psi_2 - 8\Psi_4) - 375325216 v_{2F} v_3^2 v_4^2 \cos(6\Psi_3 - 8\Psi_4) \\
& + 342787396 v_2 v_3^2 v_4^2 \cos(2\Psi_2 + 6\Psi_3 - 8\Psi_4) - 3337880 v_2 v_{2F} v_4 \cos(2\Psi_2 - 4\Psi_4)  - 307795656 v_2^3 v_{2F} v_4 \cos(2\Psi_2 - 4\Psi_4) \\&- 12517050 v_2 v_{2F}^3 v_4 \cos(2\Psi_2 - 4\Psi_4) - 1207544996 v_2 v_{2F} v_3^2 v_4 \cos(2\Psi_2 - 4\Psi_4) - 564346440 v_2 v_{2F} v_4^3 \cos(2\Psi_2 - 4\Psi_4)
\\& +7110880 v_2^2 v_4 \cos(4\Psi_2 - 4\Psi_4) + 121323664 v_2^4 v_4 \cos(4\Psi_2 - 4\Psi_4) + 74664240 v_2^2 v_{2F}^2 v_4 \cos(4\Psi_2 - 4\Psi_4) \\
& + 617328544 v_2^2 v_3^2 v_4 \cos(4\Psi_2 - 4\Psi_4) + 269241616 v_2^2 v_4^3 \cos(4\Psi_2 - 4\Psi_4) - 52989776 v_2^3 v_{2F} v_4 \cos(6\Psi_2 - 4\Psi_4) \\
& - 66446940 v_{2F} v_3^2 v_4 \cos(6\Psi_3 - 4\Psi_4) - 2356294857 v_2^2 v_{2F} v_3^2 v_4 \cos(6\Psi_3 - 4\Psi_4)  - 249176025 v_{2F}^3 v_3^2 v_4 \cos(6\Psi_3 - 4\Psi_4) \\&- 1149727236 v_{2F} v_3^4 v_4 \cos(6\Psi_3 - 4\Psi_4)  - 1343186922 v_{2F} v_3^2 v_4^3 \cos(6\Psi_3 - 4\Psi_4) \\&+ 613360314 v_2 v_{2F}^2 v_3^2 v_4 \cos(2\Psi_2 + 6\Psi_3 - 4\Psi_4)  + 132300 v_{2F}^2 v_4 \cos(4\Psi_4) + 106567650 v_2^2 v_{2F}^2 v_4 \cos(4\Psi_4)\\& + 220500 v_{2F}^4 v_4 \cos(4\Psi_4)  + 257635350 v_{2F}^2 v_3^2 v_4 \cos(4\Psi_4) +132202350 v_{2F}^2 v_4^3 \cos(4\Psi_4) + \frac{20252925}{4} v_{2F}^4 v_4^2 \cos(8\Psi_4) \\&- 5571300 v_2 v_{2F}^3 v_4 \cos(2\Psi_2 + 4\Psi_4)  + 160556060 v_2 v_3^2 v_4 \cos(2\Psi_2 - 6\Psi_3 + 4\Psi_4) + 989273880 v_2^3 v_3^2 v_4 \cos(2\Psi_2 - 6\Psi_3 + 4\Psi_4) \\
& + 1685838630 v_2 v_{2F}^2 v_3^2 v_4 \cos(2\Psi_2 - 6\Psi_3 + 4\Psi_4) + 889092050 v_2 v_3^4 v_4 \cos(2\Psi_2 - 6\Psi_3 + 4\Psi_4) \\
& + 988550136 v_2 v_3^2 v_4^3 \cos(2\Psi_2 - 6\Psi_3 + 4\Psi_4) - 1225487746 v_2^2 v_{2F} v_3^2 v_4 \cos(4\Psi_2 - 6\Psi_3 + 4\Psi_4),\\
D_{8}&= 15758400 v_2^2 + 583338784 v_2^4 + 928175360 v_2^6 + 220617600 v_2^2 v_{2F}^2 + 8166742976 v_2^4 v_{2F}^2  + 413658000 v_2^2 v_{2F}^4 + 50584464 v_3^2 \\&+ 4656556688 v_2^2 v_3^2 + 13127072536 v_2^4 v_3^2 + 708182496 v_{2F}^2 v_3^2  + 65191793632 v_2^2 v_{2F}^2 v_3^2 + 1327842180 v_{2F}^4 v_3^2 + 1998372096 v_3^4 \\&+ 18743539280 v_2^2 v_3^4  + 27977209344 v_{2F}^2 v_3^4 + 2728263840 v_3^6 + 64366400 v_4^2 + 4744175744 v_2^2 v_4^2 + 11678200416 v_2^4 v_4^2 \\
& + 901129600 v_{2F}^2 v_4^2 + 66418460416 v_2^2 v_{2F}^2 v_4^2 + 1689618000 v_{2F}^4 v_4^2 + 7332421488 v_3^2 v_4^2  + 61654919663 v_2^2 v_3^2 v_4^2 \\&+ 102653900832 v_{2F}^2 v_3^2 v_4^2 + 18717443888 v_3^4 v_4^2 + 1519608608 v_4^4+11656927648 v_2^2 v_4^4 + 21274520512 v_{2F}^2 v_4^4\\&
 +13092612712 v_3^2 v_4^4 + 925606976 v_4^6 - 3528000 v_2 v_{2F} \cos(2\Psi_2) - 974852256 v_2^3 v_{2F} \cos(2\Psi_2)  - 7260499680 v_2^5 v_{2F} \cos(2\Psi_2) \\&- 18522000 v_2 v_{2F}^3 \cos(2\Psi_2) - 5117974344 v_2^3 v_{2F}^3 \cos(2\Psi_2)  - 15435000 v_2 v_{2F}^5 \cos(2\Psi_2) - 4711743792 v_2 v_{2F} v_3^2 \cos(2\Psi_2) \\&- 76059832520 v_2^3 v_{2F} v_3^2 \cos(2\Psi_2)  - 24736654908 v_2 v_{2F}^3 v_3^2 \cos(2\Psi_2) - 58680838560 v_2 v_{2F} v_3^4 \cos(2\Psi_2) \\&- 5282520096 v_2 v_{2F} v_4^2 \cos(2\Psi_2)  - 72246858880 v_2^3 v_{2F} v_4^2 \cos(2\Psi_2) - 27733230504 v_2 v_{2F}^3 v_4^2 \cos(2\Psi_2) \\&- 203303536640 v_2 v_{2F} v_3^2 v_4^2 \cos(2\Psi_2)  - 40106316160 v_2 v_{2F} v_4^4 \cos(2\Psi_2) + 100107000 v_2^2 v_{2F}^2 \cos(4\Psi_2)\\&
 +3969918120 v_2^4 v_{2F}^2 \cos(4\Psi_2) + 250267500 v_2^2 v_{2F}^4 \cos(4\Psi_2) + 24127160400 v_2^2 v_{2F}^2 v_3^2 \cos(4\Psi_2)  \\&+ 24905673240 v_2^2 v_{2F}^2 v_4^2 \cos(4\Psi_2) - 374820600 v_2^3 v_{2F}^3 \cos(6\Psi_2) + 5099740380 v_2 v_{2F}^2 v_3^2 \cos(2\Psi_2 - 6\Psi_3) \\
& - 9962298864 v_2^2 v_{2F} v_3^2 \cos(4\Psi_2 - 6\Psi_3) + 2939171536 v_2^3 v_3^2 \cos(6\Psi_2 - 6\Psi_3) - 463270500 v_{2F}^3 v_3^2 \cos(6\Psi_3) \\
& - 8205036840 v_2 v_{2F}^3 v_4^2 \cos(2\Psi_2 - 8\Psi_4) + 21780719940 v_2^2 v_{2F}^2 v_4^2 \cos(4\Psi_2 - 8\Psi_4)  - 11603563776 v_2^3 v_{2F} v_4^2 \cos(6\Psi_2 - 8\Psi_4) \\&+ 281587824 v_2^4 v_4^2 \cos(8\Psi_2 - 8\Psi_4)  - 15686288184 v_{2F} v_3^2 v_4^2 \cos(6\Psi_3 - 8\Psi_4) + 11418742160 v_2 v_3^2 v_4^2 \cos(2\Psi_2 + 6\Psi_3 - 8\Psi_4) \\
& - 303619680 v_2 v_{2F} v_4 \cos(2\Psi_2 - 4\Psi_4)-15190639968 v_2^3 v_{2F} v_4 \cos(2\Psi_2 - 4\Psi_4)  - 1594003320 v_2 v_{2F}^3 v_4 \cos(2\Psi_2 - 4\Psi_4)\\
& -52935999984 v_2 v_{2F} v_3^2 v_4 \cos(2\Psi_2 - 4\Psi_4) - 24225082896 v_2 v_{2F} v_4^3 \cos(2\Psi_2 - 4\Psi_4)  + 395491488 v_2^2 v_4 \cos(4\Psi_2 - 4\Psi_4) \\&+ 4501681184 v_2^4 v_4 \cos(4\Psi_2 - 4\Psi_4)  + 5536880832 v_2^2 v_{2F}^2 v_4 \cos(4\Psi_2 - 4\Psi_4) + 21121353092 v_2^2 v_3^2 v_4 \cos(4\Psi_2 - 4\Psi_4)  \\&+ 9142187936 v_2^2 v_4^3 \cos(4\Psi_2 - 4\Psi_4) - 2709243936 v_2^3 v_{2F} v_4 \cos(6\Psi_2 - 4\Psi_4)  - 3469371696 v_{2F} v_3^2 v_4 \cos(6\Psi_3 - 4\Psi_4) \\&- 90489148624 v_2^2 v_{2F} v_3^2 v_4 \cos(6\Psi_3 - 4\Psi_4)  - 18214201404 v_{2F}^3 v_3^2 v_4 \cos(6\Psi_3 - 4\Psi_4) - 42032087664 v_{2F} v_3^4 v_4 \cos(6\Psi_3 - 4\Psi_4)  \\&- 49896268968 v_{2F} v_3^2 v_4^3 \cos(6\Psi_3 - 4\Psi_4)+31170903750 v_2 v_{2F}^2 v_3^2 v_4 \cos(2\Psi_2 + 6\Psi_3 - 4\Psi_4)+27694800 v_{2F}^2 v_4 \cos(4\Psi_4)\\&
 +7645869840 v_2^2 v_{2F}^2 v_4 \cos(4\Psi_4) + 69237000 v_{2F}^4 v_4 \cos(4\Psi_4) + 15428387520 v_{2F}^2 v_3^2 v_4 \cos(4\Psi_4) \\
& + 7583242800 v_{2F}^2 v_4^3 \cos(4\Psi_4) + 616562100 v_{2F}^4 v_4^2 \cos(8\Psi_4) - 673642200 v_2 v_{2F}^3 v_4 \cos(2\Psi_2 + 4\Psi_4) \\
& + 6166488664 v_2 v_3^2 v_4 \cos(2\Psi_2 - 6\Psi_3 + 4\Psi_4) + 29598015040 v_2^3 v_3^2 v_4 \cos(2\Psi_2 - 6\Psi_3 + 4\Psi_4) 
\end{align*}

\newpage
\begin{align}
& + 86330841296 v_2 v_{2F}^2 v_3^2 v_4 \cos(2\Psi_2 - 6\Psi_3 + 4\Psi_4) + 25941519584 v_2 v_3^4 v_4 \cos(2\Psi_2 - 6\Psi_3 + 4\Psi_4) \notag\\
& + 29546310624 v_2 v_3^2 v_4^3 \cos(2\Psi_2 - 6\Psi_3 + 4\Psi_4) - 46933893328 v_2^2 v_{2F} v_3^2 v_4 \cos(4\Psi_2 - 6\Psi_3 + 4\Psi_4),\notag\\
D_{9}&= 1587600 + 1225008036 v_2^2 + 26117211708 v_2^4 + 30607136928 v_2^6 + 28576800 v_{2F}^2  + 22050144648 v_2^2 v_{2F}^2\notag \\&+ 470109810744 v_2^4 v_{2F}^2 + 75014100 v_{2F}^4 + 57881629701 v_2^2 v_{2F}^4  + 41674500 v_{2F}^6 + 2959322688 v_3^2 \notag\\&+ 184426848396 v_2^2 v_3^2 + 407953797720 v_2^4 v_3^2  + 53267808384 v_{2F}^2 v_3^2 + 3319683271128 v_2^2 v_{2F}^2 v_3^2 + 139827997008 v_{2F}^4 v_3^2 \notag\\
& + 72626781816 v_3^4 + 560063908143 v_2^2 v_3^4 + 1307282072688 v_{2F}^2 v_3^4 + 79772085792 v_3^6  + 3405163104 v_4^2 
\notag \\&+ 180990383784 v_2^2 v_4^2 + 359461665783 v_2^4 v_4^2 + 61292935872 v_{2F}^2 v_4^2  + 3257826908112 v_2^2 v_{2F}^2 v_4^2 + 160893956664 v_{2F}^4 v_4^2\notag\\&
 +262689722856 v_3^2 v_4^2 + 1852643752890 v_2^2 v_3^2 v_4^2 + 4728415011408 v_{2F}^2 v_3^2 v_4^2  + 558478339839 v_3^4 v_4^2 + 54882459864 v_4^4 \notag\\&+ 358102472418 v_2^2 v_4^4 + 987884277552 v_{2F}^2 v_4^4  + 405614006232 v_3^2 v_4^4 + 30385641456 v_4^6 - 567831600 v_2 v_{2F} \cos(2\Psi_2) \notag\\
& - 61133849280 v_2^3 v_{2F} \cos(2\Psi_2) - 312136056792 v_2^5 v_{2F} \cos(2\Psi_2)  - 3974821200 v_2 v_{2F}^3 \cos(2\Psi_2) \notag\\&- 427936944960 v_2^3 v_{2F}^3 \cos(2\Psi_2) - 4968526500 v_2 v_{2F}^5 \cos(2\Psi_2) - 246838566240 v_2 v_{2F} v_3^2 \cos(2\Psi_2) \notag\\
& - 2992218938874 v_2^3 v_{2F} v_3^2 \cos(2\Psi_2) - 1727869963680 v_2 v_{2F}^3 v_3^2 \cos(2\Psi_2)  - 2169082867824 v_2 v_{2F} v_3^4 \cos(2\Psi_2)
\notag\\& -260246831040 v_2 v_{2F} v_4^2 \cos(2\Psi_2) - 2778461968752 v_2^3 v_{2F} v_4^2 \cos(2\Psi_2)  - 1821727817280 v_2 v_{2F}^3 v_4^2 \cos(2\Psi_2) \notag\\&- 7482969109935 v_2 v_{2F} v_3^2 v_4^2 \cos(2\Psi_2) - 1498299692040 v_2 v_{2F} v_4^4 \cos(2\Psi_2) + 9905353920 v_2^2 v_{2F}^2 \cos(4\Psi_2)\notag \\
& + 230787977112 v_2^4 v_{2F}^2 \cos(4\Psi_2) + 34668738720 v_2^2 v_{2F}^4 \cos(4\Psi_2) + 1233016536906 v_2^2 v_{2F}^2 v_3^2 \cos(4\Psi_2) \notag\\&+ 1222036005960 v_2^2 v_{2F}^2 v_4^2 \cos(4\Psi_2)  - 31383128700 v_2^3 v_{2F}^3 \cos(6\Psi_2) + 301069054440 v_2 v_{2F}^2 v_3^2 \cos(2\Psi_2 - 6\Psi_3) \notag\\
& - 443102115960 v_2^2 v_{2F} v_3^2 \cos(4\Psi_2 - 6\Psi_3) + 104443314084 v_2^3 v_3^2 \cos(6\Psi_2 - 6\Psi_3) - 39407237100 v_{2F}^3 v_3^2 \cos(6\Psi_3)\notag\\&
 -524319752880 v_2 v_{2F}^3 v_4^2 \cos(2\Psi_2 - 8\Psi_4) + 1061802001260 v_2^2 v_{2F}^2 v_4^2 \cos(4\Psi_2 - 8\Psi_4)  \notag\\&- 458198939520 v_2^3 v_{2F} v_4^2 \cos(6\Psi_2 - 8\Psi_4) + 11661200964 v_2^4 v_4^2 \cos(8\Psi_2 - 8\Psi_4)  \notag\\&- 631521644040 v_{2F} v_3^2 v_4^2 \cos(6\Psi_3 - 8\Psi_4) + 380010387888 v_2 v_3^2 v_4^2 \cos(2\Psi_2 + 6\Psi_3 - 8\Psi_4) \notag\\& - 21024783360 v_2 v_{2F} v_4 \cos(2\Psi_2 - 4\Psi_4) - 693847508508 v_2^3 v_{2F} v_4 \cos(2\Psi_2 - 4\Psi_4)  - 147173483520 v_2 v_{2F}^3 v_4 \cos(2\Psi_2 - 4\Psi_4) \notag\\&- 2211416439120 v_2 v_{2F} v_3^2 v_4 \cos(2\Psi_2 - 4\Psi_4) - 998886921480 v_2 v_{2F} v_4^3 \cos(2\Psi_2 - 4\Psi_4) + 19354177584 v_2^2 v_4 \cos(4\Psi_2 - 4\Psi_4)\notag \\
& + 163401429240 v_2^4 v_4 \cos(4\Psi_2 - 4\Psi_4)+348375196512 v_2^2 v_{2F}^2 v_4 \cos(4\Psi_2 - 4\Psi_4)  + 718886529684 v_2^2 v_3^2 v_4 \cos(4\Psi_2 - 4\Psi_4)\notag\\&
 +310207573152 v_2^2 v_4^3 \cos(4\Psi_2 - 4\Psi_4)  -128896329240 v_2^3 v_{2F} v_4 \cos(6\Psi_2 - 4\Psi_4)  -165813680340 v_{2F} v_3^2 v_4 \cos(6\Psi_3 - 4\Psi_4) \notag\\& -3411445191318 v_2^2 v_{2F} v_3^2 v_4 \cos(6\Psi_3 - 4\Psi_4)  -1160695762380 v_{2F}^3 v_3^2 v_4 \cos(6\Psi_3 - 4\Psi_4)  \notag\\&-1518941612772 v_{2F} v_3^4 v_4 \cos(6\Psi_3 - 4\Psi_4)  -1827709139124 v_{2F} v_3^2 v_4^3 \cos(6\Psi_3 - 4\Psi_4) \notag\\& +1491477058764 v_2 v_{2F}^2 v_3^2 v_4 \cos(2\Psi_2 + 6\Psi_3 - 4\Psi_4)  +3146358600 v_{2F}^2 v_4 \cos(4\Psi_4)  +466977495600 v_2^2 v_{2F}^2 v_4 \cos(4\Psi_4) \notag\\& +11012255100 v_{2F}^4 v_4 \cos(4\Psi_4)  +828699254460 v_{2F}^2 v_3^2 v_4 \cos(4\Psi_4)+396113616360 v_{2F}^2 v_4^3 \cos(4\Psi_4)\notag\\&
 +55991971035 v_{2F}^4 v_4^2 \cos(8\Psi_4)  -59820397560 v_2 v_{2F}^3 v_4 \cos(2\Psi_2 + 4\Psi_4)  +229566974868 v_2 v_3^2 v_4 \cos(2\Psi_2 - 6\Psi_3 + 4\Psi_4) \notag\\& +897535910118 v_2^3 v_3^2 v_4 \cos(2\Psi_2 - 6\Psi_3 + 4\Psi_4) +4132205547624 v_2 v_{2F}^2 v_3^2 v_4 \cos(2\Psi_2 - 6\Psi_3 + 4\Psi_4) \notag\\& +768860455086 v_2 v_3^4 v_4 \cos(2\Psi_2 - 6\Psi_3 + 4\Psi_4)  +894407729808 v_2 v_3^2 v_4^3 \cos(2\Psi_2 - 6\Psi_3 + 4\Psi_4) \notag\\& -1769222687166 v_2^2 v_{2F} v_3^2 v_4 \cos(4\Psi_2 - 6\Psi_3 + 4\Psi_4).
 \label{eqA2}
\end{align}

\newpage
The full results of Eq.~(\ref{eq45}) are as follows:
\begin{align*}
G_{0}&=v_3^2 v_4^2,\\
G_{1}&=  v_3^4 + v_2^2 v_4^2 + 6 v_3^2 v_4^2 + v_4^4  - 2 v_{2F} v_3^2 v_4^2 \cos(6\Psi_3 - 8\Psi_4) - 3 v_2 v_{2F} v_3^2 v_4 \cos(2\Psi_2 - 4\Psi_4) \\
& - 2 v_2 v_{2F} v_4^3 \cos(2\Psi_2 - 4\Psi_4) + 2 v_2 v_3^2 v_4 \cos(2\Psi_2 - 6\Psi_3 + 4\Psi_4)
,\\
G_{2}&= 9 v_2^2 v_3^2 + \frac{9}{2} v_2^2 v_{2F}^2 v_3^2 + 12 v_3^4 + 6 v_{2F}^2 v_3^4  + 16 v_2^2 v_4^2 + 8 v_2^2 v_{2F}^2 v_4^2 + 56 v_3^2 v_4^2 + 28 v_{2F}^2 v_3^2 v_4^2  + 12 v_4^4 + 6 v_{2F}^2 v_4^4  - 2 v_2 v_{2F} v_4^2 \cos(2\Psi_2) \\
& - 6 v_2^2 v_{2F} v_3^2 \cos(4\Psi_2 - 6\Psi_3)  + 6 v_2^2 v_{2F}^2 v_4^2 \cos(4\Psi_2 - 8\Psi_4)  - 36 v_{2F} v_3^2 v_4^2 \cos(6\Psi_3 - 8\Psi_4)  - 6 v_2^3 v_{2F} v_4 \cos(2\Psi_2 - 4\Psi_4) \\
& - 74 v_2 v_{2F} v_3^2 v_4 \cos(2\Psi_2 - 4\Psi_4)  - 42 v_2 v_{2F} v_4^3 \cos(2\Psi_2 - 4\Psi_4) - 2 v_{2F} v_3^2 v_4 \cos(6\Psi_3 - 4\Psi_4)  + \frac{5}{2} v_{2F}^2 v_3^2 v_4 \cos(4\Psi_4) \\
& + 2 v_{2F}^2 v_4^3 \cos(4\Psi_4)  + 28 v_2 v_3^2 v_4 \cos(2\Psi_2 - 6\Psi_3 + 4\Psi_4)  + 14 v_2 v_{2F}^2 v_3^2 v_4 \cos(2\Psi_2 - 6\Psi_3 + 4\Psi_4),\\
G_{3}&= 9 v_2^4 + \frac{27}{2} v_2^4 v_{2F}^2 + 180 v_2^2 v_3^2 + 270 v_2^2 v_{2F}^2 v_3^2  + 141 v_3^4 + \frac{423}{2} v_{2F}^2 v_3^4 + v_4^2 + 234 v_2^2 v_4^2 + \frac{3}{2} v_{2F}^2 v_4^2  + 351 v_2^2 v_{2F}^2 v_4^2 + 610 v_3^2 v_4^2 \\&+ 915 v_{2F}^2 v_3^2 v_4^2  + 141 v_4^4 + \frac{423}{2} v_{2F}^2 v_4^4  - 45 v_2 v_{2F} v_3^2 \cos(2\Psi_2) - \frac{45}{4} v_2 v_{2F}^3 v_3^2 \cos(2\Psi_2)  - 84 v_2 v_{2F} v_4^2 \cos(2\Psi_2) \\&- 21 v_2 v_{2F}^3 v_4^2 \cos(2\Psi_2)  + \frac{3}{2} v_2^2 v_{2F}^2 v_4^2 \cos(4\Psi_2)  + \frac{45}{2} v_2 v_{2F}^2 v_3^2 \cos(2\Psi_2 - 6\Psi_3)  - 168 v_2^2 v_{2F} v_3^2 \cos(4\Psi_2 - 6\Psi_3) \\
& - 42 v_2^2 v_{2F}^3 v_3^2 \cos(4\Psi_2 - 6\Psi_3)  - \frac{1}{4} v_{2F}^3 v_3^2 v_4^2 \cos(6\Psi_3)  - \frac{45}{2} v_2 v_{2F}^3 v_4^2 \cos(2\Psi_2 - 8\Psi_4)  + 210 v_2^2 v_{2F}^2 v_4^2 \cos(4\Psi_2 - 8\Psi_4) \\
& - 564 v_{2F} v_3^2 v_4^2 \cos(6\Psi_3 - 8\Psi_4)  - 141 v_{2F}^3 v_3^2 v_4^2 \cos(6\Psi_3 - 8\Psi_4)-204 v_2^3 v_{2F} v_4 \cos(2 \Psi_2 - 4 \Psi_4) -51 v_2^3 v_{2F}^3 v_4 \cos(2 \Psi_2 - 4 \Psi_4) \\
& - 1365 v_2 v_{2F} v_3^2 v_4 \cos(2 \Psi_2 - 4 \Psi_4)  - \frac{1365}{4} v_2 v_{2F}^3 v_3^2 v_4 \cos(2 \Psi_2 - 4 \Psi_4)  - 732 v_2 v_{2F} v_4^3 \cos(2 \Psi_2 - 4 \Psi_4) \\& - 183 v_2 v_{2F}^3 v_4^3 \cos(2 \Psi_2 - 4 \Psi_4)  + 6 v_2^2 v_4 \cos(4 \Psi_2 - 4 \Psi_4)  + 9 v_2^2 v_{2F}^2 v_4 \cos(4 \Psi_2 - 4 \Psi_4)  - 72 v_{2F} v_3^2 v_4 \cos(6 \Psi_3 - 4 \Psi_4) \\
& - 18 v_{2F}^3 v_3^2 v_4 \cos(6 \Psi_3 - 4 \Psi_4)  + \frac{3}{2} v_2 v_{2F}^2 v_3^2 v_4 \cos(2 \Psi_2 + 6 \Psi_3 - 4 \Psi_4)  + \frac{45}{2} v_2^2 v_{2F}^2 v_4 \cos(4 \Psi_4)  + 144 v_{2F}^2 v_3^2 v_4 \cos(4 \Psi_4) \\
& + 90 v_{2F}^2 v_4^3 \cos(4 \Psi_4)  - \frac{7}{4} v_2 v_{2F}^3 v_3^2 v_4 \cos(2 \Psi_2 + 4 \Psi_4)  - \frac{3}{2} v_2 v_{2F}^3 v_4^3 \cos(2 \Psi_2 + 4 \Psi_4)  + 366 v_2 v_3^2 v_4 \cos(2 \Psi_2 - 6 \Psi_3 + 4 \Psi_4) \\
& + 549 v_2 v_{2F}^2 v_3^2 v_4 \cos(2 \Psi_2 - 6 \Psi_3 + 4 \Psi_4),\\
G_{4}&= 240 v_2^4 + 720 v_2^4 v_{2F}^2 + 90 v_2^4 v_{2F}^4 + 25 v_3^2 + 2860 v_2^2 v_3^2  + 75 v_{2F}^2 v_3^2 + 8580 v_2^2 v_{2F}^2 v_3^2 + \frac{75}{8} v_{2F}^4 v_3^2  + \frac{2145}{2} v_2^2 v_{2F}^4 v_3^2 + 1720 v_3^4 \\&+ 5160 v_{2F}^2 v_3^4  + 645 v_{2F}^4 v_3^4 + 48 v_4^2 + 3328 v_2^2 v_4^2 + 144 v_{2F}^2 v_4^2  + 9984 v_2^2 v_{2F}^2 v_4^2 + 18 v_{2F}^4 v_4^2 + 1248 v_2^2 v_{2F}^4 v_4^2  + 7212 v_3^2 v_4^2 \\&+ 21636 v_{2F}^2 v_3^2 v_4^2 + \frac{5409}{2} v_{2F}^4 v_3^2 v_4^2  + 1720 v_4^4 + 5160 v_{2F}^2 v_4^4 + 645 v_{2F}^4 v_4^4  - 180 v_2^3 v_{2F} \cos(2 \Psi_2) - 135 v_2^3 v_{2F}^3 \cos(2 \Psi_2) \\
& - 1756 v_2 v_{2F} v_3^2 \cos(2 \Psi_2) - 1317 v_2 v_{2F}^3 v_3^2 \cos(2 \Psi_2)  - 2380 v_2 v_{2F} v_4^2 \cos(2 \Psi_2) - 1785 v_2 v_{2F}^3 v_4^2 \cos(2 \Psi_2) \\
& + 63 v_2^2 v_{2F}^2 v_3^2 \cos(4 \Psi_2) + \frac{21}{2} v_2^2 v_{2F}^4 v_3^2 \cos(4 \Psi_2)+ 120 v_2^2 v_{2F}^2 v_4^2 \cos(4 \Psi_2) + 20 v_2^2 v_{2F}^4 v_4^2 \cos(4 \Psi_2)\\
& + 1146 v_2 v_{2F}^2 v_3^2 \cos(2 \Psi_2 - 6 \Psi_3)  + 191 v_2 v_{2F}^4 v_3^2 \cos(2 \Psi_2 - 6 \Psi_3)  - 3420 v_2^2 v_{2F} v_3^2 \cos(4 \Psi_2 - 6 \Psi_3) \\
& - 2565 v_2^2 v_{2F}^3 v_3^2 \cos(4 \Psi_2 - 6 \Psi_3)  - 35 v_{2F}^3 v_3^2 \cos(6 \Psi_3)  - 28 v_2^2 v_{2F}^3 v_3^2 \cos(6 \Psi_3)  - 8 v_{2F}^3 v_3^4 \cos(6 \Psi_3)  - 31 v_{2F}^3 v_3^2 v_4^2 \cos(6 \Psi_3) \\
& - 1337 v_2 v_{2F}^3 v_4^2 \cos(2 \Psi_2 - 8 \Psi_4)  + 5130 v_2^2 v_{2F}^2 v_4^2 \cos(4 \Psi_2 - 8 \Psi_4) + 855 v_2^2 v_{2F}^4 v_4^2 \cos(4 \Psi_2 - 8 \Psi_4)\\&- 8600 v_{2F} v_3^2 v_4^2 \cos(6 \Psi_3 - 8 \Psi_4)  - 6450 v_{2F}^3 v_3^2 v_4^2 \cos(6 \Psi_3 - 8 \Psi_4)  - 60 v_2 v_{2F} v_4 \cos(2 \Psi_2 - 4 \Psi_4) \\&- 4620 v_2^3 v_{2F} v_4 \cos(2 \Psi_2 - 4 \Psi_4)  - 45 v_2 v_{2F}^3 v_4 \cos(2 \Psi_2 - 4 \Psi_4)  - 3465 v_2^3 v_{2F}^3 v_4 \cos(2 \Psi_2 - 4 \Psi_4) \\
& - 23004 v_2 v_{2F} v_3^2 v_4 \cos(2 \Psi_2 - 4 \Psi_4)  - 17253 v_2 v_{2F}^3 v_3^2 v_4 \cos(2 \Psi_2 - 4 \Psi_4)  - 12020 v_2 v_{2F} v_4^3 \cos(2 \Psi_2 - 4 \Psi_4) \\
& - 9015 v_2 v_{2F}^3 v_4^3 \cos(2 \Psi_2 - 4 \Psi_4)+ 224 v_2^2 v_4 \cos(4 \Psi_2 - 4 \Psi_4)  + 672 v_2^2 v_{2F}^2 v_4 \cos(4 \Psi_2 - 4 \Psi_4)+ 84 v_2^2 v_{2F}^4 v_4 \cos(4 \Psi_2 - 4 \Psi_4)\\
& - 12 v_2^3 v_{2F} v_4 \cos(6 \Psi_2 - 4 \Psi_4)  - 9 v_2^3 v_{2F}^3 v_4 \cos(6 \Psi_2 - 4 \Psi_4)  + \frac{9}{2} v_2 v_{2F}^4 v_3^2 v_4 \cos(2 \Psi_2 - 6 \Psi_3 - 4 \Psi_4) \\& - 1816 v_{2F} v_3^2 v_4 \cos(6 \Psi_3 - 4 \Psi_4)  - 1362 v_{2F}^3 v_3^2 v_4 \cos(6 \Psi_3 - 4 \Psi_4)  + 111 v_2 v_{2F}^2 v_3^2 v_4 \cos(2 \Psi_2 + 6 \Psi_3 - 4 \Psi_4) \\
& + \frac{37}{2} v_2 v_{2F}^4 v_3^2 v_4 \cos(2 \Psi_2 + 6 \Psi_3 - 4 \Psi_4)  + 1416 v_2^2 v_{2F}^2 v_4 \cos(4 \Psi_4)  + 236 v_2^2 v_{2F}^4 v_4 \cos(4 \Psi_4)  + 4770 v_{2F}^2 v_3^2 v_4 \cos(4 \Psi_4) \\
& + 795 v_{2F}^4 v_3^2 v_4 \cos(4 \Psi_4)  + 2724 v_{2F}^2 v_4^3 \cos(4 \Psi_4)  + 454 v_{2F}^4 v_4^3 \cos(4 \Psi_4)  + 35 v_{2F}^4 v_4^2 \cos(8 \Psi_4)  + 28 v_2^2 v_{2F}^4 v_4^2 \cos(8 \Psi_4) 
\end{align*}

\newpage
\begin{align*}
& + \frac{73}{8} v_{2F}^4 v_3^2 v_4^2 \cos(8 \Psi_4)  + v_{2F}^4 v_4^4 \cos(8 \Psi_4)  - 35 v_2 v_{2F}^3 v_4 \cos(2 \Psi_2 + 4 \Psi_4)  - 28 v_2^3 v_{2F}^3 v_4 \cos(2 \Psi_2 + 4 \Psi_4) \\
& - 191 v_2 v_{2F}^3 v_3^2 v_4 \cos(2 \Psi_2 + 4 \Psi_4)  - 120 v_2 v_{2F}^3 v_4^3 \cos(2 \Psi_2 + 4 \Psi_4)  + 4808 v_2 v_3^2 v_4 \cos(2 \Psi_2 - 6 \Psi_3 + 4 \Psi_4) \\
& + 14424 v_2 v_{2F}^2 v_3^2 v_4 \cos(2 \Psi_2 - 6 \Psi_3 + 4 \Psi_4)  + 1803 v_2 v_{2F}^4 v_3^2 v_4 \cos(2 \Psi_2 - 6 \Psi_3 + 4 \Psi_4),\\
G_{5}&= 225 v_2^2 + 4600 v_2^4 + 1125 v_2^2 v_{2F}^2 + 23000 v_2^4 v_{2F}^2  + \frac{3375 v_2^2 v_{2F}^4}{8} + 8625 v_2^4 v_{2F}^4 + 1050 v_3^2 + 42560 v_2^2 v_3^2  + 5250 v_{2F}^2 v_3^2 \\&+ 212800 v_2^2 v_{2F}^2 v_3^2 + \frac{7875 v_{2F}^4 v_3^2}{4} + 79800 v_2^2 v_{2F}^4 v_3^2  + 21760 v_3^4 + 108800 v_{2F}^2 v_3^4 + 40800 v_{2F}^4 v_3^4 + 1460 v_4^2 + 47001 v_2^2 v_4^2 \\
& + 7300 v_{2F}^2 v_4^2 + 235005 v_2^2 v_{2F}^2 v_4^2 + \frac{5475 v_{2F}^4 v_4^2}{2} + \frac{705015}{8} v_2^2 v_{2F}^4 v_4^2  + 89824 v_3^2 v_4^2 + 449120 v_{2F}^2 v_3^2 v_4^2 + 168420 v_{2F}^4 v_3^2 v_4^2 \\&+ 21760 v_4^4  + 108800 v_{2F}^2 v_4^4 + 40800 v_{2F}^4 v_4^4 - 7860 v_2^3 v_{2F} \cos(2 \Psi_2)  - 11790 v_2^3 v_{2F}^3 \cos(2 \Psi_2) - \frac{1965}{2} v_2^3 v_{2F}^5 \cos(2 \Psi_2) \\
& - 46060 v_2 v_{2F} v_3^2 \cos(2 \Psi_2) - 69090 v_2 v_{2F}^3 v_3^2 \cos(2 \Psi_2)  - \frac{11515}{2} v_2 v_{2F}^5 v_3^2 \cos(2 \Psi_2) - 55500 v_2 v_{2F} v_4^2 \cos(2 \Psi_2) \\
& - 83250 v_2 v_{2F}^3 v_4^2 \cos(2 \Psi_2) - \frac{13875}{2} v_2 v_{2F}^5 v_4^2 \cos(2 \Psi_2)  + 525 v_2^2 v_{2F}^2 \cos(4 \Psi_2) + 420 v_2^4 v_{2F}^2 \cos(4 \Psi_2) + \frac{525}{2} v_2^2 v_{2F}^4 \cos(4 \Psi_2) \\&+ 210 v_2^4 v_{2F}^4 \cos(4 \Psi_2)  + 4190 v_2^2 v_{2F}^2 v_3^2 \cos(4 \Psi_2) + 2095 v_2^2 v_{2F}^4 v_3^2 \cos(4 \Psi_2)  + 5625 v_2^2 v_{2F}^2 v_4^2 \cos(4 \Psi_2) + \frac{5625}{2} v_2^2 v_{2F}^4 v_4^2 \cos(4 \Psi_2) \\
& + 36540 v_2 v_{2F}^2 v_3^2 \cos(2 \Psi_2 - 6 \Psi_3) + 18270 v_2 v_{2F}^4 v_3^2 \cos(2 \Psi_2 - 6 \Psi_3)  - 61920 v_2^2 v_{2F} v_3^2 \cos(4 \Psi_2 - 6 \Psi_3) \\&- 92880 v_2^2 v_{2F}^3 v_3^2 \cos(4 \Psi_2 - 6 \Psi_3)  - 7740 v_2^2 v_{2F}^5 v_3^2 \cos(4 \Psi_2 - 6 \Psi_3) - 2730 v_{2F}^3 v_3^2 \cos(6 \Psi_3)  - 2550 v_2^2 v_{2F}^3 v_3^2 \cos(6 \Psi_3) \\&- \frac{1365}{4} v_{2F}^5 v_3^2 \cos(6 \Psi_3)  - \frac{1275}{4} v_2^2 v_{2F}^5 v_3^2 \cos(6 \Psi_3) - 705 v_{2F}^3 v_3^4 \cos(6 \Psi_3)  - \frac{705}{8} v_{2F}^5 v_3^4 \cos(6 \Psi_3) - \frac{4055}{2} v_{2F}^3 v_3^2 v_4^2 \cos(6 \Psi_3) \\
& - \frac{4055}{16} v_{2F}^5 v_3^2 v_4^2 \cos(6 \Psi_3) + \frac{525}{4} v_2 v_{2F}^4 v_3^2 \cos(2 \Psi_2 + 6 \Psi_3)  - \frac{225}{8} v_2 v_{2F}^5 v_4^3 \cos(2 \Psi_2 - 12 \Psi_4) - 48720 v_2 v_{2F}^3 v_4^2 \cos(2 \Psi_2 - 8 \Psi_4) \\
& - 6090 v_2 v_{2F}^5 v_4^2 \cos(2 \Psi_2 - 8 \Psi_4) + 108360 v_2^2 v_{2F}^2 v_4^2 \cos(4 \Psi_2 - 8 \Psi_4)  + 54180 v_2^2 v_{2F}^4 v_4^2 \cos(4 \Psi_2 - 8 \Psi_4)\\
& - 130560 v_{2F} v_3^2 v_4^2 \cos(6 \Psi_3 - 8 \Psi_4) - 195840 v_{2F}^3 v_3^2 v_4^2 \cos(6 \Psi_3 - 8 \Psi_4)  - 16320 v_{2F}^5 v_3^2 v_4^2 \cos(6 \Psi_3 - 8 \Psi_4) \\&- 3820 v_2 v_{2F} v_4 \cos(2 \Psi_2 - 4 \Psi_4)  - 89520 v_2^3 v_{2F} v_4 \cos(2 \Psi_2 - 4 \Psi_4) - 5730 v_2 v_{2F}^3 v_4 \cos(2 \Psi_2 - 4 \Psi_4) \\
& - 134280 v_2^3 v_{2F}^3 v_4 \cos(2 \Psi_2 - 4 \Psi_4) - \frac{955}{2} v_2 v_{2F}^5 v_4 \cos(2 \Psi_2 - 4 \Psi_4)  - 11190 v_2^3 v_{2F}^5 v_4 \cos(2 \Psi_2 - 4 \Psi_4) \\&- 373520 v_2 v_{2F} v_3^2 v_4 \cos(2 \Psi_2 - 4 \Psi_4)  - 560280 v_2 v_{2F}^3 v_3^2 v_4 \cos(2 \Psi_2 - 4 \Psi_4) - 46690 v_2 v_{2F}^5 v_3^2 v_4 \cos(2 \Psi_2 - 4 \Psi_4) \\
& - 192480 v_2 v_{2F} v_4^3 \cos(2 \Psi_2 - 4 \Psi_4) - 288720 v_2 v_{2F}^3 v_4^3 \cos(2 \Psi_2 - 4 \Psi_4)  - 24060 v_2 v_{2F}^5 v_4^3 \cos(2 \Psi_2 - 4 \Psi_4)\\& + 5410 v_2^2 v_4 \cos(4 \Psi_2 - 4 \Psi_4)  + 27050 v_2^2 v_{2F}^2 v_4 \cos(4 \Psi_2 - 4 \Psi_4) + \frac{40575}{4} v_2^2 v_{2F}^4 v_4 \cos(4 \Psi_2 - 4 \Psi_4)  \\&- 780 v_2^3 v_{2F} v_4 \cos(6 \Psi_2 - 4 \Psi_4) - 1170 v_2^3 v_{2F}^3 v_4 \cos(6 \Psi_2 - 4 \Psi_4)  - \frac{195}{2} v_2^3 v_{2F}^5 v_4 \cos(6 \Psi_2 - 4 \Psi_4) \\&+ \frac{2955}{4} v_2 v_{2F}^4 v_3^2 v_4 \cos(2 \Psi_2 - 6 \Psi_3 - 4 \Psi_4) - 39045 v_{2F} v_3^2 v_4 \cos(6 \Psi_3 - 4 \Psi_4) - \frac{117135}{2} v_{2F}^3 v_3^2 v_4 \cos(6 \Psi_3 - 4 \Psi_4) \\
& - \frac{39045}{8} v_{2F}^5 v_3^2 v_4 \cos(6 \Psi_3 - 4 \Psi_4)+ 5000 v_2 v_{2F}^2 v_3^2 v_4 \cos(2 \Psi_2 + 6 \Psi_3 - 4 \Psi_4)+ 2500 v_2 v_{2F}^4 v_3^2 v_4 \cos(2 \Psi_2 + 6 \Psi_3 - 4 \Psi_4)\\
& + 175 v_{2F}^2 v_4 \cos(4 \Psi_4) + 50435 v_2^2 v_{2F}^2 v_4 \cos(4 \Psi_4)  + \frac{175}{2} v_{2F}^4 v_4 \cos(4 \Psi_4) + \frac{50435}{2} v_2^2 v_{2F}^4 v_4 \cos(4 \Psi_4) \\
& + 124925 v_{2F}^2 v_3^2 v_4 \cos(4 \Psi_4) + \frac{124925}{2} v_{2F}^4 v_3^2 v_4 \cos(4 \Psi_4)  + 68290 v_{2F}^2 v_4^3 \cos(4 \Psi_4) + 34145 v_{2F}^4 v_4^3 \cos(4 \Psi_4) \\
& + \frac{12285}{4} v_{2F}^4 v_4^2 \cos(8 \Psi_4) + \frac{20745}{8} v_2^2 v_{2F}^4 v_4^2 \cos(8 \Psi_4)  + \frac{7835}{8} v_{2F}^4 v_3^2 v_4^2 \cos(8 \Psi_4) + \frac{495}{4} v_{2F}^4 v_4^4 \cos(8 \Psi_4) \\
& - 3430 v_2 v_{2F}^3 v_4 \cos(2 \Psi_2 + 4 \Psi_4) - 2840 v_2^3 v_{2F}^3 v_4 \cos(2 \Psi_2 + 4 \Psi_4)  - \frac{1715}{4} v_2 v_{2F}^5 v_4 \cos(2 \Psi_2 + 4 \Psi_4) \\&- 355 v_2^3 v_{2F}^5 v_4 \cos(2 \Psi_2 + 4 \Psi_4)  - 10565 v_2 v_{2F}^3 v_3^2 v_4 \cos(2 \Psi_2 + 4 \Psi_4) - \frac{10565}{8} v_2 v_{2F}^5 v_3^2 v_4 \cos(2 \Psi_2 + 4 \Psi_4) \\
& - 5790 v_2 v_{2F}^3 v_4^3 \cos(2 \Psi_2 + 4 \Psi_4) - \frac{2895}{4} v_2 v_{2F}^5 v_4^3 \cos(2 \Psi_2 + 4 \Psi_4)  + \frac{525}{4} v_2^2 v_{2F}^4 v_4 \cos(4 \Psi_2 + 4 \Psi_4) 
\end{align*}
\newpage
\begin{align*}
&+ 64160 v_2 v_3^2 v_4 \cos(2 \Psi_2 - 6 \Psi_3 + 4 \Psi_4)  + 320800 v_2 v_{2F}^2 v_3^2 v_4 \cos(2 \Psi_2 - 6 \Psi_3 + 4 \Psi_4) \\&+ 120300 v_2 v_{2F}^4 v_3^2 v_4 \cos(2 \Psi_2 - 6 \Psi_3 + 4 \Psi_4)  - \frac{165}{8} v_{2F}^5 v_3^2 v_4 \cos(6 \Psi_3 + 4 \Psi_4) - \frac{525}{4} v_2 v_{2F}^5 v_4^2 \cos(2 \Psi_2 + 8 \Psi_4),\\
G_{6}&= 10080 v_2^2 + 78120 v_2^4 + 75600 v_2^2 v_{2F}^2 + 585900 v_2^4 v_{2F}^2  + 56700 v_2^2 v_{2F}^4 + 439425 v_2^4 v_{2F}^4 + 3150 v_2^2 v_{2F}^6  + \frac{48825}{2} v_2^4 v_{2F}^6 \\&+ 28476 v_3^2 + 619654 v_2^2 v_3^2  + 213570 v_{2F}^2 v_3^2 + 4647405 v_2^2 v_{2F}^2 v_3^2 + \frac{320355}{2} v_{2F}^4 v_3^2  + \frac{13942215}{4} v_2^2 v_{2F}^4 v_3^2 + \frac{35595}{4} v_{2F}^6 v_3^2 \\&+ \frac{1549135}{8} v_2^2 v_{2F}^6 v_3^2  + 283584 v_3^4 + 2126880 v_{2F}^2 v_3^4 + 1595160 v_{2F}^4 v_3^4 + 88620 v_{2F}^6 v_3^4  + 35040 v_4^2 + 664800 v_2^2 v_4^2 \\&+ 262800 v_{2F}^2 v_4^2 + 4986000 v_2^2 v_{2F}^2 v_4^2  + 197100 v_{2F}^4 v_4^2 + 3739500 v_2^2 v_{2F}^4 v_4^2 + 10950 v_{2F}^6 v_4^2  + 207750 v_2^2 v_{2F}^6 v_4^2 \\&+ 1160225 v_3^2 v_4^2 + \frac{17403375}{2} v_{2F}^2 v_3^2 v_4^2  + \frac{52210125}{8} v_{2F}^4 v_3^2 v_4^2 + \frac{5801125}{16} v_{2F}^6 v_3^2 v_4^2 + 283584 v_4^4  + 2126880 v_{2F}^2 v_4^4 \\&+ 1595160 v_{2F}^4 v_4^4 + 88620 v_{2F}^6 v_4^4  - 3150 v_2 v_{2F} \cos(2 \Psi_2) - 224070 v_2^3 v_{2F} \cos(2 \Psi_2)  - 7875 v_2 v_{2F}^3 \cos(2 \Psi_2) \\&- 560175 v_2^3 v_{2F}^3 \cos(2 \Psi_2)  - \frac{7875}{4} v_2 v_{2F}^5 \cos(2 \Psi_2) - \frac{560175}{4} v_2^3 v_{2F}^5 \cos(2 \Psi_2)  - 1019844 v_2 v_{2F} v_3^2 \cos(2 \Psi_2) \\&- 2549610 v_2 v_{2F}^3 v_3^2 \cos(2 \Psi_2)  - \frac{1274805}{2} v_2 v_{2F}^5 v_3^2 \cos(2 \Psi_2)-1158948 v_2 v_{2F} v_4^2 \cos(2 \Psi_2)-2897370 v_2 v_{2F}^3 v_4^2 \cos(2 \Psi_2)\\
& -\frac{1448685}{2} v_2 v_{2F}^5 v_4^2 \cos(2 \Psi_2) + 32340 v_2^2 v_{2F}^2 \cos(4 \Psi_2) + 27600 v_2^4 v_{2F}^2 \cos(4 \Psi_2)  + 32340 v_2^2 v_{2F}^4 \cos(4 \Psi_2) \\&+ 27600 v_2^4 v_{2F}^4 \cos(4 \Psi_2) + \frac{8085}{4} v_2^2 v_{2F}^6 \cos(4 \Psi_2)  + 1725 v_2^4 v_{2F}^6 \cos(4 \Psi_2) + 168975 v_2^2 v_{2F}^2 v_3^2 \cos(4 \Psi_2) \\&+ 168975 v_2^2 v_{2F}^4 v_3^2 \cos(4 \Psi_2)  + \frac{168975}{16} v_2^2 v_{2F}^6 v_3^2 \cos(4 \Psi_2) + 197220 v_2^2 v_{2F}^2 v_4^2 \cos(4 \Psi_2) + 197220 v_2^2 v_{2F}^4 v_4^2 \cos(4 \Psi_2) \\
& + \frac{49305}{4} v_2^2 v_{2F}^6 v_4^2 \cos(4 \Psi_2) - 3150 v_2^3 v_{2F}^3 \cos(6 \Psi_2) - \frac{4725}{4} v_2^3 v_{2F}^5 \cos(6 \Psi_2)  + 942810 v_2 v_{2F}^2 v_3^2 \cos(2 \Psi_2 - 6 \Psi_3) \\&+ 942810 v_2 v_{2F}^4 v_3^2 \cos(2 \Psi_2 - 6 \Psi_3)  + \frac{471405}{8} v_2 v_{2F}^6 v_3^2 \cos(2 \Psi_2 - 6 \Psi_3) - 1060500 v_2^2 v_{2F} v_3^2 \cos(4 \Psi_2 - 6 \Psi_3) \\
& - 2651250 v_2^2 v_{2F}^3 v_3^2 \cos(4 \Psi_2 - 6 \Psi_3) - \frac{1325625}{2} v_2^2 v_{2F}^5 v_3^2 \cos(4 \Psi_2 - 6 \Psi_3)  - 120960 v_{2F}^3 v_3^2 \cos(6 \Psi_3) \\&- 130050 v_2^2 v_{2F}^3 v_3^2 \cos(6 \Psi_3) - 45360 v_{2F}^5 v_3^2 \cos(6 \Psi_3)  - \frac{195075}{4} v_2^2 v_{2F}^5 v_3^2 \cos(6 \Psi_3) - 35925 v_{2F}^3 v_3^4 \cos(6 \Psi_3) \\&- \frac{107775}{8} v_{2F}^5 v_3^4 \cos(6 \Psi_3)  - 93845 v_{2F}^3 v_3^2 v_4^2 \cos(6 \Psi_3) - \frac{281535}{8} v_{2F}^5 v_3^2 v_4^2 \cos(6 \Psi_3) + \frac{66555}{4} v_2 v_{2F}^4 v_3^2 \cos(2 \Psi_2 + 6 \Psi_3) \\
& + \frac{13311}{8} v_2 v_{2F}^6 v_3^2 \cos(2 \Psi_2 + 6 \Psi_3) - \frac{693}{4} v_2^2 v_{2F}^5 v_3^2 \cos(4 \Psi_2 + 6 \Psi_3)-\frac{29535}{8} v_2 v_{2F}^5 v_4^3 \cos(2 \Psi_2 - 12 \Psi_4)\\
& -1412415 v_2 v_{2F}^3 v_4^2 \cos(2 \Psi_2 - 8 \Psi_4) 
- \frac{4237245}{8} v_2 v_{2F}^5 v_4^2 \cos(2 \Psi_2 - 8 \Psi_4)  + 2121000 v_2^2 v_{2F}^2 v_4^2 \cos(4 \Psi_2 - 8 \Psi_4) 
\\&+ 2121000 v_2^2 v_{2F}^4 v_4^2 \cos(4 \Psi_2 - 8 \Psi_4)  + \frac{265125}{2} v_2^2 v_{2F}^6 v_4^2 \cos(4 \Psi_2 - 8 \Psi_4) 
- 1985088 v_{2F} v_3^2 v_4^2 \cos(6 \Psi_3 - 8 \Psi_4) \\
& - 4962720 v_{2F}^3 v_3^2 v_4^2 \cos(6 \Psi_3 - 8 \Psi_4) 
- 1240680 v_{2F}^5 v_3^2 v_4^2 \cos(6 \Psi_3 - 8 \Psi_4)  - 136290 v_2 v_{2F} v_4 \cos(2 \Psi_2 - 4 \Psi_4) 
\\&- 1607508 v_2^3 v_{2F} v_4 \cos(2 \Psi_2 - 4 \Psi_4)  - 340725 v_2 v_{2F}^3 v_4 \cos(2 \Psi_2 - 4 \Psi_4) 
- 4018770 v_2^3 v_{2F}^3 v_4 \cos(2 \Psi_2 - 4 \Psi_4) \\
& - \frac{340725}{4} v_2 v_{2F}^5 v_4 \cos(2 \Psi_2 - 4 \Psi_4) 
- \frac{2009385}{2} v_2^3 v_{2F}^5 v_4 \cos(2 \Psi_2 - 4 \Psi_4)  - 5959140 v_2 v_{2F} v_3^2 v_4 \cos(2 \Psi_2 - 4 \Psi_4) 
\\&- 14897850 v_2 v_{2F}^3 v_3^2 v_4 \cos(2 \Psi_2 - 4 \Psi_4)  - \frac{7448925}{2} v_2 v_{2F}^5 v_3^2 v_4 \cos(2 \Psi_2 - 4 \Psi_4) 
- 3045594 v_2 v_{2F} v_4^3 \cos(2 \Psi_2 - 4 \Psi_4) \\
& - 7613985 v_2 v_{2F}^3 v_4^3 \cos(2 \Psi_2 - 4 \Psi_4) 
- \frac{7613985}{4} v_2 v_{2F}^5 v_4^3 \cos(2 \Psi_2 - 4 \Psi_4)  + 109752 v_2^2 v_4 \cos(4 \Psi_2 - 4 \Psi_4) 
\\&+ 823140 v_2^2 v_{2F}^2 v_4 \cos(4 \Psi_2 - 4 \Psi_4)  + 617355 v_2^2 v_{2F}^4 v_4 \cos(4 \Psi_2 - 4 \Psi_4) 
+ \frac{68595}{2} v_2^2 v_{2F}^6 v_4 \cos(4 \Psi_2 - 4 \Psi_4) \\
& - 29310 v_2^3 v_{2F} v_4 \cos(6 \Psi_2 - 4 \Psi_4) 
- 73275 v_2^3 v_{2F}^3 v_4 \cos(6 \Psi_2 - 4 \Psi_4)  - \frac{73275}{4} v_2^3 v_{2F}^5 v_4 \cos(6 \Psi_2 - 4 \Psi_4) 
\end{align*}
\newpage
\begin{align*}
&+ \frac{429075}{8} v_2 v_{2F}^4 v_3^2 v_4 \cos(2 \Psi_2 - 6 \Psi_3 - 4 \Psi_4)  + \frac{85815}{16} v_2 v_{2F}^6 v_3^2 v_4 \cos(2 \Psi_2 - 6 \Psi_3 - 4 \Psi_4)\\
& -768042 v_{2F} v_3^2 v_4 \cos(6 \Psi_3 - 4 \Psi_4) 
- 1920105 v_{2F}^3 v_3^2 v_4 \cos(6 \Psi_3 - 4 \Psi_4)  - \frac{1920105}{4} v_{2F}^5 v_3^2 v_4 \cos(6 \Psi_3 - 4 \Psi_4) 
\\&+ 173475 v_2 v_{2F}^2 v_3^2 v_4 \cos(2 \Psi_2 + 6 \Psi_3 - 4 \Psi_4)  + 173475 v_2 v_{2F}^4 v_3^2 v_4 \cos(2 \Psi_2 + 6 \Psi_3 - 4 \Psi_4) 
\\&+ \frac{173475}{16} v_2 v_{2F}^6 v_3^2 v_4 \cos(2 \Psi_2 + 6 \Psi_3 - 4 \Psi_4)  + 16380 v_{2F}^2 v_4 \cos(4 \Psi_4) 
+ 1398240 v_2^2 v_{2F}^2 v_4 \cos(4 \Psi_4) 
\\&+ 16380 v_{2F}^4 v_4 \cos(4 \Psi_4)  + 1398240 v_2^2 v_{2F}^4 v_4 \cos(4 \Psi_4) 
+ \frac{4095}{4} v_{2F}^6 v_4 \cos(4 \Psi_4) 
+ 87390 v_2^2 v_{2F}^6 v_4 \cos(4 \Psi_4) \\
& + 2875770 v_{2F}^2 v_3^2 v_4 \cos(4 \Psi_4) 
+ 2875770 v_{2F}^4 v_3^2 v_4 \cos(4 \Psi_4) 
+ \frac{1437885}{8} v_{2F}^6 v_3^2 v_4 \cos(4 \Psi_4) \\
& + 1532280 v_{2F}^2 v_4^3 \cos(4 \Psi_4) 
+ 1532280 v_{2F}^4 v_4^3 \cos(4 \Psi_4)+ \frac{191535}{2} v_{2F}^6 v_4^3 \cos(4 \Psi_4)  + 151200 v_{2F}^4 v_4^2 \cos(8 \Psi_4)\\
& + \frac{548925}{4} v_2^2 v_{2F}^4 v_4^2 \cos(8 \Psi_4) 
+ 15120 v_{2F}^6 v_4^2 \cos(8 \Psi_4) 
+ \frac{109785}{8} v_2^2 v_{2F}^6 v_4^2 \cos(8 \Psi_4)  + \frac{236895}{4} v_{2F}^4 v_3^2 v_4^2 \cos(8 \Psi_4) 
\\&+ \frac{47379}{8} v_{2F}^6 v_3^2 v_4^2 \cos(8 \Psi_4) 
+ 8550 v_{2F}^4 v_4^4 \cos(8 \Psi_4) 
+ 855 v_{2F}^6 v_4^4 \cos(8 \Psi_4)  + \frac{1485}{8} v_{2F}^6 v_4^3 \cos(12 \Psi_4) 
\\&- 170520 v_2 v_{2F}^3 v_4 \cos(2 \Psi_2 + 4 \Psi_4) 
- 148500 v_2^3 v_{2F}^3 v_4 \cos(2 \Psi_2 + 4 \Psi_4)  - 63945 v_2 v_{2F}^5 v_4 \cos(2 \Psi_2 + 4 \Psi_4) 
\\&- \frac{111375}{2} v_2^3 v_{2F}^5 v_4 \cos(2 \Psi_2 + 4 \Psi_4) 
- 423425 v_2 v_{2F}^3 v_3^2 v_4 \cos(2 \Psi_2 + 4 \Psi_4)  - \frac{1270275}{8} v_2 v_{2F}^5 v_3^2 v_4 \cos(2 \Psi_2 + 4 \Psi_4) 
\\&- 214740 v_2 v_{2F}^3 v_4^3 \cos(2 \Psi_2 + 4 \Psi_4) 
- \frac{161055}{2} v_2 v_{2F}^5 v_4^3 \cos(2 \Psi_2 + 4 \Psi_4)  + 18720 v_2^2 v_{2F}^4 v_4 \cos(4 \Psi_2 + 4 \Psi_4) 
\\&+ 1872 v_2^2 v_{2F}^6 v_4 \cos(4 \Psi_2 + 4 \Psi_4) 
- \frac{693}{4} v_2^3 v_{2F}^5 v_4 \cos(6 \Psi_2 + 4 \Psi_4)  + 870182 v_2 v_3^2 v_4 \cos(2 \Psi_2 - 6 \Psi_3 + 4 \Psi_4) 
\\&+ 6526365 v_2 v_{2F}^2 v_3^2 v_4 \cos(2 \Psi_2 - 6 \Psi_3 + 4 \Psi_4)  + \frac{19579095}{4} v_2 v_{2F}^4 v_3^2 v_4 \cos(2 \Psi_2 - 6 \Psi_3 + 4 \Psi_4) 
\\&+ \frac{2175455}{8} v_2 v_{2F}^6 v_3^2 v_4 \cos(2 \Psi_2 - 6 \Psi_3 + 4 \Psi_4)  - 4752 v_{2F}^5 v_3^2 v_4 \cos(6 \Psi_3 + 4 \Psi_4) 
+ \frac{429}{8} v_2 v_{2F}^6 v_3^2 v_4 \cos(2 \Psi_2 + 6 \Psi_3 + 4 \Psi_4) \\
& - \frac{131571}{8} v_2 v_{2F}^5 v_4^2 \cos(2 \Psi_2 + 8 \Psi_4) 
+ \frac{693}{4} v_2^2 v_{2F}^6 v_4^2 \cos(4 \Psi_2 + 8 \Psi_4),\\
G_{7}&= 1225 + 287924 v_2^2 + 1255429 v_2^4 + \frac{25725 v_{2F}^2}{2} + 3023202 v_2^2 v_{2F}^2 + \frac{26364009 v_2^4 v_{2F}^2}{2}  + \frac{128625 v_{2F}^4}{8} + \frac{7558005 v_2^2 v_{2F}^4}{2} \\&+ \frac{131820045 v_2^4 v_{2F}^4}{8} + \frac{42875 v_{2F}^6}{16}  + \frac{2519335 v_2^2 v_{2F}^6}{4} + \frac{43940015 v_2^4 v_{2F}^6}{16} + 634676 v_3^2 + 8966692 v_2^2 v_3^2  + 6664098 v_{2F}^2 v_3^2 \\&+ 94150266 v_2^2 v_{2F}^2 v_3^2 + \frac{16660245 v_{2F}^4 v_3^2}{2} + \frac{235375665}{2} v_2^2 v_{2F}^4 v_3^2  + \frac{5553415 v_{2F}^6 v_3^2}{4} + \frac{78458555}{4} v_2^2 v_{2F}^6 v_3^2 + 3784397 v_3^4 \\&+ \frac{79472337 v_{2F}^2 v_3^4}{2}  + \frac{397361685 v_{2F}^4 v_3^4}{8} + \frac{132453895 v_{2F}^6 v_3^4}{16} + 733264 v_4^2 + 9450616 v_2^2 v_4^2  + 7699272 v_{2F}^2 v_4^2 + 99231468 v_2^2 v_{2F}^2 v_4^2 \\&+ 9624090 v_{2F}^4 v_4^2 + 124039335 v_2^2 v_{2F}^4 v_4^2  + 1604015 v_{2F}^6 v_4^2 + \frac{41346445}{2} v_2^2 v_{2F}^6 v_4^2 + 15396856 v_3^2 v_4^2 + 161666988 v_{2F}^2 v_3^2 v_4^2 \\
& + 202083735 v_{2F}^4 v_3^2 v_4^2 + \frac{67361245}{2} v_{2F}^6 v_3^2 v_4^2 + 3784334 v_4^4 + 39735507 v_{2F}^2 v_4^4  + \frac{198677535 v_{2F}^4 v_4^4}{4} + \frac{66225845 v_{2F}^6 v_4^4}{8} \\&- 189140 v_2 v_{2F} \cos(2 \Psi_2)  - 5305888 v_2^3 v_{2F} \cos(2 \Psi_2) - 709275 v_2 v_{2F}^3 \cos(2 \Psi_2) - 19897080 v_2^3 v_{2F}^3 \cos(2 \Psi_2) \\
& - \frac{709275}{2} v_2 v_{2F}^5 \cos(2 \Psi_2) - 9948540 v_2^3 v_{2F}^5 \cos(2 \Psi_2) - \frac{236425}{16} v_2 v_{2F}^7 \cos(2 \Psi_2)  - \frac{829045}{2} v_2^3 v_{2F}^7 \cos(2 \Psi_2) \\&- 20609624 v_2 v_{2F} v_3^2 \cos(2 \Psi_2)  - 77286090 v_2 v_{2F}^3 v_3^2 \cos(2 \Psi_2) - 38643045 v_2 v_{2F}^5 v_3^2 \cos(2 \Psi_2) - \frac{12881015}{8} v_2 v_{2F}^7 v_3^2 \cos(2 \Psi_2)\\
& -22615488 v_2 v_{2F} v_4^2 \cos(2 \Psi_2) - 84808080 v_2 v_{2F}^3 v_4^2 \cos(2 \Psi_2) - 42404040 v_2 v_{2F}^5 v_4^2 \cos(2 \Psi_2)  - 1766835 v_2 v_{2F}^7 v_4^2 \cos(2 \Psi_2) \\&+ 1237005 v_2^2 v_{2F}^2 \cos(4 \Psi_2) + 1093365 v_2^4 v_{2F}^2 \cos(4 \Psi_2)  + 2061675 v_2^2 v_{2F}^4 \cos(4 \Psi_2) + 1822275 v_2^4 v_{2F}^4 \cos(4 \Psi_2) \\&+ \frac{6185025}{16} v_2^2 v_{2F}^6 \cos(4 \Psi_2)  + \frac{5466825}{16} v_2^4 v_{2F}^6 \cos(4 \Psi_2) + 5331564 v_2^2 v_{2F}^2 v_3^2 \cos(4 \Psi_2) + 8885940 v_2^2 v_{2F}^4 v_3^2 \cos(4 \Psi_2) 
\end{align*}
\newpage
\begin{align*}
& + \frac{6664455}{4} v_2^2 v_{2F}^6 v_3^2 \cos(4 \Psi_2) + \frac{11540865}{2} v_2^2 v_{2F}^2 v_4^2 \cos(4 \Psi_2)  + \frac{19234775}{2} v_2^2 v_{2F}^4 v_4^2 \cos(4 \Psi_2) + \frac{57704325}{32} v_2^2 v_{2F}^6 v_4^2 \cos(4 \Psi_2) \notag\\
& - 266910 v_2^3 v_{2F}^3 \cos(6 \Psi_2) - \frac{400365}{2} v_2^3 v_{2F}^5 \cos(6 \Psi_2) - \frac{80073}{8} v_2^3 v_{2F}^7 \cos(6 \Psi_2)  + \frac{24255}{4} v_2^4 v_{2F}^4 \cos(8 \Psi_2) \notag\\&+ \frac{14553}{8} v_2^4 v_{2F}^6 \cos(8 \Psi_2) + 21599802 v_2 v_{2F}^2 v_3^2 \cos(2 \Psi_2 - 6 \Psi_3)  + 35999670 v_2 v_{2F}^4 v_3^2 \cos(2 \Psi_2 - 6 \Psi_3) \notag\\&+ \frac{53999505}{8} v_2 v_{2F}^6 v_3^2 \cos(2 \Psi_2 - 6 \Psi_3)  - 17627988 v_2^2 v_{2F} v_3^2 \cos(4 \Psi_2 - 6 \Psi_3) - 66104955 v_2^2 v_{2F}^3 v_3^2 \cos(4 \Psi_2 - 6 \Psi_3) \notag\\
& - \frac{66104955}{2} v_2^2 v_{2F}^5 v_3^2 \cos(4 \Psi_2 - 6 \Psi_3) - \frac{22034985}{16} v_2^2 v_{2F}^7 v_3^2 \cos(4 \Psi_2 - 6 \Psi_3)  - \frac{8145795}{2} v_{2F}^3 v_3^2 \cos(6 \Psi_3) \notag\\&- \frac{9909865}{2} v_2^2 v_{2F}^3 v_3^2 \cos(6 \Psi_3)  - \frac{24437385}{8} v_{2F}^5 v_3^2 \cos(6 \Psi_3) - \frac{29729595}{8} v_2^2 v_{2F}^5 v_3^2 \cos(6 \Psi_3)  - \frac{4887477}{32} v_{2F}^7 v_3^2 \cos(6 \Psi_3) \notag\\&- \frac{5945919}{32} v_2^2 v_{2F}^7 v_3^2 \cos(6 \Psi_3) - 1384285 v_{2F}^3 v_3^4 \cos(6 \Psi_3)-\frac{4152855}{4} v_{2F}^5 v_3^4 \cos(6 \Psi_3) - \frac{830571}{16} v_{2F}^7 v_3^4 \cos(6 \Psi_3)\notag\\
& -\frac{13908335}{4} v_{2F}^3 v_3^2 v_4^2 \cos(6 \Psi_3) - \frac{41725005}{16} v_{2F}^5 v_3^2 v_4^2 \cos(6 \Psi_3)  - \frac{8345001}{64} v_{2F}^7 v_3^2 v_4^2 \cos(6 \Psi_3) + \frac{3003}{8} v_{2F}^6 v_3^4 \cos(12 \Psi_3) \notag\\
& + 1097250 v_2 v_{2F}^4 v_3^2 \cos(2 \Psi_2 + 6 \Psi_3) + 329175 v_2 v_{2F}^6 v_3^2 \cos(2 \Psi_2 + 6 \Psi_3)  - \frac{132363}{4} v_2^2 v_{2F}^5 v_3^2 \cos(4 \Psi_2 + 6 \Psi_3) \notag\\&- \frac{44121}{16} v_2^2 v_{2F}^7 v_3^2 \cos(4 \Psi_2 + 6 \Psi_3)  - \frac{1040193}{4} v_2 v_{2F}^5 v_4^3 \cos(2 \Psi_2 - 12 \Psi_4)-\frac{346731}{16} v_2 v_{2F}^7 v_4^3 \cos(2 \Psi_2 - 12 \Psi_4)
\notag\\& -\frac{71679475}{2} v_2 v_{2F}^3 v_4^2 \cos(2 \Psi_2 - 8 \Psi_4) 
- \frac{215038425}{8} v_2 v_{2F}^5 v_4^2 \cos(2 \Psi_2 - 8 \Psi_4)  - \frac{43007685}{32} v_2 v_{2F}^7 v_4^2 \cos(2 \Psi_2 - 8 \Psi_4) 
\notag\\&+ \frac{79318953}{2} v_2^2 v_{2F}^2 v_4^2 \cos(4 \Psi_2 - 8 \Psi_4)  + \frac{132198255}{2} v_2^2 v_{2F}^4 v_4^2 \cos(4 \Psi_2 - 8 \Psi_4) 
+ \frac{396594765}{32} v_2^2 v_{2F}^6 v_4^2 \cos(4 \Psi_2 - 8 \Psi_4)\notag \\
& - 30274727 v_{2F} v_3^2 v_4^2 \cos(6 \Psi_3 - 8 \Psi_4) 
- \frac{454120905}{4} v_{2F}^3 v_3^2 v_4^2 \cos(6 \Psi_3 - 8 \Psi_4)  - \frac{454120905}{8} v_{2F}^5 v_3^2 v_4^2 \cos(6 \Psi_3 - 8 \Psi_4) 
\notag\\&- \frac{151373635}{64} v_{2F}^7 v_3^2 v_4^2 \cos(6 \Psi_3 - 8 \Psi_4) - 3739820 v_2 v_{2F} v_4 \cos(2 \Psi_2 - 4 \Psi_4) 
- 27680968 v_2^3 v_{2F} v_4 \cos(2 \Psi_2 - 4 \Psi_4) \notag\\
& - 14024325 v_2 v_{2F}^3 v_4 \cos(2 \Psi_2 - 4 \Psi_4) 
- 103803630 v_2^3 v_{2F}^3 v_4 \cos(2 \Psi_2 - 4 \Psi_4)  - \frac{14024325}{2} v_2 v_{2F}^5 v_4 \cos(2 \Psi_2 - 4 \Psi_4) 
\notag\\&- 51901815 v_2^3 v_{2F}^5 v_4 \cos(2 \Psi_2 - 4 \Psi_4)  - \frac{4674775}{16} v_2 v_{2F}^7 v_4 \cos(2 \Psi_2 - 4 \Psi_4) 
- \frac{17300605}{8} v_2^3 v_{2F}^7 v_4 \cos(2 \Psi_2 - 4 \Psi_4)\notag \\
& - 94226888 v_2 v_{2F} v_3^2 v_4 \cos(2 \Psi_2 - 4 \Psi_4) 
- 353350830 v_2 v_{2F}^3 v_3^2 v_4 \cos(2 \Psi_2 - 4 \Psi_4)  - 176675415 v_2 v_{2F}^5 v_3^2 v_4 \cos(2 \Psi_2 - 4 \Psi_4) 
\notag\\&- \frac{58891805}{8} v_2 v_{2F}^7 v_3^2 v_4 \cos(2 \Psi_2 - 4 \Psi_4)  - 47901756 v_2 v_{2F} v_4^3 \cos(2 \Psi_2 - 4 \Psi_4) 
- 179631585 v_2 v_{2F}^3 v_4^3 \cos(2 \Psi_2 - 4 \Psi_4)\notag \\
& - \frac{179631585}{2} v_2 v_{2F}^5 v_4^3 \cos(2 \Psi_2 - 4 \Psi_4) 
- \frac{59877195}{16} v_2 v_{2F}^7 v_4^3 \cos(2 \Psi_2 - 4 \Psi_4)  + 2040164 v_2^2 v_4 \cos(4 \Psi_2 - 4 \Psi_4)\notag\\
& + 21421722 v_2^2 v_{2F}^2 v_4 \cos(4 \Psi_2 - 4 \Psi_4) 
+ \frac{53554305}{2} v_2^2 v_{2F}^4 v_4 \cos(4 \Psi_2 - 4 \Psi_4)  + \frac{17851435}{4} v_2^2 v_{2F}^6 v_4 \cos(4 \Psi_2 - 4 \Psi_4) 
\notag\\&- 849072 v_2^3 v_{2F} v_4 \cos(6 \Psi_2 - 4 \Psi_4)  - 3184020 v_2^3 v_{2F}^3 v_4 \cos(6 \Psi_2 - 4 \Psi_4) 
- 1592010 v_2^3 v_{2F}^5 v_4 \cos(6 \Psi_2 - 4 \Psi_4) \notag\\
& - \frac{265335}{4} v_2^3 v_{2F}^7 v_4 \cos(6 \Psi_2 - 4 \Psi_4) 
+ \frac{21121485}{8} v_2 v_{2F}^4 v_3^2 v_4 \cos(2 \Psi_2 - 6 \Psi_3 - 4 \Psi_4)\notag  \\&+ \frac{12672891}{16} v_2 v_{2F}^6 v_3^2 v_4 \cos(2 \Psi_2 - 6 \Psi_3 - 4 \Psi_4) 
- 14310471 v_{2F} v_3^2 v_4 \cos(6 \Psi_3 - 4 \Psi_4)  - \frac{214657065}{4} v_{2F}^3 v_3^2 v_4 \cos(6 \Psi_3 - 4 \Psi_4) 
\notag\\&- \frac{214657065}{8} v_{2F}^5 v_3^2 v_4 \cos(6 \Psi_3 - 4 \Psi_4)  - \frac{71552355}{64} v_{2F}^7 v_3^2 v_4 \cos(6 \Psi_3 - 4 \Psi_4) 
+ \frac{10190145}{2} v_2 v_{2F}^2 v_3^2 v_4 \cos(2 \Psi_2 + 6 \Psi_3 - 4 \Psi_4)\notag \\
& + \frac{16983575}{2} v_2 v_{2F}^4 v_3^2 v_4 \cos(2 \Psi_2 + 6 \Psi_3 - 4 \Psi_4) 
+ \frac{50950725}{32} v_2 v_{2F}^6 v_3^2 v_4 \cos(2 \Psi_2 + 6 \Psi_3 - 4 \Psi_4)  + 776160 v_{2F}^2 v_4 \cos(4 \Psi_4) 
\notag\\&+ 33753447 v_2^2 v_{2F}^2 v_4 \cos(4 \Psi_4)  + 1293600 v_{2F}^4 v_4 \cos(4 \Psi_4) 
+ 56255745 v_2^2 v_{2F}^4 v_4 \cos(4 \Psi_4)  + 242550 v_{2F}^6 v_4 \cos(4 \Psi_4) 
\notag\\&+ \frac{168767235}{16} v_2^2 v_{2F}^6 v_4 \cos(4 \Psi_4)  + 61108740 v_{2F}^2 v_3^2 v_4 \cos(4 \Psi_4) 
+ 101847900 v_{2F}^4 v_3^2 v_4 \cos(4 \Psi_4)  + \frac{76385925}{4} v_{2F}^6 v_3^2 v_4 \cos(4 \Psi_4) 
\notag\\&+ 31999527 v_{2F}^2 v_4^3 \cos(4 \Psi_4)  + 53332545 v_{2F}^4 v_4^3 \cos(4 \Psi_4) 
+ \frac{159997635}{16} v_{2F}^6 v_4^3 \cos(4 \Psi_4)  + \frac{44660385}{8} v_{2F}^4 v_4^2 \cos(8 \Psi_4)
\end{align*}

\newpage
\begin{align}
& + \frac{43876525}{8} v_2^2 v_{2F}^4 v_4^2 \cos(8 \Psi_4) 
+ \frac{26796231}{16} v_{2F}^6 v_4^2 \cos(8 \Psi_4) 
+ \frac{26325915}{16} v_2^2 v_{2F}^6 v_4^2 \cos(8 \Psi_4)  + \frac{21261275}{8} v_{2F}^4 v_3^2 v_4^2 \cos(8 \Psi_4) 
\notag\\&+ \frac{12756765}{16} v_{2F}^6 v_3^2 v_4^2 \cos(8 \Psi_4) 
+ \frac{855085}{2} v_{2F}^4 v_4^4 \cos(8 \Psi_4)  + \frac{513051}{4} v_{2F}^6 v_4^4 \cos(8 \Psi_4) 
+ 31031 v_{2F}^6 v_4^3 \cos(12 \Psi_4) 
\notag\\&- 6202980 v_2 v_{2F}^3 v_4 \cos(2 \Psi_2 + 4 \Psi_4)  - 5653445 v_2^3 v_{2F}^3 v_4 \cos(2 \Psi_2 + 4 \Psi_4) 
- 4652235 v_2 v_{2F}^5 v_4 \cos(2 \Psi_2 + 4 \Psi_4) \notag\\
& - \frac{16960335}{4} v_2^3 v_{2F}^5 v_4 \cos(2 \Psi_2 + 4 \Psi_4) 
- \frac{930447}{4} v_2 v_{2F}^7 v_4 \cos(2 \Psi_2 + 4 \Psi_4)  - \frac{3392067}{16} v_2^3 v_{2F}^7 v_4 \cos(2 \Psi_2 + 4 \Psi_4) 
\notag\\&- \frac{27959925}{2} v_2 v_{2F}^3 v_3^2 v_4 \cos(2 \Psi_2 + 4 \Psi_4)  - \frac{83879775}{8} v_2 v_{2F}^5 v_3^2 v_4 \cos(2 \Psi_2 + 4 \Psi_4) 
- \frac{16775955}{32} v_2 v_{2F}^7 v_3^2 v_4 \cos(2 \Psi_2 + 4 \Psi_4)\notag \\
& - \frac{13488615}{2} v_2 v_{2F}^3 v_4^3 \cos(2 \Psi_2 + 4 \Psi_4) 
- \frac{40465845}{8} v_2 v_{2F}^5 v_4^3 \cos(2 \Psi_2 + 4 \Psi_4)  - \frac{8093169}{32} v_2 v_{2F}^7 v_4^3 \cos(2 \Psi_2 + 4 \Psi_4) 
\notag\\&+ \frac{5053125}{4} v_2^2 v_{2F}^4 v_4 \cos(4 \Psi_2 + 4 \Psi_4)  + \frac{3031875}{8} v_2^2 v_{2F}^6 v_4 \cos(4 \Psi_2 + 4 \Psi_4)-\frac{67221}{2} v_2^3 v_{2F}^5 v_4 \cos(6 \Psi_2 + 4 \Psi_4)\notag\\&
-\frac{22407}{8} v_2^3 v_{2F}^7 v_4 \cos(6 \Psi_2 + 4 \Psi_4) + 
11976608 v_2 v_3^2 v_4 \cos(2 \Psi_2 - 6 \Psi_3 + 4 \Psi_4) + 
125754384 v_2 v_{2F}^2 v_3^2 v_4 \cos(2 \Psi_2 - 6 \Psi_3 + 4 \Psi_4)\notag \\&+ 
157192980 v_2 v_{2F}^4 v_3^2 v_4 \cos(2 \Psi_2 - 6 \Psi_3 + 4 \Psi_4) + 
26198830 v_2 v_{2F}^6 v_3^2 v_4 \cos(2 \Psi_2 - 6 \Psi_3 + 4 \Psi_4)\notag \\&- 
\frac{7083153}{16} v_{2F}^5 v_3^2 v_4 \cos(6 \Psi_3 + 4 \Psi_4) - 
\frac{2361051}{64} v_{2F}^7 v_3^2 v_4 \cos(6 \Psi_3 + 4 \Psi_4) + 
\frac{273273}{16} v_2 v_{2F}^6 v_3^2 v_4 \cos(2 \Psi_2 + 6 \Psi_3 + 4 \Psi_4)\notag \\&- 
\frac{2203509}{2} v_2 v_{2F}^5 v_4^2 \cos(2 \Psi_2 + 8 \Psi_4) - 
\frac{734503}{8} v_2 v_{2F}^7 v_4^2 \cos(2 \Psi_2 + 8 \Psi_4) + 
\frac{933933}{32} v_2^2 v_{2F}^6 v_4^2 \cos(4 \Psi_2 + 8 \Psi_4)  \notag\\&-
\frac{30459}{64} v_{2F}^7 v_3^2 v_4^2 \cos(6 \Psi_3 + 8 \Psi_4) - 
\frac{21021}{32} v_2 v_{2F}^7 v_4^3 \cos(2 \Psi_2 + 12 \Psi_4).
\label{eqA3}
\end{align}

\begin{figure}[H]
\centering
\includegraphics[scale=0.4]
{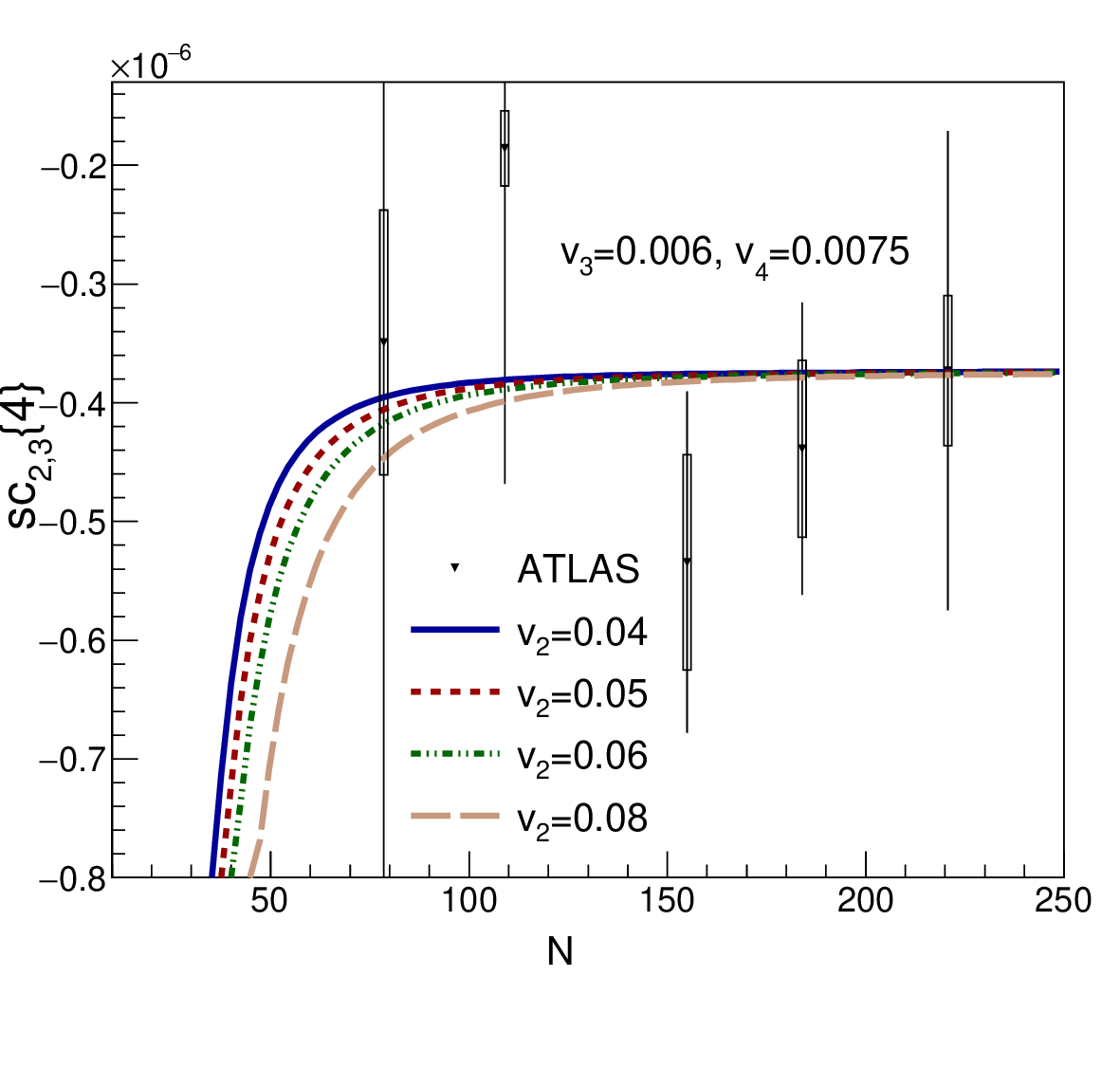}
\includegraphics[scale=0.4]{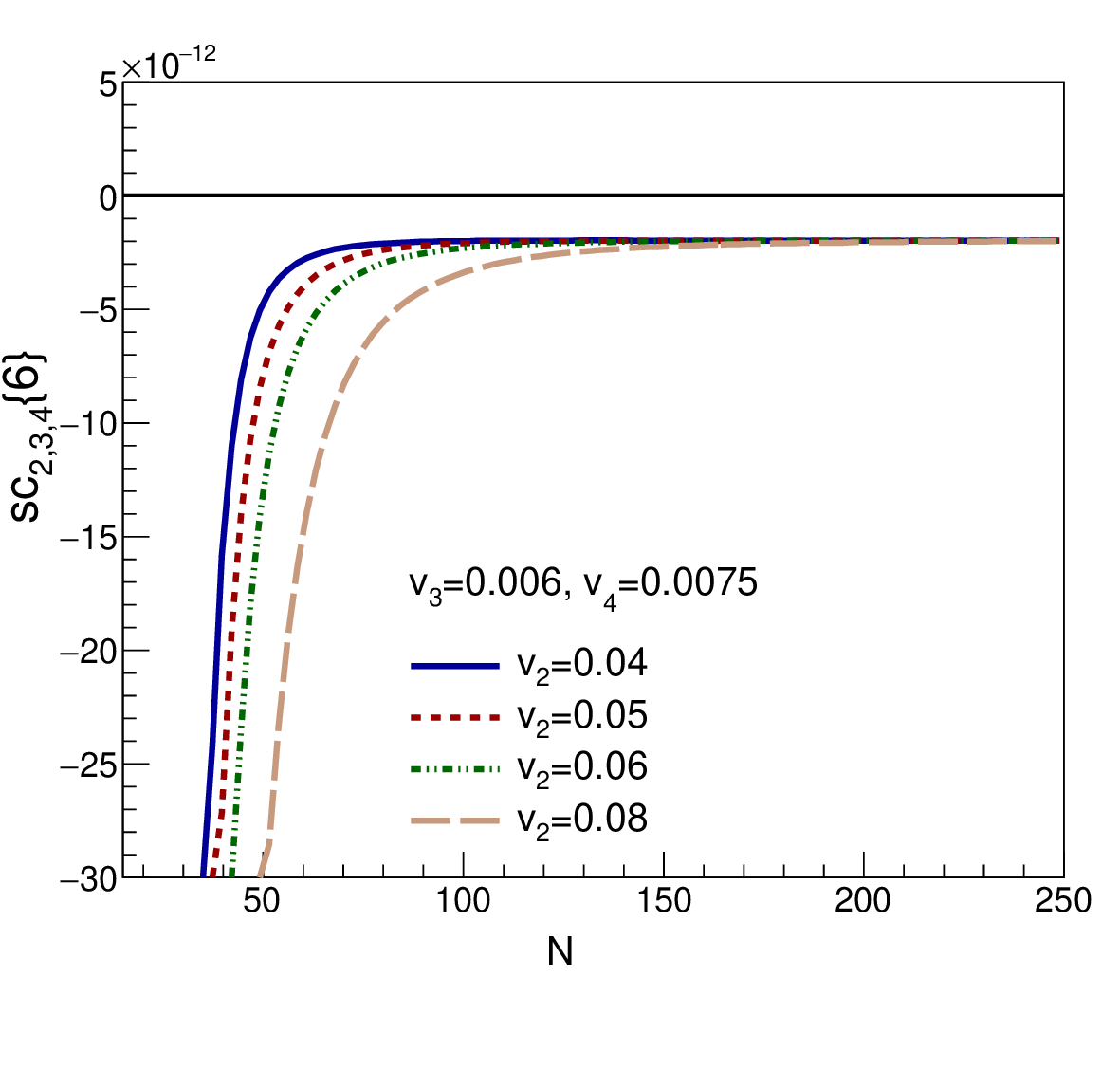}
\includegraphics[scale=0.4]{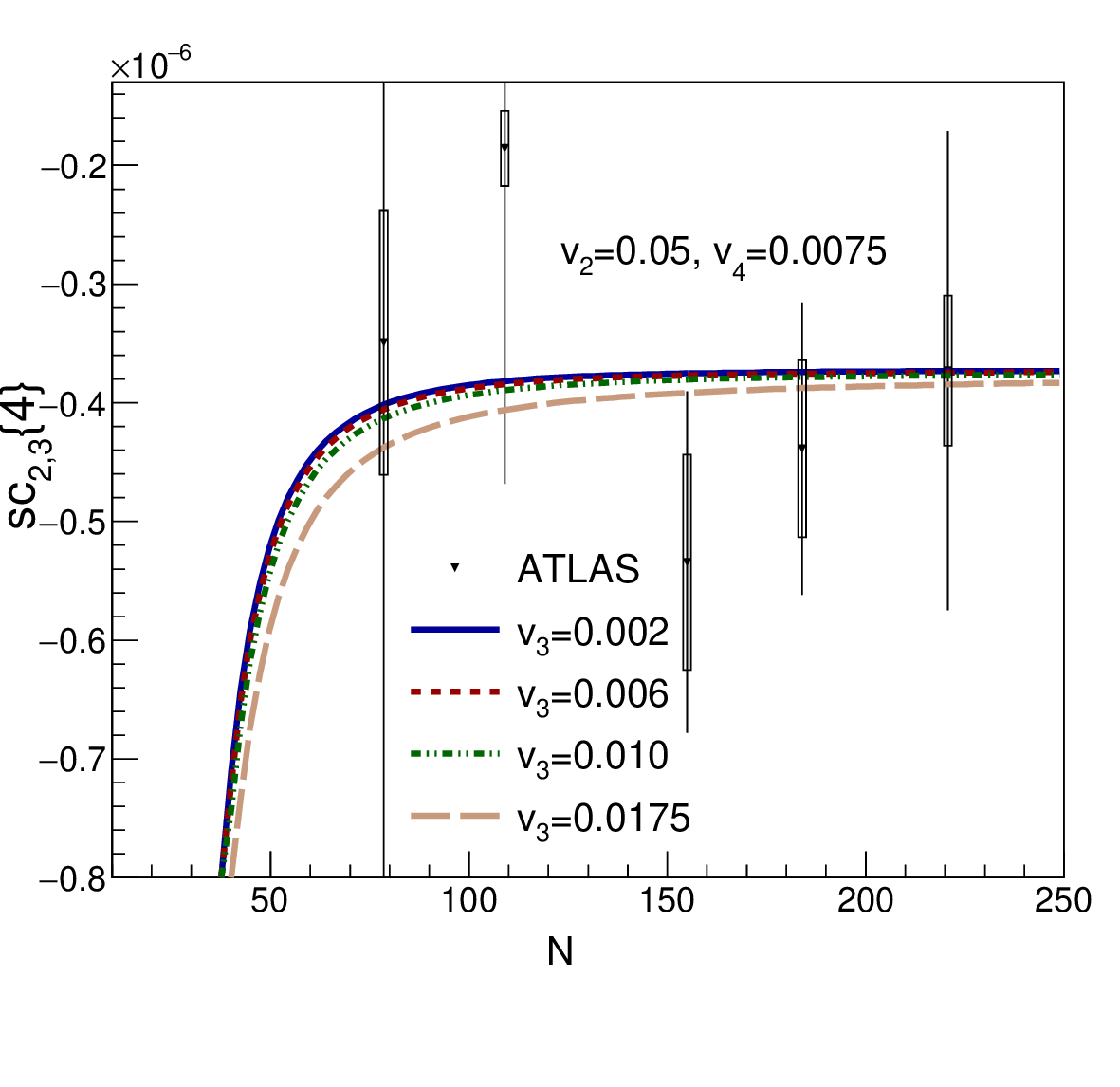}
\includegraphics[scale=0.4]{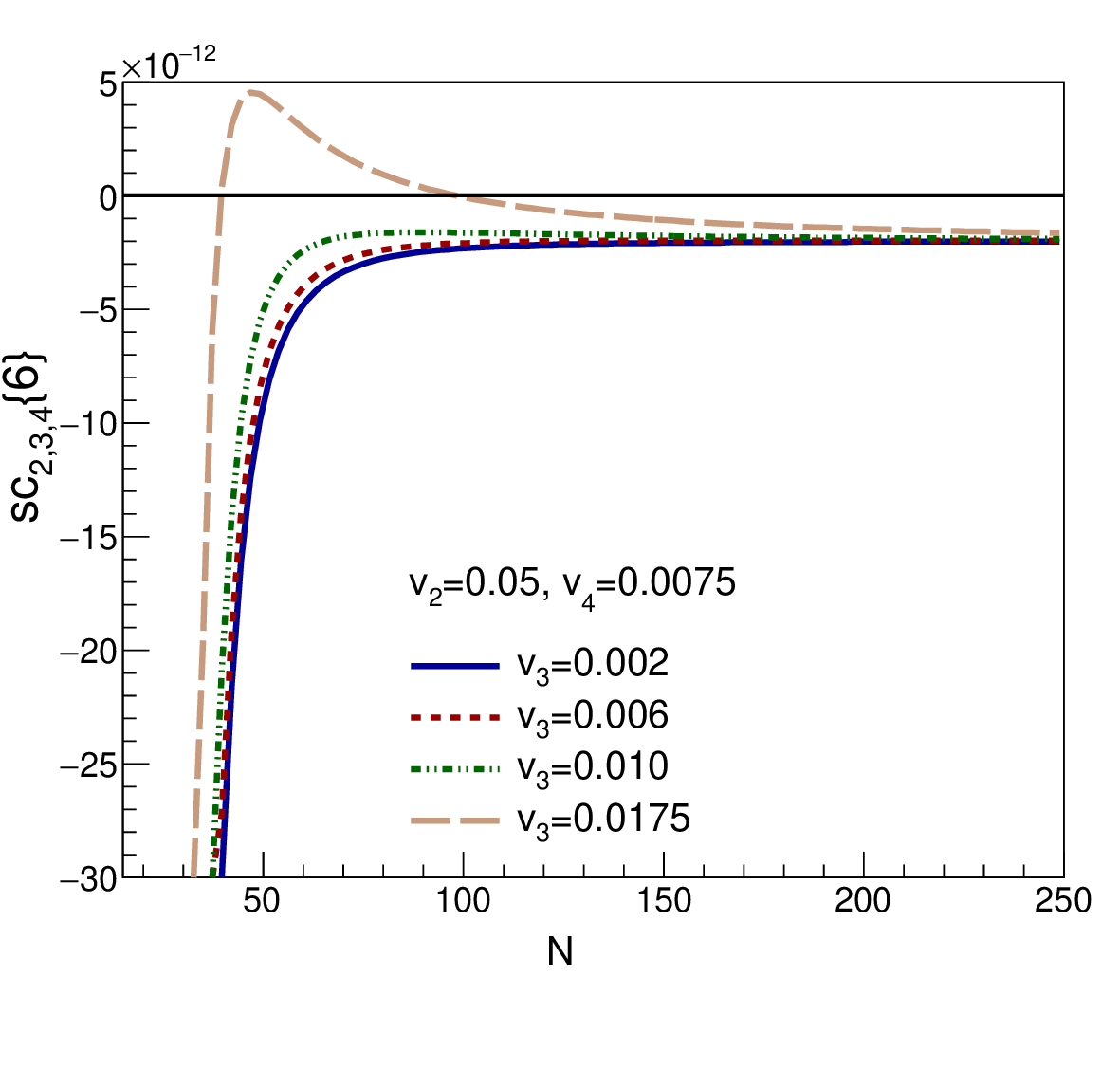}
\includegraphics[scale=0.4]{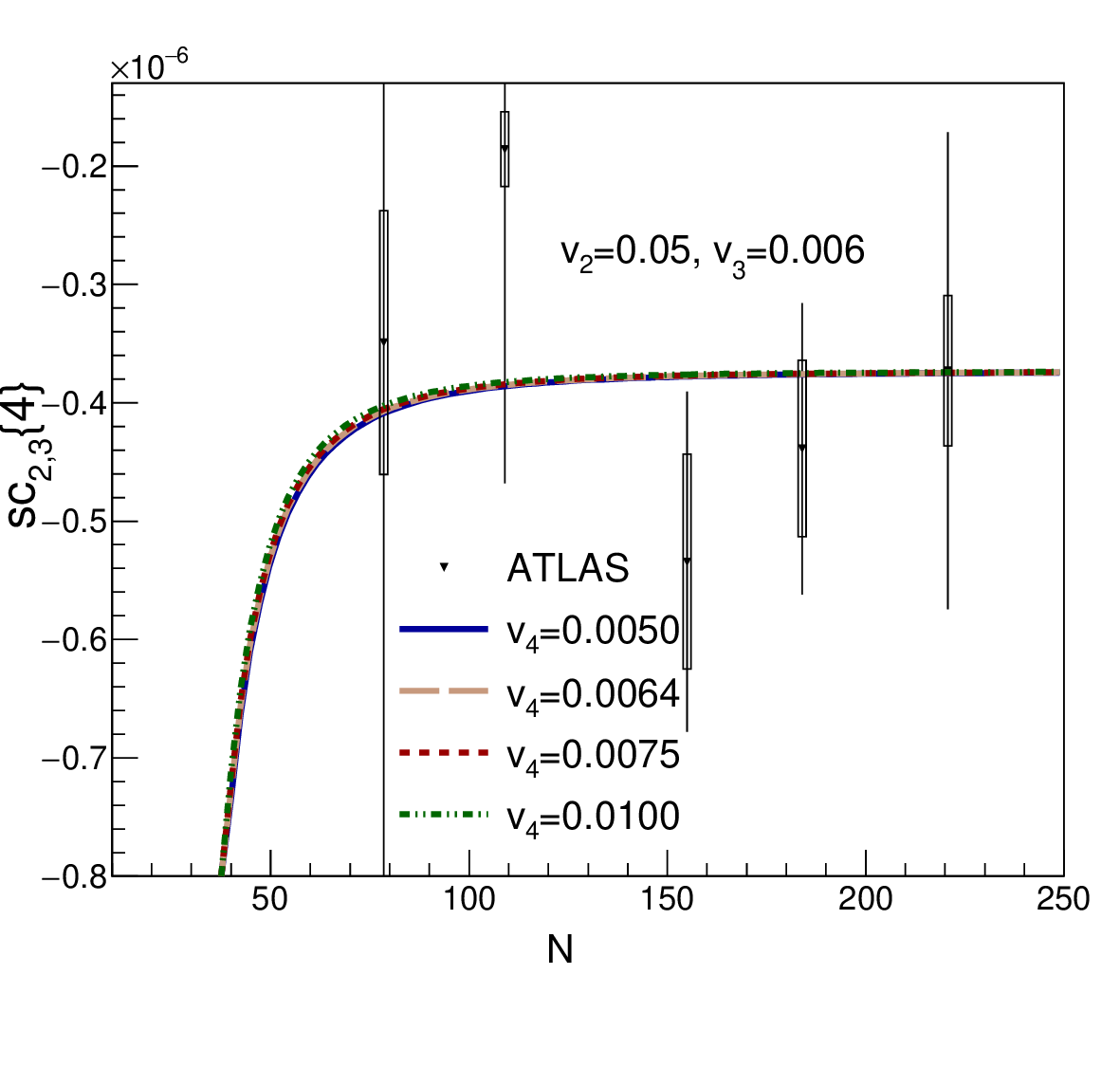}
\includegraphics[scale=0.4]{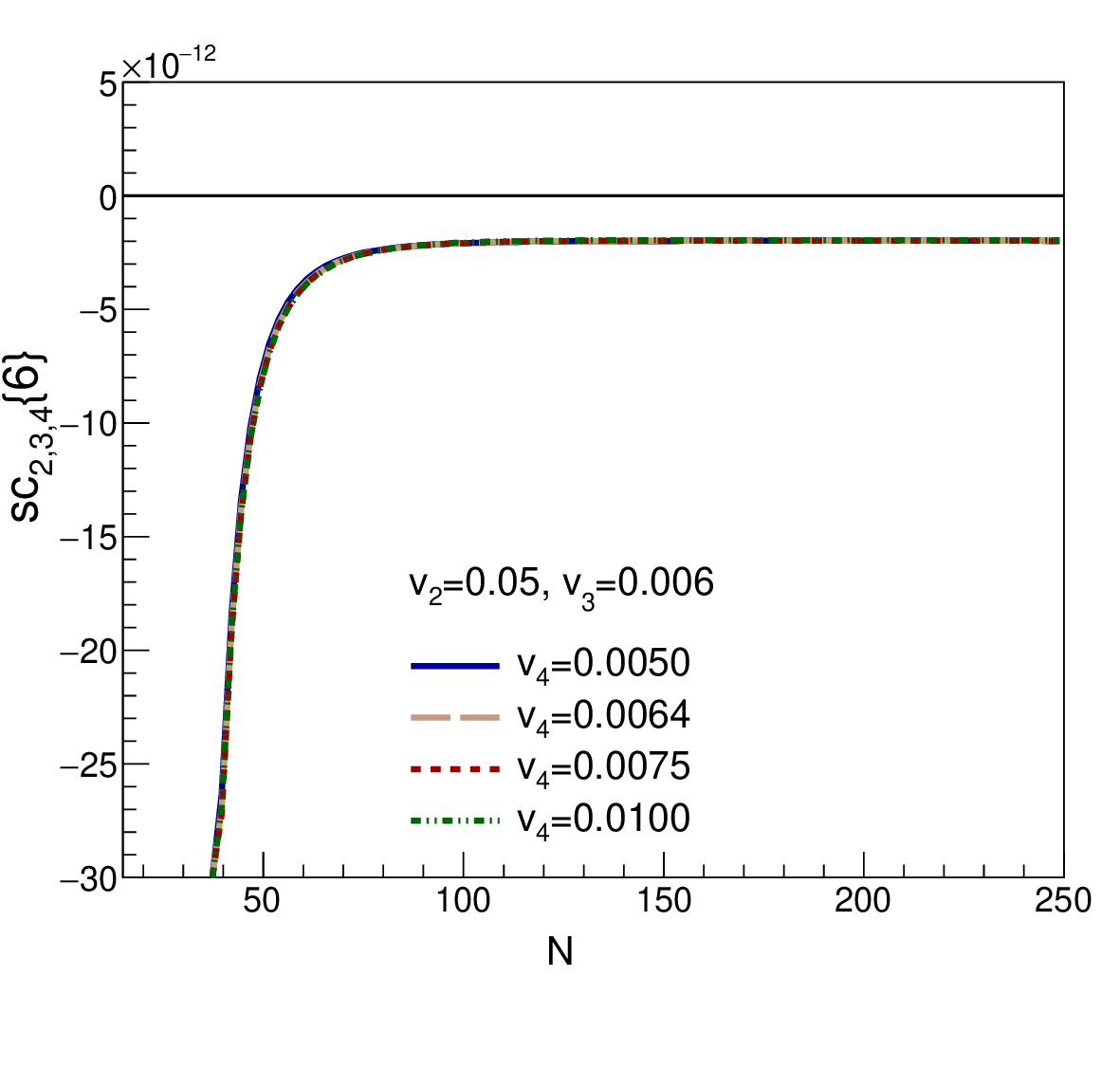}
\vspace{-8mm}
\caption{The symmetric cumulants $sc_{2,3} \left \{ 4 \right \}$ (left column) and $sc_{2,3,4} \left \{ 6 \right \}$ (right column) as a function of the number of particles $N$ for various values of $v_n$ at momentum $p=0.6$ GeV. When varying $v_2$, the values of $v_3$ and $v_4$ are kept constant at 0.006 and 0.0075, respectively. When varying $v_3$, the values of $v_2$ and $v_4$ are kept constant at 0.05 and 0.0075, respectively. When varying $v_4$, the values of $v_2$ and $v_3$ are kept constant at 0.05 and 0.006, respectively. The ATLAS data for $0.3< p_{T} < 3$ GeV in \textit{p}+\textit{p} collisions at 13 TeV using the four-subevent cumulant method are shown for comparisons, where the error bars and boxes represent the statistical and systematic uncertainties, respectively \cite{data1}.} 
\label{changevn}
\end{figure}
\begin{figure}[H]
\centering
\includegraphics[scale=0.4]
{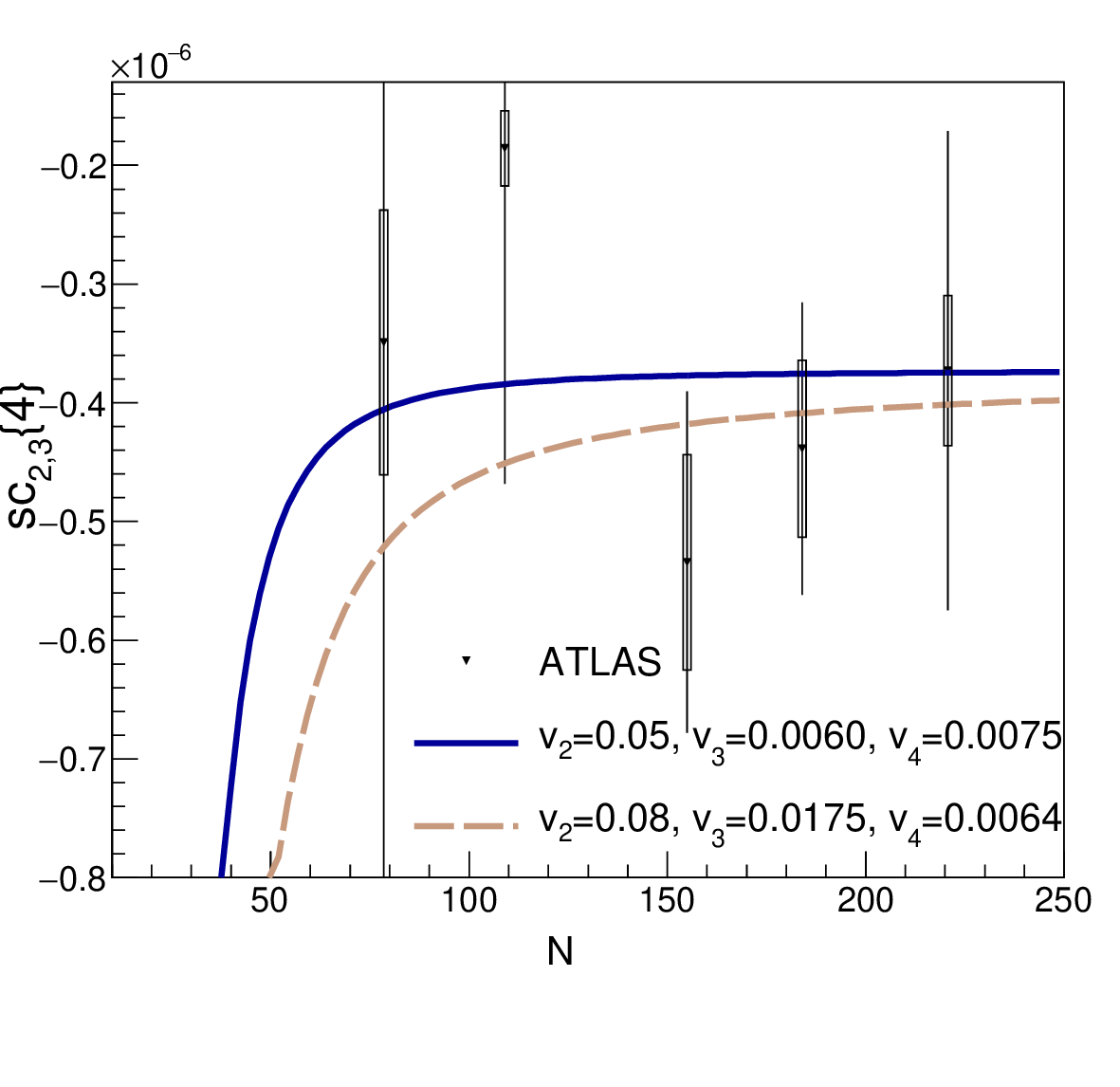}
\includegraphics[scale=0.4]
{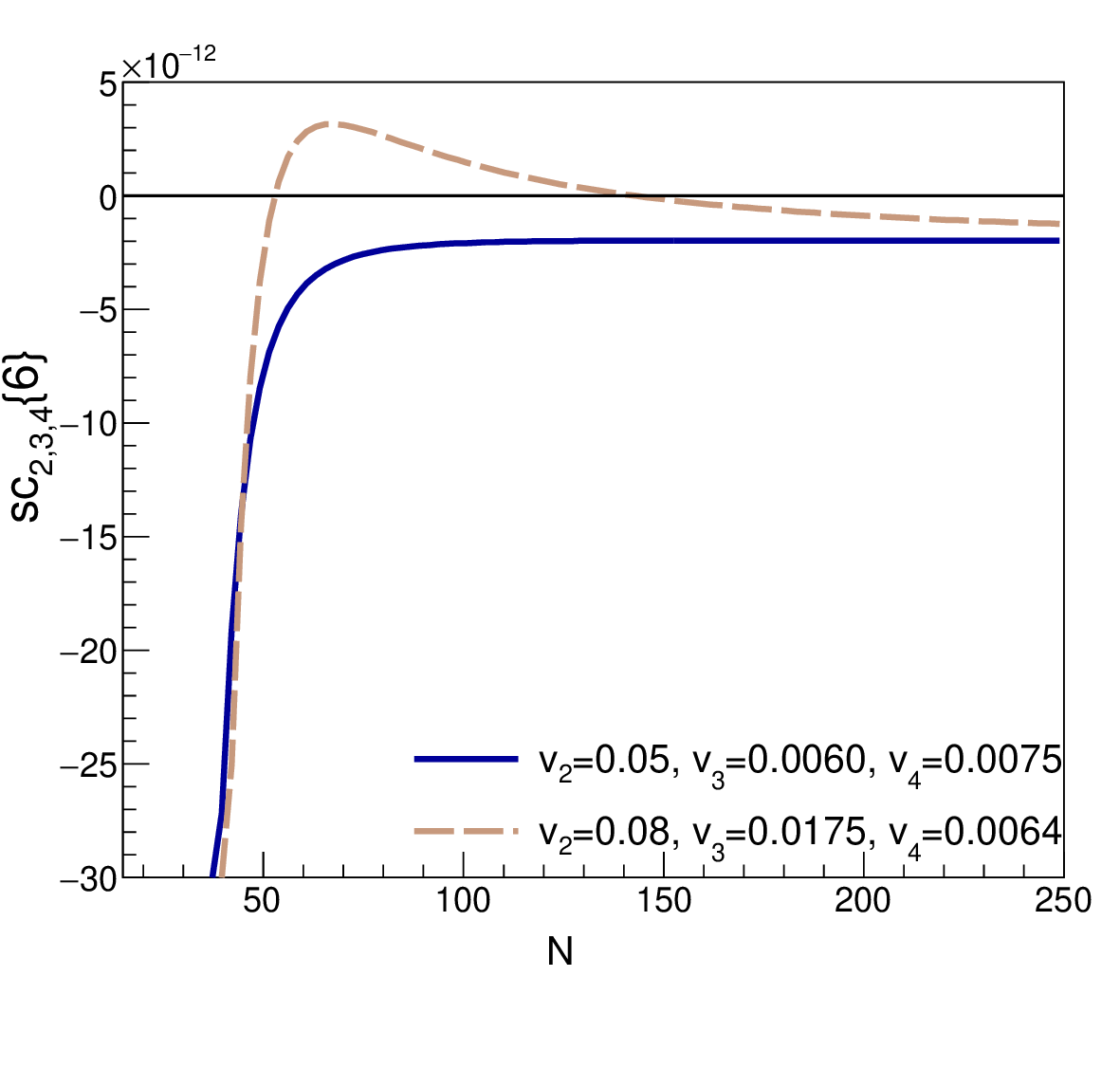}
\caption{The symmetric cumulants $sc_{2,3} \left \{ 4 \right \}$ (left) and $sc_{2,3,4} \left \{ 6 \right \}$ (right) as a function of the number of particles $N$ for various values of $v_n$ at momentum $p=0.6$ GeV. The ATLAS data for $0.3< p_{T} < 3$ GeV in \textit{p}+\textit{p} collisions at 13 TeV using the four-subevent cumulant method are shown for comparisons, where the error bars and boxes represent the statistical and systematic uncertainties, respectively \cite{data1}.}
\label{newold}
\end{figure}
In Fig.~\ref{changevn}, the values for the solid line, dash line, and dot-dash line are taken from the lower limit, intermediate value, and upper limit of the experimental results of the \textit{p}+\textit{p} system, respectively, while the value of the long-dash line is derived from the experimental results of the \textit{p}+Pb system \cite{vnATLAS}. When varying $v_2$, the values of $v_3$ and $v_4$ are kept constant at 0.006 and 0.0075, respectively. When varying $v_3$, the values of $v_2$ and $v_4$ are kept constant at 0.05 and 0.0075, respectively. When varying $v_4$, the values of $v_2$ and $v_3$ are kept constant at 0.05 and 0.006, respectively. It can be seen that for $sc_{2,3} \left \{ 4 \right \}$, the change of parameters does not affect the fitting of experimental data, but for $sc_{2,3,4} \left \{ 6 \right \}$, the increase of $v_3$ will lead to a sign change from negative to positive. In Fig.~\ref{newold}, we present the results of $sc_{2,3} \left \{ 4 \right \}$ and $sc_{2,3,4} \left \{ 6 \right \}$ under two different sets of $v_n$ parameters. The parameter values for the solid line are derived from experimental results in \textit{p}+\textit{p} system, while the parameter values for the long-dash line are from experimental results in \textit{p}+Pb system \cite{vnATLAS}.

\bibliography{ref}
\end{document}